\documentclass[twocolumn,prb,superscriptaddress]{revtex4}

\usepackage{graphicx}
\usepackage{bm}

\begin{document}

\title{Complete Spin and Valley Polarization by Total External Reflection
  from Potential Barriers in Bilayer Graphene and Monolayer
  Transition Metal Dichalcogenides
}

\author{P. A. Maksym}
\affiliation{Department of Physics, University of Tokyo, Hongo, Tokyo
 113-0033, Japan}
\affiliation{School of Physics and Astronomy, University of Leicester, 
Leicester LE1 7RH, UK}

\author{H. Aoki}
\affiliation{Department of Physics, University of Tokyo, Hongo, Tokyo
 113-0033, Japan}
\affiliation{National Institute of Advanced Industrial Science and
  Technology (AIST), Tsukuba 305-8568, Japan}

\date{\today}

\begin{abstract}
It is shown that potential barriers in bilayer graphene (BLG) and monolayer
transition metal dichalcogenides (TMDs) can split a valley unpolarized
incident current into reflected and transmitted currents with opposite
valley polarization. Valley asymmetric transmission inevitably occurs
because of the low symmetry of the total Hamiltonian and when total
external reflection occurs the transmission is 100\% valley polarized in
BLG and 100\% spin and valley polarized in TMDs, except for exponentially
small corrections. By adjusting the potential, 100\% polarization can be
obtained regardless of the crystallographic orientation of the barrier. A
valley polarizer can be realized by arranging for a collimated beam of
carriers to be incident on a barrier within the range of angles for total
external reflection. The transmission coefficients of barriers with a
relative rotation of $\pm\pi/3$ are related by symmetry. This allows two
barriers to be used to demonstrate that the current is valley polarized. A
soft-walled potential is used to model the barrier and the method used to
find the transmission coefficients is explained. In the case of monolayer
TMDs, a 4-band $\mathbf{k}\cdot\mathbf{p}$ Hamiltonian is used and the
$\mathbf{k}\cdot\mathbf{p}$ parameters are obtained by fitting to
\textit{ab-initio} band structures.
\end{abstract}

\maketitle

\section{Introduction}
\label{IntroSection}

One of the important objectives of valleytronics
\cite{Vitale18,Amet15,Bussolotti18,Schaibley16,Shkolnikov02,Gunawan06,Takashina06,Zhu12}
is to generate and detect valley polarized currents, that is currents
restricted to one valley of a two-valley material. There are many proposals for
electrical control of valley polarization in 2D materials
\cite{Rycerz07, Xiao07, Garcia08, Abergel09, Pereira09, Schomerus10,
  Park12, Park15, Chen16, Cresti08, Tkachov09, Gunlycke11, Wu11, Koshino13}
but fabrication of the necessary devices remains a challenge. This work is
about an alternative approach which may be easy to realize as it only
depends on components that have already been demonstrated. The idea is to
arrange for carriers in one valley to be completely reflected from a
potential barrier while carriers in the other valley are transmitted. This
can be used to realize a valley polarizer in bilayer graphene (BLG) and
a spin and valley polarizer in transition metal dichalcogenides (TMDs).

Existing proposals for valley polarizers are difficult to realize because
they require structures with precise crystallographic orientation and in
some cases very small size. In addition, the proposed designs typically
separate the current into two streams whose direction is valley dependent.
However a polarizer should produce one output stream and it is necessary
to find a way of collecting the desired one. This is complicated by the
strong trigonal warping of the constant energy contours in many 2D materials.

The difficulty is that the outgoing current streams typically emerge from a
system of gates and because of trigonal warping the stream directions depend
on the crystallographic orientation of the gates \cite{OrientationExample}.
However existing fabrication methods for 2D material based devices do not
allow the crystallographic orientation of the 2D material to be controlled.
Hence the gate orientation is unknown and in effect random. This means
the desired current stream is difficult to collect because its direction
is also unknown.

This difficulty does not arise in our approach because there is only one
current stream on the output side of the device. The main idea is to use
total external reflection to ensure that carriers in the undesired valley
are reflected from the incident side of a potential barrier. This results
in one current stream of practically 100\% polarization in the desired
valley on the transmitted side of the barrier.

Total external reflection occurs when there are propagating waves on one
side of an interface but only evanescent waves occur on the other
side. This situation occurs only in a certain range of incidence angles and
when the energy contours are warped, this range is different in the two
valleys. This results in a large angular region where the reflection is
practically 100\% in one of the valleys. When valley unpolarized carriers
are incident within this region, the transmitted current is 100\% valley
polarized except for an exponentially small correction due to quantum
tunneling. The valley polarization is insensitive to the barrier
orientation because the barrier potential can be adjusted to optimize the
region width.

These effects enable a valley polarizer to be realized by using an electron
collimator to produce an incident current stream centered on the angular
region required for total external reflection. The necessary collimator has
been fabricated in monolayer graphene (MLG) \cite{Barnard17} and its
beamwidth is similar to the angular widths of total external reflection
regions in BLG and TMDs. In addition, the barrier can be realized with a
structure similar to a FET. Hence the polarizer can be assembled from
existing components.

In monolayer semiconducting TMDs strong spin-orbit (SO) coupling ensures that
states of the same energy in opposite valleys have opposite spin. 
Consequently a valley polarizer made from a monolayer semiconducting TMD
is also a spin polarizer.

While this work is centered on the large valley asymmetry that results from
total external reflection, valley asymmetric transmission itself is
inevitable because of the low symmetry of the Hamiltonian. For most barrier
orientations time reversal is the only symmetry of the total Hamiltonian of
the barrier and 2D material. This has the consequence that the barrier
transmission coefficient is valley asymmetric but special conditions, for
example total external reflection, are needed to make the asymmetry large.

The symmetry properties of the transmission coefficient are also relevant to
detection of valley polarization. Although the Hamiltonian at an arbitrary
barrier orientation has low symmetry, the trigonal symmetry of the constant
energy contours leads to symmetry relations between the transmission
coefficients for barriers of different orientation. The most important one
is that the transmitted valley swaps when a barrier is rotated by $\pm
\pi/3$. This means two valley polarizers may be used to demonstrate 
valley polarization in the same way that crossed Polaroid filters are
used to demonstrate polarization of light.

In summary, the objectives of this work are first, to show that valley
asymmetric transmission is a consequence of the low symmetry of the total
Hamiltonian. Secondly, to show that the barrier transmission coefficient in
the regime of total external reflection exhibits large valley asymmetry in
both BLG and TMDs. Thirdly, to show that it should be feasible to use this
effect to realize a valley polarizer and finally to show that it should be
feasible to demonstrate valley polarization with a crossed pair of valley
polarizers.

Existing work on valley polarization in 2D materials started with
pioneering theoretical studies of valleytronics in MLG
\cite{Rycerz07, Xiao07}. Subsequently, ways of realizing a valley polarizer
were explored theoretically in a wide range of geometries in BLG and MLG, refs.
\cite{Garcia08, Abergel09, Pereira09, Schomerus10, Park12, Park15, Chen16, Cresti08, Tkachov09, Gunlycke11, Wu11, Koshino13}
for example. Other valley dependent effects are also known
\cite{Low10, Peterfalvi12, Zhang13}. In addition experimental studies of
valley polarization in BLG have been published recently \cite{Li20, Gold21}
and steps have been taken towards realizing a valley polarizer \cite{Chen20}.
In TMDs, optically induced valley polarization has been achieved in MoS$_2$
\cite{Mak14} and valley-sensitive photocurrents have been observed
\cite{Zhang19}. In addition there are theoretical predictions of
spin-dependent refraction at domain boundaries \cite{Habe15} and small
valley polarization in crystallographically oriented potential barriers
\cite{Hsieh18}. However total external reflection in graphene and TMDs
has not been investigated. The present work centers on BLG and TMDs where
the predicted effects should be easy to observe. Higher energies and
potentials would be needed in the case of MLG \cite{Pereira09, Chen16}.

This paper begins with an outline of the physics
(Section \ref{OverviewSection}) where we explain why total external
reflection is valley asymmetric and give examples of
valley asymmetric transmission in BLG and TMDs. Next we show that the
valley asymmetric transmission is a consequence of the low
symmetry of the total Hamiltonian
(Section \ref{TheorySection}). This requires a careful discussion because
the velocity and momentum of the carriers are not parallel when trigonal
warping occurs. We consider incident carriers selected by both velocity and
momentum and show that valley asymmetric transmission occurs in both cases.
Symmetry relations between the transmission coefficients of barriers of
different orientation are also derived in this section. This is followed by
an outline of the numerical methods used in this work
(Section \ref{NumSection}). Valley asymmetric transmission is detailed in
sections \ref{BLGSection} (BLG) and \ref{TMDSection} (TMDs). In the BLG
section we first explain the device model used to obtain a realistic
barrier potential, then show that potential can be adjusted to
make the width of the single valley region large for all barrier
orientations and finally discuss experimental feasibility. The TMD
section begins with an explanation of the fitting procedure used to
obtain a $\mathbf{k}\cdot\mathbf{p}$ Hamiltonian that reproduces the
trigonal warping in \textit{ab-initio} band structures. The remaining
discussion parallels that for BLG. The feasibility of realizing a valley
polarizer is examined in section \ref{PolariserSection} with the conclusion
that it should be possible provided that trigonal warping is strong enough
and the device can be operated in the ballistic transport regime. An
example of polarization detection with two crossed polarizers is given
in section \ref{DetectionSection} and our results are summarized and
discussed in section \ref{DiscussionSection}.

\section{Valley polarization by total external reflection}
\label{OverviewSection}

\begin{figure}
\begin{center}  
  \includegraphics[width=8.5cm]{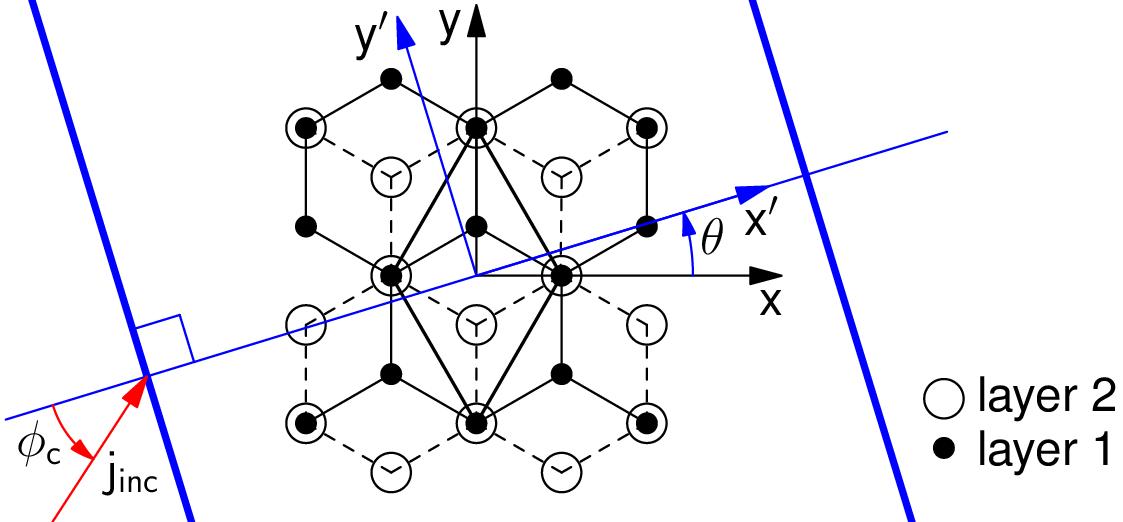}\\[2ex]
  \includegraphics[width=4.2cm]{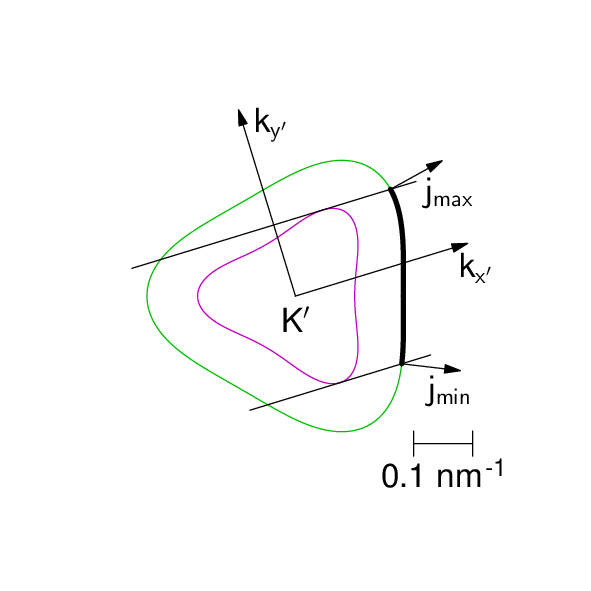}
  \includegraphics[width=4.2cm]{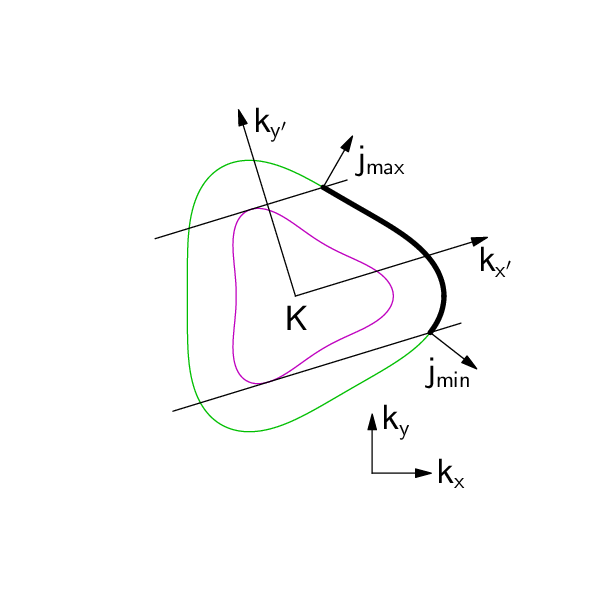}
  \caption{(Color online). Top: plan view of bilayer graphene and top
    gate (schematic). Bold lines (blue):
    gate edges; medium lines: unit cell; feint lines: bonds;
    open and filled circles: carbon atoms. In-plane current (\textsf{j})
    is incident at angle $\phi_c$. The axis rotation angle,
    $\theta = 17^\circ$. Bottom: scale drawing of
    energy contours and critical angles. Light outer line:
    contact contour (green); dark inner line: barrier contour (magenta).
    States on the bold part of each contact contour are transmitted.
    Arrows normal to the contours: critical current directions.
    The $KK'$ distance is reduced so all the contours fit into one
    figure. }
\label{axisfig}
\end{center}
\end{figure}

Fig.~\ref{axisfig} illustrates how total external reflection and trigonal
warping lead to valley polarized transmission through a potential barrier.
The top part of the figure shows a potential barrier that is generated
by a uniform bottom gate and a finite-width top gate. The top gate is
rotated by an angle $\theta$ relative to the crystallographic co-ordinates,
$x, y$. The external potential is expressed in co-ordinates, $x', y'$,
fixed to the gate and is taken to be independent of $y'$. The crystal
structure is that of BLG; the geometry is similar in the case of TMDs.

The bottom part of the figure shows constant energy contours inside the
barrier and the contact regions outside the barrier. The barrier contours
are inside the contact contours because the potential in the barrier is
higher than in the contacts so the band energy there is lower and so is the
wave number, $k$. However $k_{y'}$, the component of $\mathbf{k}$ parallel
to the barrier, is a conserved quantity that is identical in the contacts
and barrier.

The states in the barrier may be propagating or evanescent. Propagating
states occur only in the $k_{y'}$ range delimited by the lines that are
tangential to the barrier contour and parallel to the $k_{x'}$ axis. In
this range there are two real $k_{x'}$ values at each $k_{y'}$ but outside
the range all the $k_{x'}$ values are complex and all the barrier states
are evanescent. Then the current through the barrier is limited by
tunneling and can be made exponentially small by making the barrier width
sufficiently large. This is the regime of total external reflection.

Because of trigonal warping, the critical angles for total
external reflection are very different in the two valleys. Propagating
barrier states occur only when the contact states are in the limited range
indicated by the bold lines in Fig.~\ref{axisfig}. The critical angles for
total external reflection occur at the ends of this range. The current
carried by a contact state with wave vector $\mathbf{k}$ is in the
direction normal to the contact contour. The normals are shown in the
figure and it is clear that the critical angles are very different in the two
valleys.

\begin{figure}
  \begin{center}
  \includegraphics[width=4.4cm, angle=-90]{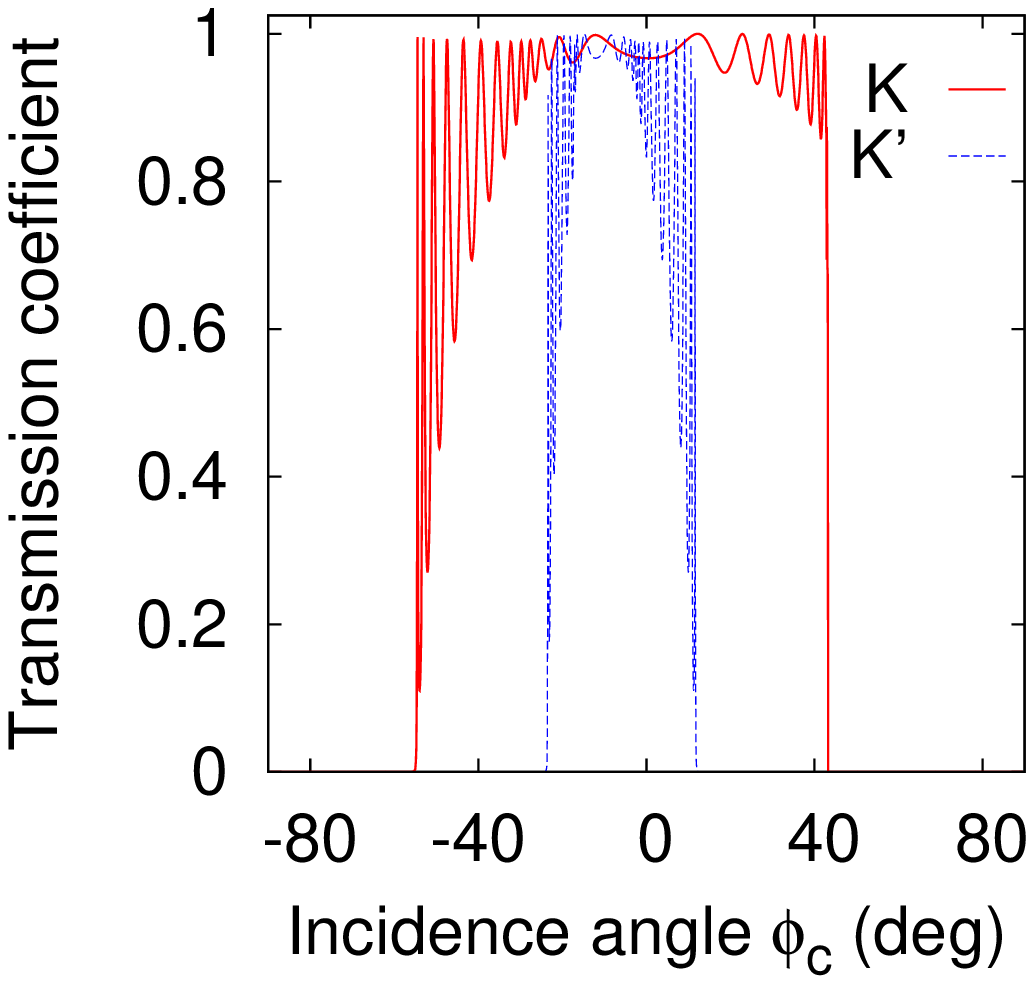}\hspace{1mm}
  \includegraphics[width=4.4cm, angle=-90]{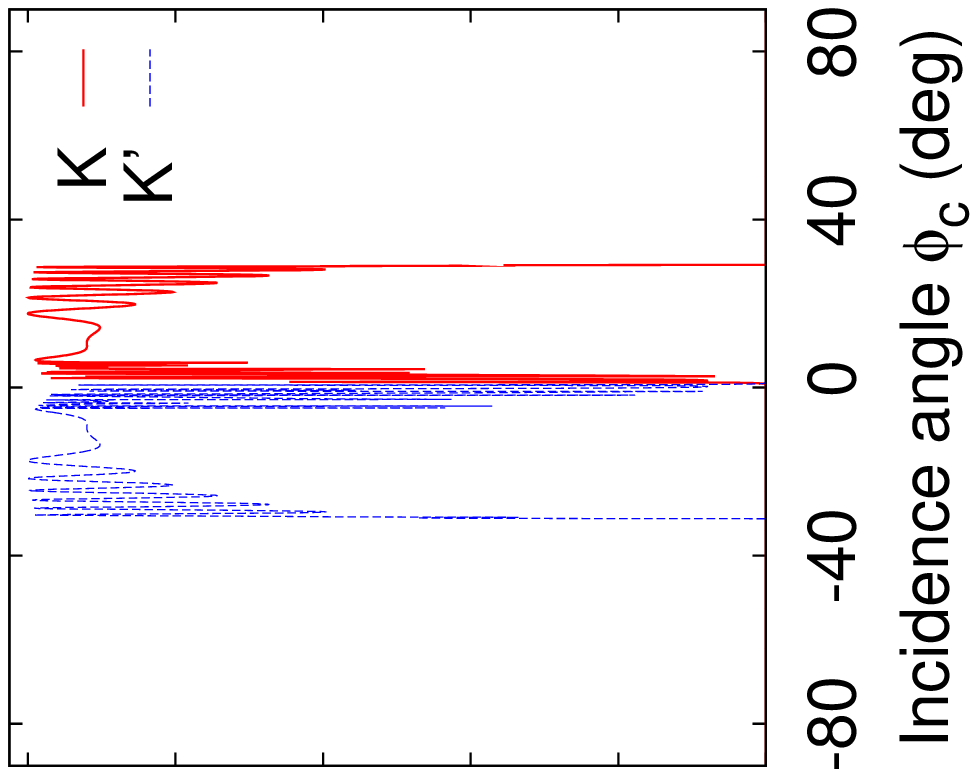}
  \caption{(Color online). Transmission coefficient for a potential
    barrier in BLG. The electron energy
    is 56 meV and the top gate width is 300 nm.
    Left: same-valley case, $V_i + U_i = $
    47.61 meV in layer 1, 5.586 meV in layer 2, $\theta = 17^\circ$.
    Right: different-valley case, $V_i + U_i = $
    53.51 meV in layer 1, 9.124 meV in layer 2, $\theta = 31^\circ$.
    See section \ref{BLGSection} for details of potential.}
\label{bgTfig}
\end{center}
\end{figure}

\begin{figure}
\begin{center}  
  \includegraphics[width=4.4cm, angle=-90]{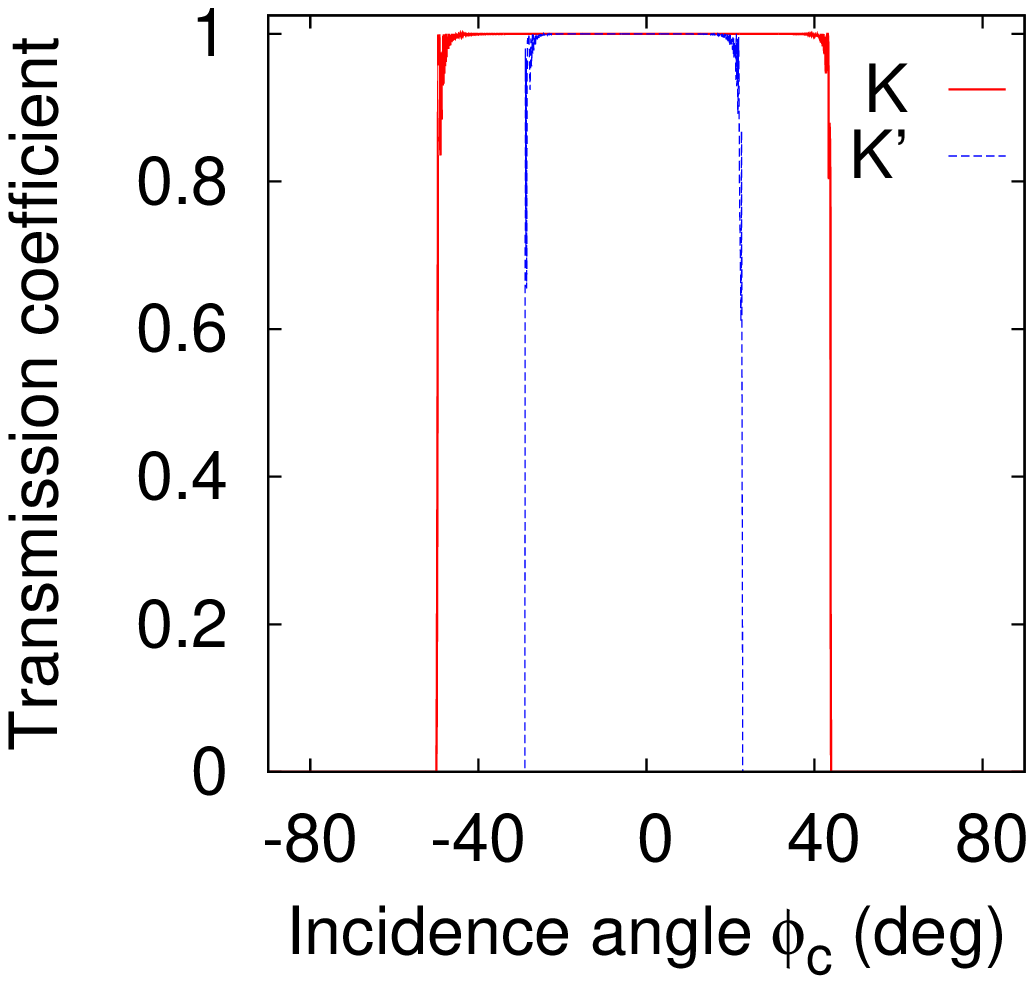}\hspace{1mm}
  \includegraphics[width=4.4cm, angle=-90]{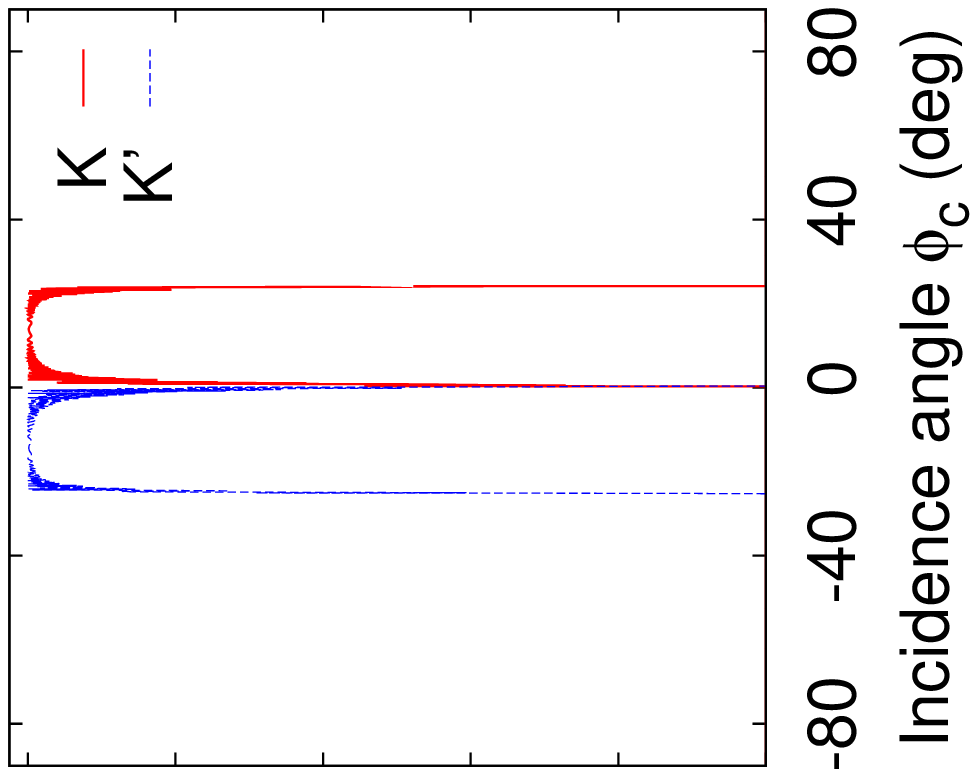}
  \caption{(Color online). Transmission coefficient for a potential
    barrier in MoTe$_2$.
    The hole energy is 116.9 meV and the top gate width is 300 nm.
    Left: same-valley case, $V = 66.55$ meV, $\theta = 17^\circ$.
    Right: different-valley case, $V = 106.7$ meV, $\theta = 31^\circ$.
  See section \ref{TMDSection} for details of potential.}
\label{tmdTfig}
\end{center}
\end{figure}

The transmission coefficient as a function of current incidence angle is
shown in Fig.~\ref{bgTfig} for electrons in BLG and Fig.~\ref{tmdTfig} for
holes in MoTe$_2$. In both cases the transmission coefficient approaches
zero rapidly near the critical angles for total external reflection. This
results in wide angular regions where the transmission coefficient is $\simeq
1$ in one valley and $\simeq 0$ in the other valley. If a collimated, valley
unpolarized beam of carriers is incident on a barrier in one of these
regions, the carriers in one valley are fully reflected while those in the
other valley are transmitted. Hence a valley polarized current emerges on
the exit side of the barrier.

In general, there is one region of single valley transmission at positive
angles of incidence and one at negative angles of incidence. These regions
may be in the same valley or in different valleys. Figs.~\ref{bgTfig} and
~\ref{tmdTfig} show examples of both cases. No critical energy or potential
is needed to observe these regions. They occur over a wide range of
potentials and energies (Figs.~\ref{optvfig}, \ref{allsvtfig},
\ref{bgendepfig} and \ref{tmdendepfig}) and the top gate voltage can be
adjusted so they are observable at all crystallographic orientations of
the barrier (Figs.~\ref{svtfig} and \ref{tmdsvtfig}).

\section{Theory}
\label{TheorySection}

\subsection{Overview}
\label{OvthSection}

In this section we show that valley asymmetric transmission is
\textit{inevitable} in the presence of trigonal warping and low symmetry.
The effect of trigonal warping on transmission has been investigated for
some special cases \cite{Peterfalvi12,Park12} but the properties we need,
particularly the role of points of inflection, have not been detailed and
we derive them from first principles. We then show how the low symmetry of
a potential barrier oriented at an arbitrary angle to the crystallographic
axes inevitably leads to valley asymmetric transmission.  Finally we show
that the trigonal symmetry of the energy contours leads to useful symmetry
relations between the transmission coefficients of barriers at different
orientations.

To obtain the transmission coefficient theoretically one has to specify the
direction of incidence. Experimentally, the incident particles can be
selected by velocity or momentum but when the energy contours are warped,
these vectors are not parallel. Hence the theoretical direction of
incidence must be chosen to match the expected experimental
conditions. Throughout this work the incident beam is taken to be
collimated, i.e. selected by velocity, and the direction of incidence is
specified by the polar angle of the incident current, $\phi_c$
(Fig.~\ref{axisfig}). However in this section we show that valley
asymmetric transmission is inevitable regardless of whether the incident
particles are selected by velocity or momentum.

This conclusion follows from straightforward but lengthy analysis. We state
the Hamiltonians for BLG and TMDs in section \ref{HSection}, then detail
the symmetry properties of plane wave states (\ref{PWSection}) and currents
(\ref{JSection}). Next (\ref{MultipleJSection}) we show how to find all the
plane wave states that contribute to the incident current at angle $\phi_c$
(there may be more than one when there are points of inflection on the
energy contour). In section \ref{vaphicSection} we show that $k_{y'}$
expressed as a function of $\phi_c$ is not the same in each valley and
hence that the transmission coefficient expressed as a function of $\phi_c$
is valley asymmetric. Then (\ref{vakySection}) we prove that the
transmission coefficient expressed as a function of $k_{y'}$ is valley
asymmetric because of the low symmetry of the total Hamiltonian.  Finally
we detail the symmetry relations between the transmission coefficients of
barriers at different orientations (\ref{symrelSection}).

\subsection{Hamiltonians}
\label{HSection}

The total Hamiltonians in each valley are obtained from band Hamiltonians
in the literature by rotating co-ordinates anti-clockwise by
an angle $\theta$ and applying a unitary transformation that reduces the
$\theta$ dependence to factors of the form $\exp(\pm 3i\theta)$.

In the case of BLG and the $K$-valley, the unitary transformation is
$\mathrm{diag}(e^{-i\theta}, 1, 1, e^{i\theta})$ and the band Hamiltonian in
ref. \cite{McCann13} becomes
\begin{equation}
  H_{0K} = 
    \left( \begin{array}{cccc}
    U_1 & v_0\pi_K^\dagger & -v_4\pi_K^\dagger  & v_3\pi_K e^{3i\theta}\\
    v_0\pi_K & U_1 + \Delta' & t & -v_4\pi_K^\dagger \\
    -v_4\pi_K  & t & U_2  + \Delta' & v_0\pi_K^\dagger\\
    v_3\pi_K^\dagger e^{-3i\theta} & -v_4\pi_K & v_0\pi_K & U_2 \\
    \end{array} \right), \label{hblg}
 \end{equation}
where $\pi_K = p_{x'} + ip_{y'}$, $p_{x'}$ and $p_{y'}$ are momentum
components and the parameter values are detailed in section
\ref{BLGSection}. The spatially uniform potentials $U_i$ in layer $i$
result from the bottom gate bias and the total Hamiltonian, $H_K=
H_{0K}+V$, where $V=\mathrm{diag}(V_1(x'), V_1(x'), V_2(x'), V_2(x'))$
and $V_i$ is the top gate bias in layer $i$. In $K'$, $\pi_K$ is replaced
by $\pi_{K'} \equiv -p_{x'} + ip_{y'}$ and $\theta$ by $-\theta$.

In the case of semiconducting TMDs, the band Hamiltonian for one
monolayer is obtained by applying the unitary transformation 
$\mathrm{diag}(1, e^{i\theta}, e^{-i\theta}, e^{-i\theta})$ to
the Hamiltonian in ref. \cite{Kormanyos13}. This gives
\begin{equation}
  H_{0Ks_z} = 
    \left( \begin{array}{cccc}
    \epsilon_v +  \lambda s_z & c_3\pi_K^\dagger & c_2\pi_K  & c_4\pi_K\\
    c_3\pi_K & \epsilon_c & c_5\pi_K^\dagger e^{-3i\theta} & c_6\pi_K^\dagger e^{-3i\theta}\\
    c_2\pi_K^\dagger  & c_5\pi_K e^{3i\theta} & \epsilon_{v-3} & 0\\
    c_4\pi_K^\dagger & c_6\pi_K e^{3i\theta} & 0 & \epsilon_{c+2} -\lambda s_z\\
    \end{array} \right), \label{htmd}
 \end{equation}
where the spin index, $s_z=\pm 1$. The band edge energies are $\epsilon_v$
in the valence band,
$\epsilon_c$ in the conduction band, $\epsilon_{v-3}$ three bands below the
valence band  and $\epsilon_{c+2}$ two bands above the conduction band.
$c_i = \gamma_i / \hbar$ and the $\gamma_i$ are parameters defined in
ref. \cite{Kormanyos13}. The SO splitting of the
valence band is $2|\lambda|$ and the small SO splitting of the conduction band
is neglected. In $K'$, $-\lambda$ replaces $\lambda$. The parameter values
are detailed in section \ref{TMDSection}.

The potential in the full TMD Hamiltonian is taken to be a scalar function,
$V(x')$ instead of the diagonal matrix that appears in the BLG Hamiltonian.
This is because hole states are of interest and the effect of a
perpendicular electric field on the valence bands appears to be small
\cite{Rostami13}. However it is difficult to quantify this as there is no
electrostatic model similar to the one in ref. \cite{McCann06} for BLG.

To analyze the symmetry of the transmission coefficients it is only
necessary to consider the Hamiltonian without SO coupling. The reason is
that SO coupling is negligible in BLG while in TMDs spin-valley locking
occurs \cite{Xiao12,Liu13}. That is, the main effect of SO coupling in TMDs is to associate
opposite spins with opposite valleys. For example, if there is a spin up
state of energy $E$ in the $K$ valley, then there is a spin down state of
energy $E$ in the $K'$ valley.
In the following, this spin reversal will be taken as understood. This
allows the notation to be simplified and means the same analysis applies to
BLG and TMDs.

\subsection{Plane wave states}
\label{PWSection}
Plane wave states of energy $E$ and wave vector $\mathbf{k}$ in valley
$\alpha$ satisfy
\begin{equation}
  H_{0\alpha} \mathbf{e}_{\alpha}(\mathbf{k})
  \exp(i\mathbf{k}\cdot\mathbf{r}')
  = E\mathbf{e}_{\alpha}(\mathbf{k})\exp(i\mathbf{k}\cdot\mathbf{r}'),
\end{equation}
where the band Hamiltonians, $H_{0\alpha}$, are defined in Eqs.~(\ref{hblg})
(BLG) and (\ref{htmd}) (TMDs) and $\mathbf{e}_\alpha(\mathbf{k})$
is a 4-component polarization vector.

There are 4 distinct $\mathbf{k}$-vectors for each energy unless the energy
coincides with a band extremum. Other than in this exceptional case, two of
the plane wave states are propagating and two are evanescent, in the energy
range considered here \cite{StateNote}. However the evanescent part of the
wave function vanishes at infinity. Therefore the symmetry of the
transmission coefficients depends only on how the propagating wave
polarization vectors transform under the symmetry operations of the full
band Hamiltonian.

The full band Hamiltonian without SO coupling is the $8\times 8$ matrix
\begin{equation}
    H_0 = \left( \begin{array}{cc}
    \tilde{H}_{0K} & 0 \\
    0 & \tilde{H}_{0K'} \\
    \end{array} \right), \label{h8x8}
\end{equation}
where $\tilde{H}_{0\alpha}$ is the band Hamiltonian in valley $\alpha$
without SO coupling. The symmetries used in the present work are time
reversal and $x'$ inversion. The relevant invariance operators are:

\noindent\textit{Time reversal}. For all $\theta$, $H_0$ is invariant under
\begin{equation}
    \left( \begin{array}{cc}
    0 & E_4 \\
    E_4 & 0 \\
    \end{array} \right) \Theta, \label{trev}
\end{equation}
where $\Theta$ is the complex conjugation operator and $E_4$ is the $4\times
4$ unit matrix.

\noindent$x'$\textit{inversion}, $x'\rightarrow -x'$. When
$\theta\equiv 0 \pmod{2\pi/3}$,
the $y'$ axis is in a mirror plane of the atomic lattice and $H_0$ is
invariant under
\begin{equation}
    \left( \begin{array}{cc}
    0 & E_4 \\
    E_4 & 0 \\
    \end{array} \right) I_{x'}, \label{xinv}
\end{equation}
where $I_{x'}$ is the $x'$ inversion operator. It follows that
the polarization vectors in $K$ and $K'$ are related as follows.

\noindent\textit{Time reversal}. For all $\theta$,
\begin{equation}
  \mathbf{e}_{K}(k_{x'}, k_{y'}) = \mathbf{e}_{K'}^*(-k_{x'}, -k_{y'}).
  \label{esymtrev}
\end{equation}

\noindent$x'$\textit{inversion}, $x'\rightarrow -x'$. When
$\theta\equiv 0 \pmod{2\pi/3}$,
\begin{equation}
  \mathbf{e}_{K}(k_{x'}, k_{y'}) = \mathbf{e}_{K'}(-k_{x'}, k_{y'}).
  \label{esymxinv}
\end{equation}

\subsection{Currents carried by plane wave states}
\label{JSection}

The current density $\mathbf{j}_\alpha = (1/A)
\langle \mathbf{v}_\alpha \rangle$ where $\langle \mathbf{v}_\alpha
\rangle$ is the expectation value of the
velocity in valley $\alpha$ and $A$ is the system area. The velocity
operators are given by the matrix coefficients of the momentum terms in the
band Hamiltonians. In the case of BLG 
\begin{eqnarray}
  v_{x'K} &=& 
    \left( \begin{array}{cccc}
    0 & v & -v_4  & v_3 e^{3i\theta}\\
    v & 0 & 0 & -v_4 \\
    -v_4  & 0 & 0 & v\\
    v_3 e^{-3i\theta} & -v_4 & v & 0 \\
    \end{array} \right), \nonumber \\
  v_{y'K} &=& 
    \left( \begin{array}{cccc}
    0 & -iv & iv_4  & iv_3 e^{3i\theta}\\
    iv & 0 & 0 & iv_4 \\
    -iv_4  & 0 & 0 & -iv\\
    -iv_3 e^{-3i\theta} & -iv_4 & iv & 0 \\
    \end{array} \right).
    \label{vblg}
\end{eqnarray}

In the case of TMDs
\begin{eqnarray}
  v_{x'K} &=&
    \left( \begin{array}{cccc}
    0 & c_3 & c_2  & c_4\\
    c_3  & 0 & c_5 e^{-3i\theta} & c_6 e^{-3i\theta}\\
    c_2  & c_5 e^{3i\theta} & 0 & 0\\
    c_4  & c_6 e^{3i\theta} & 0 & 0\\
    \end{array} \right), \nonumber\\
  v_{y'K} &=& 
    \left( \begin{array}{cccc}
    0 & -ic_3 & ic_2  & ic_4\\
    c_3  & 0 & -ic_5 e^{-3i\theta} & -ic_6 e^{-3i\theta}\\
    -ic_2  & ic_5 e^{3i\theta} & 0 & 0\\
    -ic_4  & ic_6 e^{3i\theta} & 0 & 0\\
    \end{array} \right). \label{vtmd}
 \end{eqnarray}
In the $K'$ valley in both BLG and TMDs $-\theta$ replaces $\theta$ and
in $v_{x'K'}$ the sign of the velocity parameters changes. The full
velocity operator is an $8\times 8$ matrix with the same block structure as
the full band Hamiltonian, Eq.~(\ref{h8x8}).

The current densities in both BLG and TMDs satisfy symmetry relations that
result from the polarization vector relations,
Eqs.~(\ref{esymtrev}) and (\ref{esymxinv}), and the facts that the velocity 
changes sign under time reversal, its $x'$ component changes sign under
$x'$ inversion and both operations swap the blocks of the full velocity
operator. The resulting symmetry relations are:

\noindent\textit{Time reversal}. For all $\theta$,
\begin{eqnarray}
  j_{x'K}(k_{x'}, k_{y'}) &=& -j_{x'K'}(-k_{x'}, -k_{y'})\nonumber\\
  j_{y'K}(k_{x'}, k_{y'}) &=& -j_{y'K'}(-k_{x'}, -k_{y'}). \label{jtrev}
\end{eqnarray}

\noindent$x'$\textit{inversion}, $x'\rightarrow -x'$. When
$\theta\equiv 0 \pmod{2\pi/3}$,
\begin{eqnarray}
  j_{x'K}(k_{x'}, k_{y'}) &=& -j_{x'K'}(-k_{x'}, k_{y'})\nonumber\\
  j_{y'K}(k_{x'}, k_{y'}) &=&  j_{y'K'}(-k_{x'}, k_{y'}).
\end{eqnarray}

\subsection{States that carry current at angle $\phi_c$}
\label{MultipleJSection}

To find the transmission coefficient for current incident at angle $\phi_c$
one has to find all the incident states that carry current at this angle.
The current carried by a state with wave vector $\mathbf{k}$ flows in the
direction of the group velocity $\mathbf{v}_g = (1/\hbar)\mbox{\boldmath
$\nabla$} E(\mathbf{k})$. At
constant energy, $\mathbf{v}_g$ only depends on the polar angle, $\phi_k$
of the $\mathbf{k}$-vector. The state or states that carry current at angle
$\phi_c$, are found from the solution of
\begin{equation}
  \phi_v(\phi_k, E) = \phi_c,
  \label{phieq}
\end{equation}
where $\phi_v$ is the polar angle of the group velocity. The group velocity
may be found either from the gradient of $E(\mathbf{k})$ or from the
expectation value of the velocity operator.

The number of solutions of Eq.~(\ref{phieq}) depends on whether there are
points of inflection on the energy contour. In the case of TMDs there are
no points of inflection and there is only one solution. However points of
inflection occur in the case of BLG and three solutions occur in a small
range of $\phi_c$.

\begin{figure}
\begin{center}  
  \includegraphics[width=4cm, angle=-90]{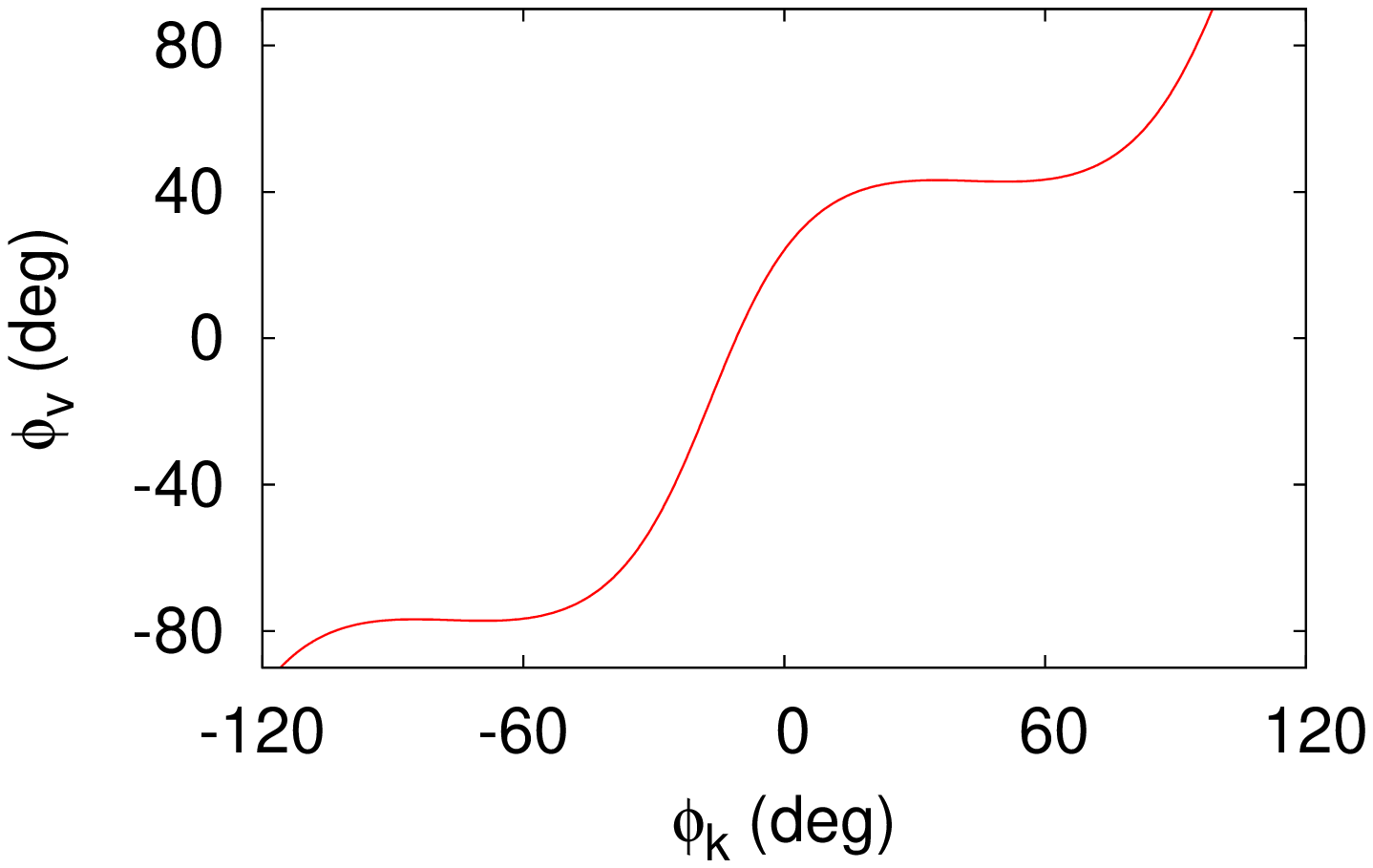}\\[1em]
  \includegraphics[width=4cm, angle=-90]{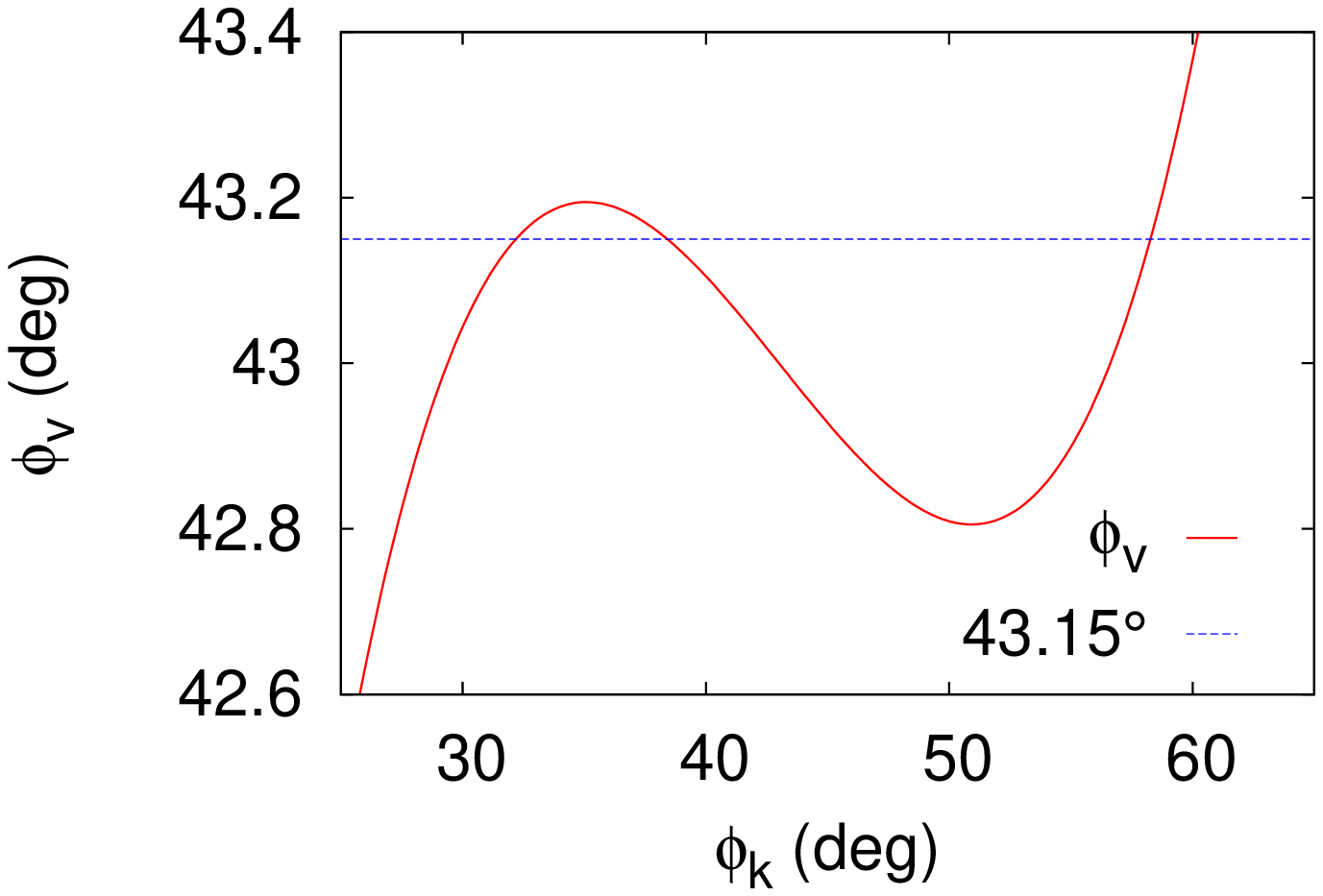}
  \caption{(Color online). Polar angle of group velocity for the
    $K$ valley (BLG) contact contour in Fig.~1. Upper: 
    $-90^\circ \le \phi_v \le 90^\circ$. Lower: expanded view
    near $\phi_v = 43^\circ$.}
\label{janglefig}
\end{center}
\end{figure}

Fig.~\ref{janglefig} shows $\phi_v(\phi_k)$ for the case of BLG and
the $K$ valley contact contour shown in Fig.~1. In most of the
$\phi_k$ range, $\phi_v$ increases monotonically and $\phi_v(\phi_k, E) =
\phi_c$ has only one solution. However multiple solutions occur near
$\phi_v=43^\circ$ and $-77^\circ$. These parts of the curve correspond to
the nearly straight parts of the $K$ valley contour. $\phi_v(\phi_k)$ near
$\phi_v=43^\circ$ is shown on an enlarged scale in the lower part of
Fig.~\ref{janglefig}. Three solutions occur between the maximum and
minimum, for example at $\phi_c = 43.15^\circ$. Then three distinct states
carry current in the same direction and one has to sum over the currents
carried by these states to find the transmission coefficient. In this work,
each state is taken to contribute to the total current with equal weight.

The existence of the maximum and minimum is necessary for the multiple
solutions to occur and the geometrical interpretation of this condition is
that there are points of inflection on the energy contour. The vector
$(-v_{gy'} ,v_{gx'})$ is tangential to the contour and by using this fact
and taking the contour to be parametrized by $\phi_k$, the condition for a
point of inflection (vanishing curvature) can be reduced to
$d\phi_v/d\phi_k = 0$. Hence points of inflection on the contour are
necessary for multiple current carrying states to occur.

\subsection{Valley Asymmetry of $T_\alpha(\phi_{c})$}
\label{vaphicSection}

\begin{figure}
\begin{center}  
  \includegraphics[width=4.2cm]{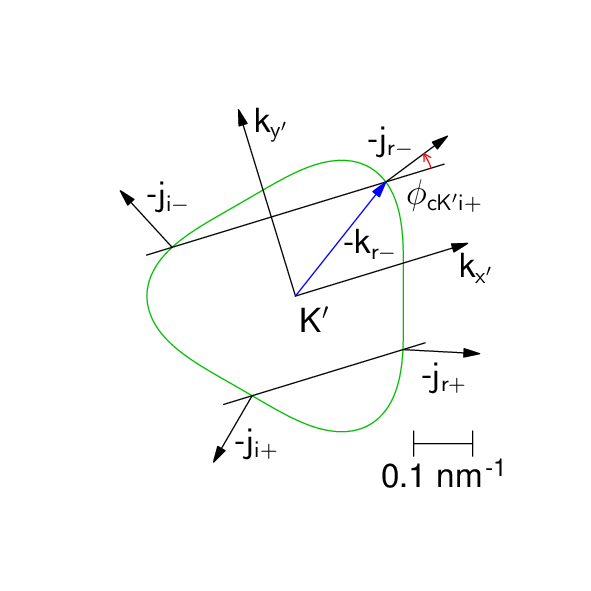}
  \includegraphics[width=4.2cm]{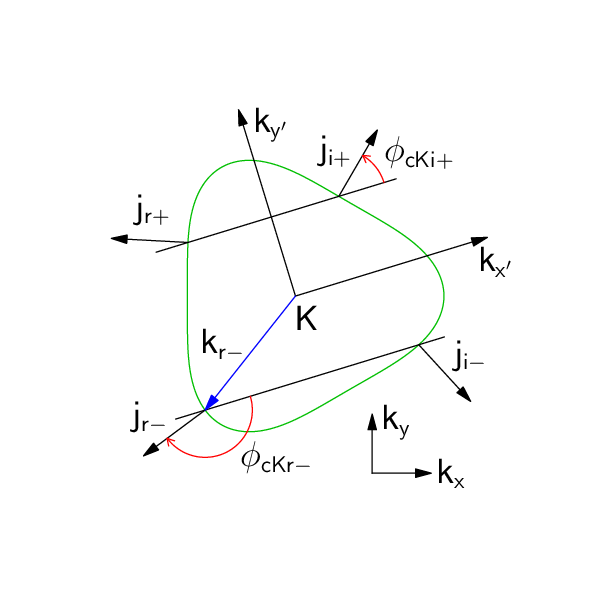}
  \caption{(Color online). Current directions (\textsf{j}) for all states
    with the same value of $|k_{y'}|$ in the $K$ and $K'$ valleys in BLG.
    The lines parallel to the $k_{x'}$ axis indicate $|k_{y'}|$ and the states
    occur at the intersections of these lines with the contact energy contours
    (green). Right: Current directions (black arrows) for incident (i) and
    reflected (r) states in $K$. The $\pm$ subscripts indicate the sign of
    $k_{y'}$. Left: Current directions in the $K'$ valley, obtained
    by time-reversing those in $K$. On time reversal, reflected states
    ($j_{x'}<0$) in $K$ become incident states ($j_{x'}>0$) in $K'$.
    Arrows (blue) labeled $\pm k_{r-}$ and arcs (red) respectively show
    $\mathbf{k}$-vectors and current angles of the states discussed in
    section \ref{vaphicSection}. The axis rotation angle,
    $\theta = 17^\circ$, and contours are as in Fig.~1.}
\label{jfig}
\end{center}
\end{figure}

The transmission coefficient $T_\alpha$ in valley $\alpha$ is the ratio of
the total transmitted current to the total incident current. It can be
obtained from the sum of the currents
transmitted by each state that contributes to current incident at angle
$\phi_c$. The solution of the scattering problem for each incident state
gives the transmitted current as a function of $k_{y'}$, then
$T_\alpha(\phi_{c})$ is found by substituting $k_{y'}(\phi_c)$ for
$k_{y'}$. $T_\alpha(\phi_{c})$ is valley asymmetric because
$k_{y'}(\phi_c)$ is not the same in each valley.

This is illustrated in Fig.~\ref{jfig} for a $\phi_c$ value for which there
is only one current carrying state. There are four states with the same
value of $|k_{y'}|$ in each valley. The directions of the currents carried
by these states are indicated by the arrows normal to the contours. The
subscripts on the arrow labels indicate the current direction (i, incident
or r, reflected) and sign of $k_{y'}$ in the $K$ valley. The currents in
$K'$ are obtained by time-reversing those in $K$, see Eq.~(\ref{jtrev}).
For example, the current carried by the state incident with $k_{y'} > 0$
in $K'$ is obtained by time reversing $\mathsf{j}_{r-}$ in $K$. The polar
angles of the currents carried by states incident with $k_{y'} > 0$ are
$\phi_{cKi+}$ in the $K$ valley and $\phi_{cK'i+}$  in the $K'$ valley.
These angles are clearly different and this demonstrates $k_{y'}(\phi_c)$
is not the same in each valley.

The reason for this asymmetry is that the $\mathbf{k}$-vectors and currents
in the $K'$ valley are related to those in the $K$ valley by time reversal
but the contours in each valley are not inversion symmetric about the
valley center. Because of time reversal symmetry, the $\mathbf{k}$-vector of
the incident state with $k_{y'} > 0$ in $K'$ is $-\mathbf{k}_{r-}$ where
$\mathbf{k}_{r-}$ is the $\mathbf{k}$-vector of the reflected state with
$k_{y'} < 0$ in $K$.  Hence the angle of incidence $\phi_{cK'i+}$ in the
$K'$ valley is related by symmetry to the angle of reflection $\phi_{cKr-}$
in the $K$ valley. Thus $\phi_{cK'i+} = \pi + \phi_{cKr-}$. However the
angles $\phi_{cKr-}$ and $\phi_{cKi+}$ in the $K$ valley are not
symmetry related because the contour is not inversion symmetric. Consequently
there is in general no symmetry relation between the angles of incidence,
$\phi_{cK'i+}$ and $\phi_{cKi+}$. However the special case of
$\theta \equiv 0 \pmod{\pi/3}$ is an exception (section \ref{vakySection}).

It follows that $k_{y'}(\phi_c)$ is not the same function in each valley.
Hence the transmitted current as a function of $\phi_c$ is valley
asymmetric except for special incidence conditions where the curves of
$T_K(\phi_c)$ and $T_{K'}(\phi_c)$ cross. This remains true in the
case of multiple current-carrying states but in this case additional
asymmetry may occur because the number of current-carrying states may be
different in the two valleys. Hence $T_\alpha(\phi_c)$ is valley
asymmetric, except for possible crossings.

\subsection{Valley Asymmetry of $T_\alpha(k_{y'})$}
\label{vakySection}

The transmission coefficient is also valley asymmetric when it is
expressed as a function of $k_{y'}$. Hence valley asymmetric transmission
occurs when particles are selected by momentum as well as by velocity.

The origin of the valley asymmetry with respect to $k_{y'}$ is that the
energy contours are not mirror symmetric about the $k_{x'}$ axis unless
$\theta \equiv 0 \pmod{\pi/3}$. Hence $T_\alpha(k_{y'})$ in each valley is
not a symmetric function of $k_{y'}$.  That is $T_\alpha(k_{y'}) \ne
T_\alpha(-k_{y'})$ unless $\theta \equiv 0 \pmod{\pi/3}$. However by using
time reversal symmetry it can be proved that
\begin{displaymath}
T_K(k_{y'}) = T_{K'}(-k_{y'}).
\end{displaymath}
Therefore $T_K(k_{y'}) \ne T_{K'}(k_{y'})$, except for
special incidence conditions where the curves of $T_K(k_{y'})$ and
$T_{K'}(k_{y'})$ cross.

It remains to prove that $T_K(k_{y'}) = T_{K'}(-k_{y'})$. This is done by
using the $S$-matrix description of the asymptotic regime where the
evanescent wave amplitudes are negligible. When $x'$ approaches $-\infty$,
the scattering states in the $K$ valley have the general form
\begin{equation}
  \psi_{K-} = \left[i_0\mathbf{e}_{K}(k_i, k_{y'})e^{ik_ix'} +
    r\mathbf{e}_{K}(k_r, k_{y'})e^{ik_rx'}\right] e^{ik_{y'}y'}
  \label{asypsip}
\end{equation}
and when $x'$ approaches $+\infty$, the form is
\begin{equation}
  \psi_{K+} = \left[t\mathbf{e}_{K}(k_i, k_{y'})e^{ik_ix'} +
    x_0\mathbf{e}_{K}(k_r, k_{y'})e^{ik_rx'}\right] e^{ik_{y'}y'}.
  \label{asypsim}
\end{equation}
Here the $x'$-component of the current is positive for the state with
$k_{x'} = k_i$ and negative for the state 
with $k_{x'} = k_r$. $i_0$ is the amplitude of the
incident wave, $r$ is the amplitude of the reflected wave, $t$ is the
amplitude of the transmitted wave and $x_0$ is the amplitude of a wave
incident from the right.

The asymptotic wave amplitudes are related by the
$S$-matrix defined by
\begin{equation}
    \left( \begin{array}{c}
    r \\
    t \\
    \end{array} \right) =
    \left( \begin{array}{cc}
    S_{Ka} & S_{Kb}\\
    S_{Kc} & S_{Kd}\\
    \end{array} \right)
    \left( \begin{array}{c}
    i_0 \\
    x_0 \\
    \end{array} \right).  \label{sdef}
\end{equation}
The $S$-matrix is unitary provided the propagating state polarization
vectors are normalized to unit current. If this normalization is not used,
current
conservation still constrains the form of the $S$-matrix but does not
constrain it to be unitary because when the energy contours are warped, the
currents carried by the incident and reflected states are not of the same
magnitude. For example, the normalization
$\mathbf{e}_\alpha(\mathbf{k})^\dagger\mathbf{e}_\alpha(\mathbf{k}) = 1$
is convenient for
numerical calculations but with this normalization, the $S$-matrix
satisfies the generalized unitarity relation $S^\dagger J S = \tilde{J}$ where 
$J = \mathrm{diag}(|j_r|, |j_i|)$ and $\tilde{J} = \mathrm{diag}(|j_i|, |j_r|)$.

The proof of the relation $T_K(k_{y'}) = T_{K'}(-k_{y'})$ is simplest when
the $S$-matrices are unitary. We detail this case then state the 
change that is introduced by generalized unitarity. To prove the relation
the $S$-matrix in the $K$ valley is related to the one in the $K'$ valley.
Application of the time reversal operator to the asymptotic states in
Eqs.~(\ref{asypsip}) and (\ref{asypsim}) gives
\begin{eqnarray}
  \psi_{K'-} &=& \bigg[i_0^*\mathbf{e}_{K'}(-k_i, -k_{y'})e^{-ik_ix'} +
   \nonumber\\
    &+& r^*\mathbf{e}_{K'}(-k_r, -k_{y'})e^{-ik_rx'}\bigg] e^{-ik_{y'}y'}
\end{eqnarray}
when $x'$ approaches $-\infty$ and
\begin{eqnarray}
  \psi_{K'+} &=& \bigg[t^*\mathbf{e}_{K'}(-k_i, -k_{y'})e^{-ik_ix'}
    +\nonumber\\
    &+&
    x_0^*\mathbf{e}_{K'}(-k_r, -k_{y'})e^{-ik_rx'}\bigg] e^{-ik_{y'}y'}
\end{eqnarray}
when $x'$ approaches $+\infty$. However because the sign of the current
changes under time reversal, Eq.~(\ref{jtrev}), the state that carries
positive current in the $x'$ direction has $k_{x'} = -k_r$ and the state
that carries negative current has $k_{x'} = -k_i$. Consequently, the wave
amplitudes in the time-reversed state are related by
\begin{equation}
    \left( \begin{array}{c}
    i_0^* \\
    x_0^* \\
    \end{array} \right) =
    \left( \begin{array}{cc}
    S_{K'a} & S_{K'b}\\
    S_{K'c} & S_{K'd}\\
    \end{array} \right)
    \left( \begin{array}{c}
    r^* \\
    t^* \\
    \end{array} \right).  \label{strev}
\end{equation}
Then after using the unitarity of $S$ and complex conjugating the resulting
equation, it can be seen that the $S$-matrices in the two valleys satisfy
$S_K(k_{y'})= S_{K'}^T(-k_{y'})$, where the sign change results from
time reversal.

Next, this relation is used to prove that $T_K(k_{y'}) = T_{K'}(-k_{y'})$.
The transmitted amplitude for a unit amplitude wave incident from
the left is $S_{\alpha c}$ and the transmitted amplitude for a unit
amplitude wave incident from the right  is $S_{\alpha b}$. These amplitudes
are related by $S_{Kc}(k_{y'})=S_{K'b}(-k_{y'})$. In addition,
$|S_{\alpha c}(k_{y'})|=|S_{\alpha b}(k_{y'})|$ because of unitarity.
Hence $T_K(k_{y'}) = |S_{Kc}(k_{y'})|^2 = |S_{K'b}(-k_{y'})|^2 =
|S_{K'c}(-k_{y'})|^2= T_{K'}(-k_{y'})$. These relations remain valid when
the $S$-matrices satisfy the generalized unitarity relation but in this
case the $S$-matrices in the two valleys are related by
$J_K S_K(k_{y'}) \tilde{J}_{K}^{-1}= S_{K'}^T(-k_{y'})$.

In the special case of $\theta \equiv 0 \pmod{\pi/3}$, the transmission
coefficient expressed as a function of $k_{y'}$ has
higher symmetry: when the potential is symmetric under $I_{x'}$,
$T_\alpha(k_{y'}) = T_\alpha(-k_{y'})$ hence
$T_K(k_{y'}) = T_{K'}(k_{y'})$. This can be proved by applying $I_{x'}$ to
the asymptotic states in Eqs.~(\ref{asypsip})
and (\ref{asypsim}) and then using the definition of the $S$-matrix.

When multiple current-carrying states occur, the transmission is valley
asymmetric for each state, except when $\theta \equiv 0 \pmod{\pi/3}$.
Hence $T_\alpha(k_{y'})$ is valley asymmetric except for possible crossings
and except when $\theta \equiv 0 \pmod{\pi/3}$.

\subsection{Transmission Coefficient Relations}
\label{symrelSection}

The transmission coefficients at different values of $\theta$ are related
by symmetry and we have found two particularly useful symmetry relations.
The first is 
\begin{equation}
  T_K(\phi_c, \theta) = \hat{T}_{K'}(\phi_c, \theta\pm \pi/3),
  \label{symrel1}
\end{equation}
where $\hat{T}$ is the transmission coefficient for a barrier with the
spatially inverted potential, $V(-x')$. The second symmetry relation is
\begin{equation}
  T_K(\phi_c, \theta) = T_{K'}(-\phi_c, \pm\pi/3 - \theta).
  \label{symrel2}
\end{equation}
In both relations it is understood that the spins are opposite in the
case of TMDs.

The symmetry relations occur because there are
operators that transform the band Hamiltonian $H_{0K}(\theta)$ into
$H_{0K'}(\theta\pm \pi/3)$ and $H_{0K'}(\pi/3 - \theta)$. To show this
$T_\alpha$ is first taken to be a function of $k_{y'}$.

Then the relation equivalent to Eq.~(\ref{symrel1}) is 
\begin{equation}
  T_K(k_{y'}, \theta) = \hat{T}_{K'}(k_{y'}, \theta\pm \pi/3).
  \label{symrel1k}
\end{equation}
This is a consequence of the way the band Hamiltonians transform
under the product of spatial inversion, $I_{x'y'}$, and
complex conjugation, $\Theta$. The momentum transforms as
$I_{x'y'} \Theta \pi_K \Theta I_{x'y'} = -\pi_{K'}$ and the Hamiltonians satisfy
\begin{equation}
  D I_{x'y'} \Theta H_{0K}(\theta) \Theta I_{x'y'} D = H_{0K'}(\theta \pm \pi / 3),
  \label{h0transform}
\end{equation}
where $D = \mathrm{diag}(1, -1, -1, 1)$ in the case of BLG and
$D = \mathrm{diag}(-1, 1, 1, 1)$ in the case of TMDs. Hence if $\psi$ is
an eigenstate of $H_{0K}(\theta) + V(x')$, then
$D I_{xy} \Theta \psi$ is an eigenstate of $H_{0K'}(\theta \pm \pi / 3) + V(-x')$ and
the symmetry relation, Eq.~(\ref{symrel1k}), is proved. The relation holds
for any $y'$-independent potential.

The second symmetry relation, Eq.~(\ref{symrel2}), is equivalent to
\begin{equation}
  T_K(k_{y'}, \theta) = T_{K'}(-k_{y'}, \pm\pi/3 - \theta).
  \label{symrel2k}
\end{equation}
This is a consequence of the
transformation of the Hamiltonians under inversion of the $y'$ co-ordinate,
$I_y'$. In this case $I_{y'} \pi_K I_{y'} = -\pi_{K'}$ and the Hamiltonians
satisfy
\begin{equation}
  D I_{y'} H_{0K}(\theta) I_{y'} D = H_{0K'}(\pm\pi / 3 - \theta),
  \label{h0transform2}
\end{equation}
which leads to Eqs.~(\ref{symrel2k}).

The symmetry relations expressed as a function of $\phi_c$,
Eqs.~(\ref{symrel1}) and (\ref{symrel2}), follow from relations between
the current components. By transforming the polarization vectors with
$D I_{x'y'} \Theta$, it can be shown that
$\mathbf{j}_K(\mathbf{k},\theta) =
\mathbf{j}_{K'}(\mathbf{k},\theta\pm\pi/3)$.
Hence $\phi_{cK}(\theta) = \phi_{cK'}(\theta\pm\pi/3)$ and this together with
Eq.~(\ref{symrel1k}) leads to Eq.~(\ref{symrel1}).
Similarly 
${j}_{x'K}(k_{x'}, k_{y'}, \theta) =
j_{x'K'}(k_{x'}, -k_{y'},\pm\pi/3-\theta)$ and ${j}_{y'K}(k_{x'}, k_{y'}, \theta) =
-j_{y'K'}(k_{x'}, -k_{y'},\pm\pi/3-\theta)$. Hence
$\phi_{cK}(\theta) = -\phi_{cK'}(\pm\pi/3-\theta)$ and this together with
Eq.~(\ref{symrel2k}) leads to Eq.~(\ref{symrel2}). These symmetry relations
are valid for all $\phi_c$ hence remain valid when there is more than one
current carrying state.

\section{Numerical Methods}
\label{NumSection}

\subsection{Transmission Coefficients}
\label{TnumSection}

The transmission coefficients are found numerically because we need to
consider soft-walled barriers (Section \ref{BLGVSection}) for which the
transmission coefficient cannot be found analytically. The numerical
procedure is based on an S-matrix method \cite{Maksym97} that is used in
surface science and is numerically stable when evanescent waves are
present. In brief, the system is divided into short segments and the
segment S-matrices are combined to find the system S-matrix and
transmission coefficient. The segments must be short enough to allow the
segment S-matrix to be computed from the segment transfer matrix to the
required accuracy. The only difference between the procedure used in
surface science and the present one is the method used to compute the
transfer matrices.

The transfer matrices are obtained from the numerical solution of
\begin{equation}
  (H_{0\alpha} + V_c + \Delta V)\psi_\alpha = E\psi_\alpha,
  \label{psieq}
\end{equation}
where
$H_{0\alpha}$ is the band Hamiltonian in valley $\alpha$ and $V_c$ is the
position independent potential in the contacts. $\Delta V(x') = V(x') -
V_c$ where $V(x')$ is the total potential. The 4-component wave function
is expressed in the form
\begin{equation}
\psi_\alpha = \exp(ik_{y'} y')\sum_{j=1}^4 Q_j(x') \mathbf{e}_{\alpha}(k_j, k_{y'})
  \exp(ik_j x').
\label{qdef}
\end{equation}
The transfer matrix, $M$, for a segment of
length $l$ with right boundary at position, $x'$ is defined by
\begin{equation}
  D(x'-l) \mathbf{Q}(x'-l) = M(x'-l, x') D(x') \mathbf{Q}(x'),
  \label{mdef}
\end{equation}
where
$\mathbf{Q}(x')$ is a vector whose elements are $Q_j(x')$ and $D(x')$ is a
diagonal matrix whose diagonal elements are $\exp(ik_j x')$.

$M$ is found by solving a differential equation for $\mathbf{Q}(x')$. By
substituting the form of $\psi_\alpha$ given by Eq.~(\ref{qdef}) into
Eq.~(\ref{psieq}) and using the fact that the polarization vectors are
orthogonal with respect to $v_x$ \cite{Maksym18} it can be shown that
\begin{equation}
  \frac{d}{dx'}\mathbf{Q} = \frac{-i}{\hbar c}
  D^{-1} P_L \Delta V P_R D \mathbf{Q},
  \label{qeq}
\end{equation}
where $P_R$ is a matrix whose columns are $\mathbf{e}_{\alpha}(k_j, k_{y'})$,
$P_L$ is a matrix whose rows are
$A_j\mathbf{e}^\dagger_{\alpha}(k_j^*, k_{y'})$
and $c = v_0$ in the case of BLG and $c=c_3$ in the case of TMDs. The
constants $A_j$ are chosen so that
$A_j\mathbf{e}^\dagger_{\alpha}(k_j^*, k_{y'}) v_x
\mathbf{e}_{\alpha}(k_j, k_{y'}) = c$.

The calculation of the segment S-matrix requires half of the transfer
matrix and half of its inverse \cite{Maksym97}. Each column of both
matrices is found by solving Eq.~(\ref{qeq}) numerically with a fourth order
Runge-Kutta method and appropriate initial conditions. The relative error
in the transmission coefficients is $<10^{-6}$. The procedure defined by
Eq.~(\ref{qeq}) is not the only way of finding the transfer matrix but we
have not investigated the alternatives. The procedure can be generalized to
deal with the case of $y'$-dependent potentials.

\subsection{Propagating Region Boundaries}
\label{PRnumSection}

To find the range of current angles where single valley transmission occurs
one has to find the range of current angles, $\phi_c$, on the propagating
part of the energy contour in each valley, see bold lines and vectors
normal to the contours in Fig.~\ref{axisfig}. The extrema of the current
angle occur either at the end points of the propagating range or at points
of inflection in the propagating range. Hence to find the range of current
angles in the propagating part of the contour, one has to compute the
current angles at the end points of the propagating part and at the points
of inflection, then choose the angles that give the largest range. This
method was used to find the propagating part of each contact contour in
Fig.~\ref{axisfig} and the regions of single valley transmission detailed
in sections \ref{BLGSection} and \ref{TMDSection}. In the case of 
Fig.~\ref{axisfig}, points of inflection occur within the propagating part
but the extrema of the current angle occur at the end points.

\section{Single valley transmission in BLG}
\label{BLGSection}

\subsection{Overview}
\label{OvblgSection}

In this section we explain the device model used to find the barrier
potential in BLG, detail the valley asymmetric transmission and show
that the barrier potential can be adjusted so that large valley
asymmetry occurs for all barrier orientations. We also give an
extended discussion of the feasibility of observing the predicted valley
asymmetry.

The device model and potential are detailed in section \ref{BLGVSection}.
We then explain the valley asymmetric transmission section
(\ref{VablgSection}) and show that large valley asymmetry can be obtained
for all barrier orientations by adjusting the potential
(\ref{OptblgSection}). Next we show that the valley asymmetry persists over
a range of Hamiltonian parameters (\ref{BLGParameterSection}) and is
insensitive to substrate interactions (\ref{SubstrateSection}).
The section closes with a discussion of the requirements for observing the
predicted valley asymmetry (\ref{BLGFeasibilitySection}).

\subsection{Model device and potential}
\label{BLGVSection}

\begin{figure}
\begin{center}  
  \includegraphics[width=4.4cm, angle=0]{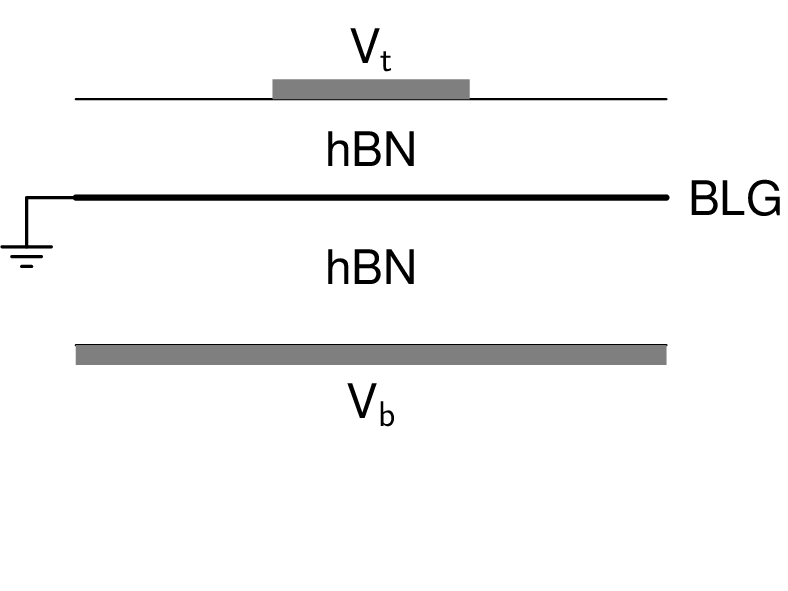}
  \caption{Cross section of device used to find model potential
    (schematic). The filled rectangles represent the top and bottom gates.}
\label{devfig}
\end{center}
\end{figure}

The barrier is taken to be generated by a device that has a narrow top gate
and a wide bottom gate (Fig.~\ref{devfig}). The bottom gate is 16 nm below
the BLG and the top gate is 4 nm above it. The BLG is grounded and the
space between the BLG and the gates is occupied with hBN. The electrostatic
potentials in this device are estimated from a self-consistent solution of
the Laplace equation based on the theory in ref. \cite{McCann06} and
the numerical method in ref. \cite{Maksym19}. This allows the potential to
be found for a realistic device structure but the potential is approximate
because of approximations made in the theory.

The self-consistent potential is constant underneath the top gate and
approaches a different constant in the contacts. Near the top gate edges it
has the form of a soft step that varies monotonically between the two
constant values. The self-consistent calculation of the potential is
expensive but we only need potentials for a small range of device
parameters. We therefore fit a model potential to the self-consistent one.

The contact potentials, $U_1 = 14$ meV, $U_2 = -14$ meV, are obtained with
a bottom gate voltage $\sim 2000$ mV. All calculations are done for these
fixed values. The top gate voltage is varied over a small range to
optimize the widths of the single valley regions. Within this range,
the total potential is taken to vary linearly with top gate voltage,
$V_t$,
\begin{eqnarray}
  U_1 + V_1 &=& V_{01} + \frac{dV_1}{dV_t} \Delta V_t,\nonumber \\
  U_2 + V_2 &=& V_{02} + \frac{dV_2}{dV_t}\Delta V_t,\nonumber 
\end{eqnarray}
where $V_{01} = 37$ meV, $V_{02} = -0.78$ meV are the total potentials when
$V_t \sim -200$ mV and $dV_1/dV_t = -0.12$, $dV_2/dV_t = -0.072$ are
estimated by numerical differentiation of the self-consistent potential.

The model step function is chosen to reproduce the self-consistent potential. 
We have found that the self-consistent potential varies rapidly near the
gate edges and slowly far from the gate edges. This variation cannot be
reproduced well with a function that depends only on one length parameter,
however a reasonable approximation is
\begin{eqnarray}
  F(x') &=& \frac{1}{2}[1 + \tanh(x'/a)], \quad x' > x_0,\nonumber \\
  &=& \frac{\alpha}{(x'-\beta)^2 + \gamma}, \quad x' \le x_0, \label{walleq}
\end{eqnarray}
where two of the parameters are constrained by the requirements that $F$ and
$dF/dx'$ are continuous at $x'=x_0$. The parameters $a, x_0$ and $\gamma$ are
used to adjust the shape of $F$ while $\alpha$ and $\beta$ are used to
enforce continuity. The continuity requirement can be satisfied with two
different values of $\alpha$ and $\beta$; the values that give the best fit
are chosen. The resulting parameter set is $a = 3$ nm, $x_0 = 2.5$ nm,
$\gamma = 20$ nm$^2$, $\alpha = 17.89$ nm$^2$ and $\beta = 3.626$ nm.
$F(x')$ defined in Eq.~(\ref{walleq}) gives an upward step, the downward step
is modeled with $F(-x')$ so the barrier is symmetric.

\subsection{Valley asymmetric transmission}
\label{VablgSection}

\begin{table}
\begin{tabular}{rrrr}
\hline
& Set 1 & Set 2 & Set 3\\\hline
$\gamma_0$ & 3160  & 3000  & 2900 \\
$\gamma_3$ & 380  & 300  & 100 \\
$\gamma_4$& 140  & 150  & 120 \\
$t$ & 381  & 400  & 300 \\
$\Delta'$ & 22  & 18  & 0 \\
\hline
\end{tabular}
\caption{Bilayer graphene Hamiltonian parameters in meV. The velocity
  parameters are related to the $\gamma$ parameters by
  $v_i=a\gamma_i\sqrt{3}/2\hbar$, where $a=0.246$ nm is the lattice constant.
  The parameters are taken from Table 1 in ref. \cite{McCann13}.
  In parameter set 3, $\Delta'$ is not given in ref.
  \cite{McCann13} and is assumed to be 0.}
\label{BLGParameterTable}
\end{table}

The valley asymmetric transmission is shown in Fig.~\ref{bgTfig}. The
transmission coefficients are computed as described in section
\ref{TnumSection} and with Hamiltonian parameter set 1 in
Table \ref{BLGParameterTable}.
Multiple current carrying states occur only in a narrow $\phi_c$ range,
$\sim 0.4^\circ$. The sensitivity to the parameter values is discussed in
section \ref{BLGParameterSection}.

Two types of valley asymmetry occur.
When $\theta$ is not close to $\pi/6 \pmod{2\pi/3}$ and 
not close to $\pi/2 \pmod{2\pi/3}$ (Fig.~\ref{bgTfig}, left) single
valley transmission occurs in the same valley at both positive and negative
$\phi_c$. But when $\theta$ is close to $\pi/6 \pmod{2\pi/3}$ or close
to $\pi/2 \pmod{2\pi/3}$ (Fig.~\ref{bgTfig}, right) the single valley
transmission at positive and negative $\phi_c$ is in different valleys. 

In both cases the transmission coefficient is large in a central region and
goes to zero abruptly at two critical angles. In the central region the
barrier states are propagating and transmission resonances occur. The
resonances are sharp because the barrier is wide. Beyond the two critical
angles, the barrier states are evanescent. Then tunneling occurs in the
barrier and when the barrier width is large, the transmission coefficient
is exponentially small. With the 300 nm gate width assumed in this work,
the transmission coefficient is typically between $10^{-4}$ - $10^{-3}$ at
about 0.1$^\circ$ into the tunneling regime and several orders of
magnitude smaller a few degrees into it. Then the transmission coefficient is
practically zero and valley asymmetric total external reflection occurs as
explained in section \ref{OverviewSection}. In addition, Fig.~\ref{axisfig}
shows that the range of incident current angles is larger in $K$ than in
$K'$. This explains why the width of $T_K$ in Fig.~\ref{bgTfig} (left) is
larger than the width of $T_{K'}$.

\begin{figure}
\begin{center}  
  \includegraphics[width=4.4cm, angle=-90]{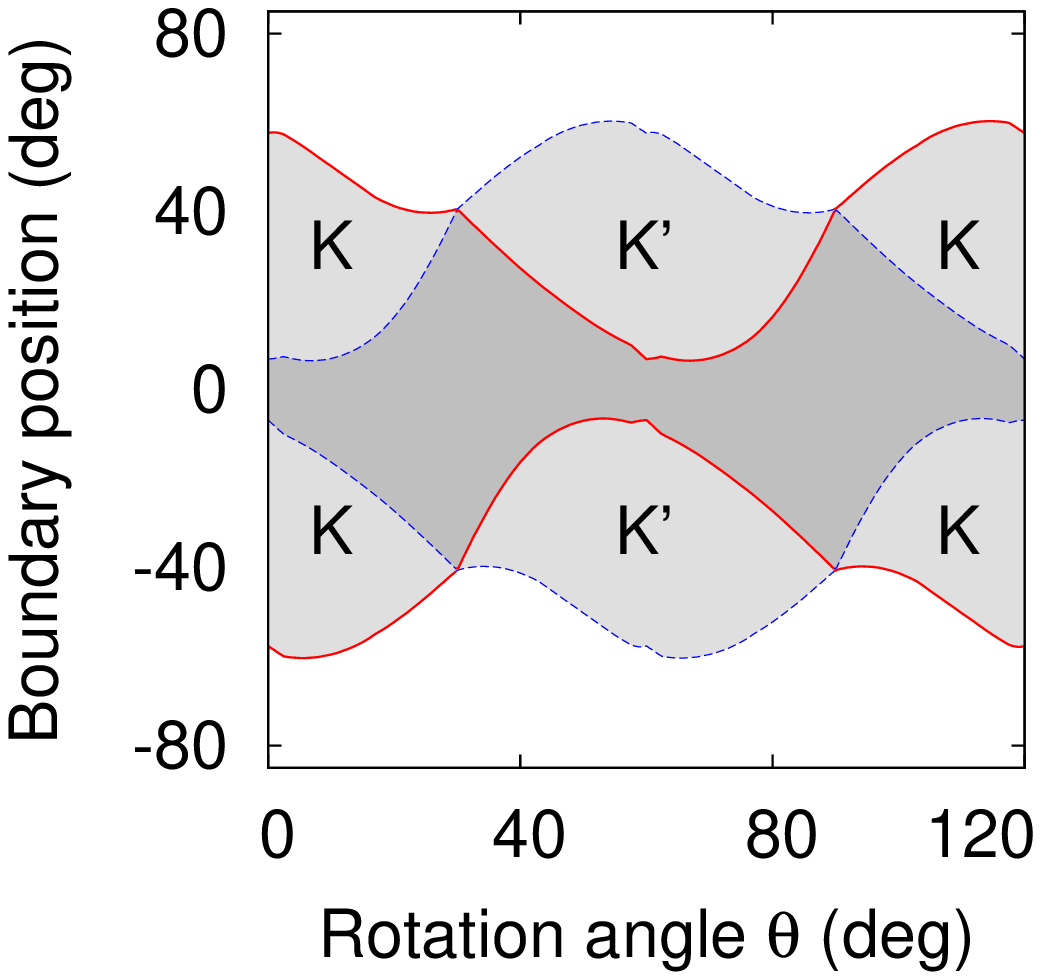}
  \includegraphics[width=4.4cm, angle=-90]{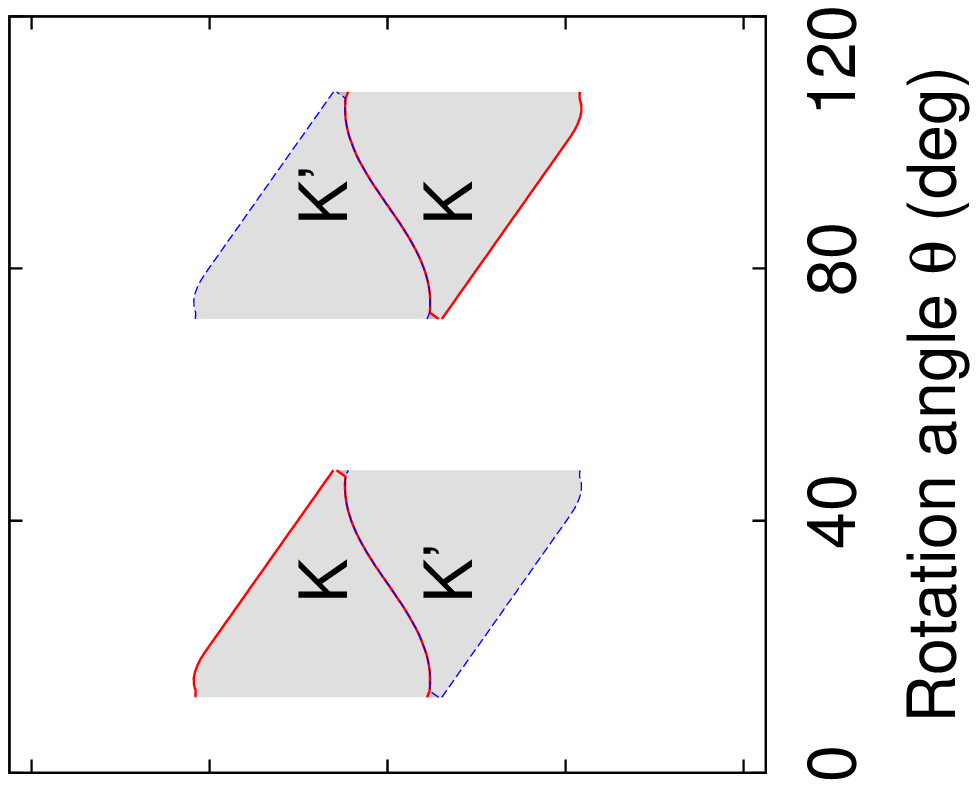}
  \caption{(Color online). Optimized single valley regions for 56 meV
    electrons in BLG.
    Solid red lines: $K$ transmission boundaries, dashed blue lines: $K'$
    boundaries. Light fill: single valley transmission, dark
    fill: two valley transmission.
    Left: same valley at $\pm$ incidence. Right: different valleys
    at $\pm$ incidence; in this case sharp cut-offs occur and,
    for clarity, some tiny regions are not shown; see section
    \ref{OptblgSection} and Fig.~\ref{svtp2fig}.}
\label{svtfig}
\end{center}
\end{figure}

\subsection{Optimized single valley transmission}
\label{OptblgSection}

Fig.~\ref{bgTfig} shows that single valley transmission occurs in two
regions that are bounded by critical angles for total external reflection.
It is desirable to maximize the angular width of the single valley regions.
We do this by varying the top gate voltage: the region
boundaries are computed as described in section \ref{PRnumSection} and
the gate voltage is adjusted to maximize the region widths.

Fig.~\ref{svtfig} shows this leads to large angular widths. By choosing the
case when the valleys are the same on both sides of $0^\circ$ (left) or
different (right), single valley regions of width $\sim 17.3 - 53.0^\circ$
can be obtained for all $\theta$. The $\theta$ dependence of the single
valley regions is quite different in the same-valley and different-valley
cases. The explanation is as follows.

In the same-valley case, single valley regions of finite width are found
for all $\theta$ except $\theta = \pi / 6$ and $\theta = \pi / 2$. When
$\theta = \pi / 6$, it follows from Eq.~(\ref{symrel2}) that
$T_K(\phi_c, \pi /6) = T_{K'}(-\phi_c, \pi/6)$. Hence if single valley
transmission occurs in one valley $\alpha$ for $\phi_c > 0$, it must occur in
the other valley when $\phi_c < 0$. Therefore the same-valley case
cannot occur at $\theta = \pi / 6$ for any value of the potential. 
In the case of $\theta = \pi / 2$, the fact that the barrier is symmetric
gives $\hat{T}_\alpha = T_\alpha$ and it then follows from Eq.~(\ref{symrel1}) 
that $T_K(\phi_c, \pi /6) = T_{K'}(\phi_c, \pi/2)$ and
$T_{K'}(-\phi_c, \pi /6) = T_{K}(-\phi_c, \pi/2)$. These relations together
with the relation $T_K(\phi_c, \pi /6) = T_{K'}(-\phi_c, \pi/6)$ lead to
$T_K(\phi_c, \pi /2) = T_{K'}(-\phi_c, \pi/2)$. Again, the same-valley case
cannot occur for any value of the potential. These arguments explain
why the single valley regions widths in the same-valley case shrink to zero
at $\theta = \pi / 6$ and $\theta = \pi / 2$.

In the different-valley case, single valley regions of large width occur
near $\theta = \pi / 6$ and $\theta = \pi / 2$ but sharp cut-offs occur as
$\theta$ departs from these values. Beyond the cut-offs it is difficult or
even impossible to find potentials that lead to different-valley
behavior. This is a consequence of the $\theta$ dependence of the
propagating part of each energy contour. When $\theta$ changes, one end
point of a propagating part may go around a corner of a contour. When this
happens, $\phi_c$ changes rapidly with $\theta$ and the propagating range
broadens rapidly. If the propagating range in the other valley remains
narrow, a crossover from different-valley to same-valley behavior may occur.

For example, consider the cut-off near $\theta = 12^\circ$ in
Fig.~\ref{svtfig}. If $\theta$ decreases from $\sim 30^\circ$, the end point of
the propagating range in the $K$ valley moves around the right hand corner
of the contour. Then a crossover to same-valley behavior occurs but the
different-valley behavior can be restored by raising the barrier
height. This shrinks the barrier contour (Fig.~\ref{axisfig}) hence
shrinks the propagating part of the contact contours in both valleys and
restores the different-valley behavior. However the barrier height cannot
be raised above $E$ as the transmission becomes exponentially small. This
condition corresponds to the cut-off near $\theta = 12^\circ$ and all the other
cut-offs.

Beyond the cut-offs tiny regions of different-valley behavior can be found
by changing the potential drastically. These regions correspond to 
$\theta$ and $\phi_c$ ranges of only a few degrees which is too small to be
of practical use. For this reason and for clarity they are not shown in
Fig.~\ref{svtfig} but are detailed in section \ref{BLGParameterSection}.

The rapid variation of $\phi_c$ when the end point of a propagating part
goes around a corner of an energy contour also affects the same-valley
behavior. This is why small peaks and dips occur in the same-valley
boundary lines near $\theta = 0$, $60$ and $120^\circ$.

\begin{figure}
\begin{center}  
  \includegraphics[width=4.4cm, angle=-90]{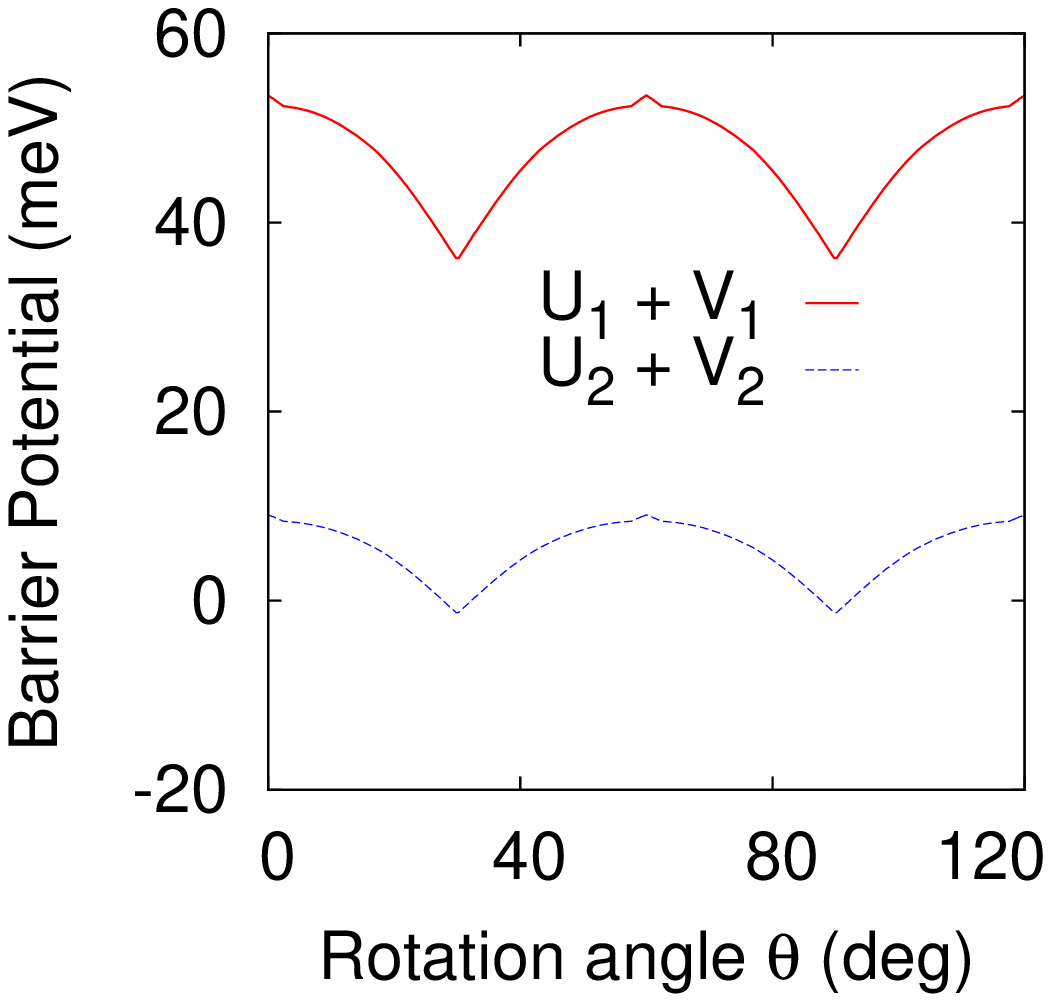}
  \includegraphics[width=4.4cm, angle=-90]{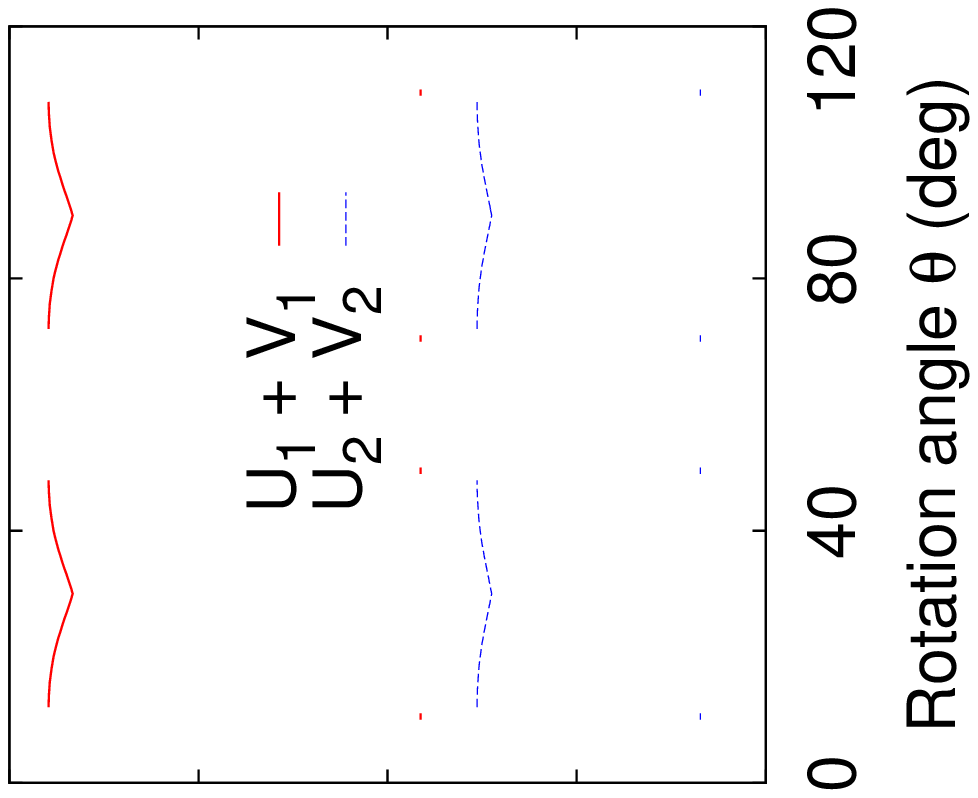}
  \caption{(Color online). Optimal barrier potentials in BLG.
    The corresponding single valley
    regions are shown in Fig.~\ref{svtfig}.
    Left: same-valley case. Right: different-valley case.}
\label{optvfig}
\end{center}
\end{figure}

Fig.~\ref{optvfig} shows the optimal barrier potentials used to compute the
single valley region boundaries shown in Fig.~\ref{svtfig}. The range of
potentials needed for two single valley regions in the same valley does not
overlap with the range needed for two single valley regions in different
valleys. In addition, if the region widths are calculated with a
$\theta$-independent potential, equal to the mid-range optimal potential,
they shrink by $\sim 10$-$20$\%. These observations confirm it is necessary
to adjust the top gate voltage to get single valley regions of large width
for all $\theta$.

The required bias is modest and outside the range needed for a Lifshitz
transition. A Lifshitz transition occurs in BLG when the interlayer bias
is either very low \cite{McCann13} or very high \cite{Boswami13}. This
does not affect the valley asymmetry but causes the energy contours to
become disconnected and may be inconvenient because the transmitted
current may be reduced. However the bias values in Fig.~\ref{optvfig} are
not in the range where disconnected contours occur.

\subsection{Sensitivity to Hamiltonian Parameters}
\label{BLGParameterSection}

A wide range of Hamiltonian parameters appears in the literature so it is
important to check the sensitivity of the single valley region widths to
the parameter values (Table~\ref{BLGParameterTable}).  This is done by
repeating the calculations with parameter sets 2 and 3.

\begin{figure}
\begin{center}  
  \includegraphics[width=4.4cm, angle=-90]{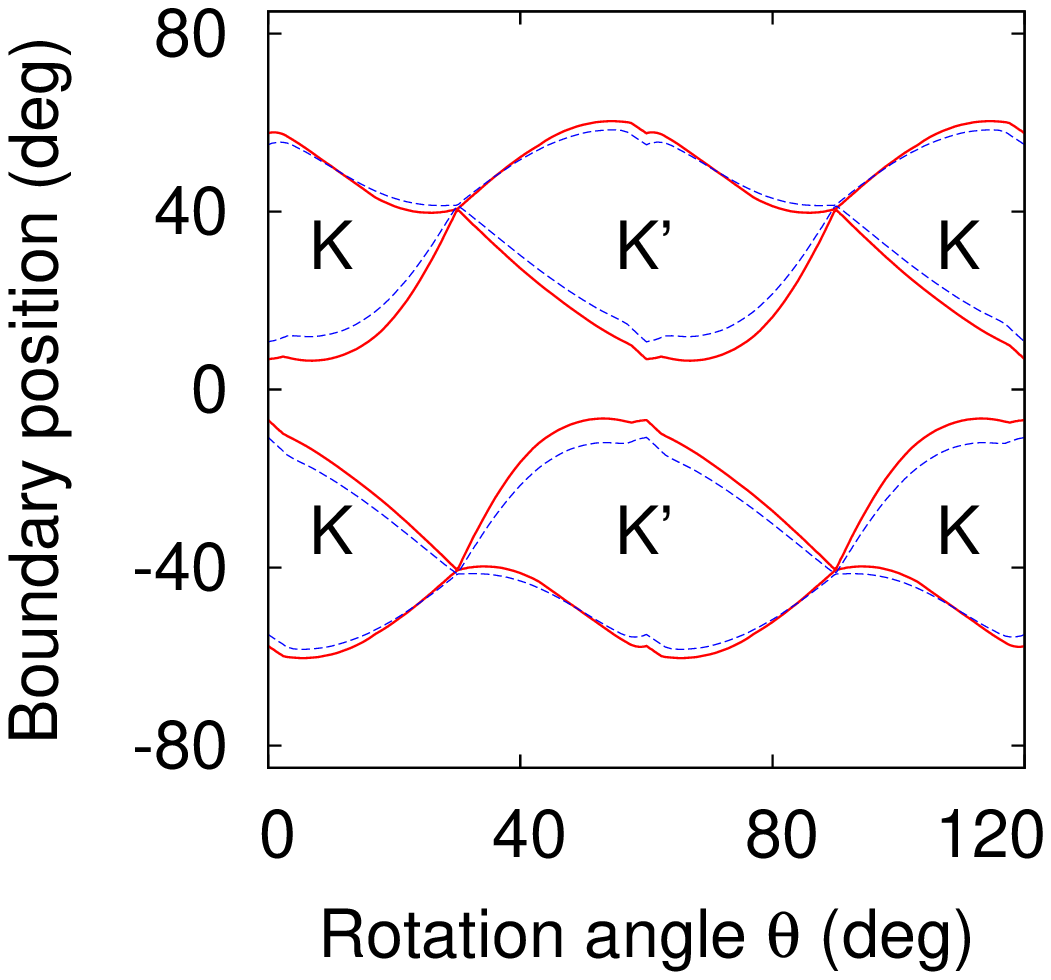}
  \includegraphics[width=4.4cm, angle=-90]{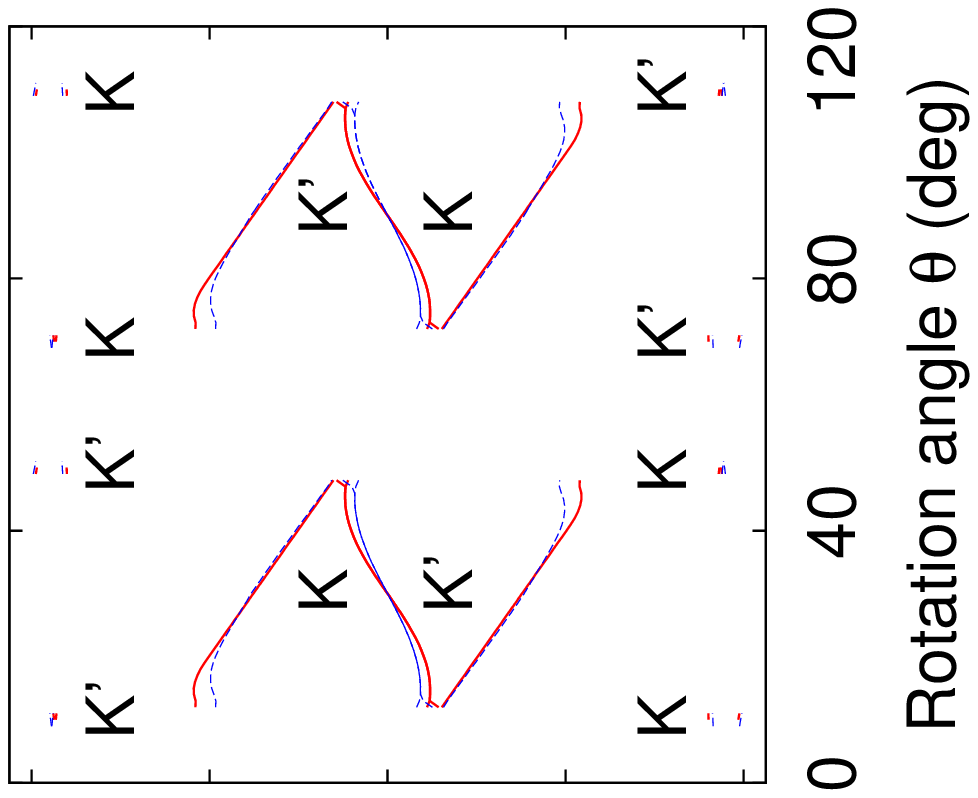}
  \caption{(Color online). Optimized single valley regions in BLG.
    Solid red lines: $K$ and $K'$ region
    boundaries computed with parameter set 1 as in Fig.~\ref{svtfig}; dashed
    blue lines: $K$ and $K'$ boundaries computed with parameter set 2.
    Left: same-valley case. Right: different-valley case.
    See Table \ref{BLGParameterTable} for parameter values.}
\label{svtp2fig}
\end{center}
\end{figure}

\begin{figure}
\begin{center}  
  \includegraphics[width=4.4cm, angle=-90]{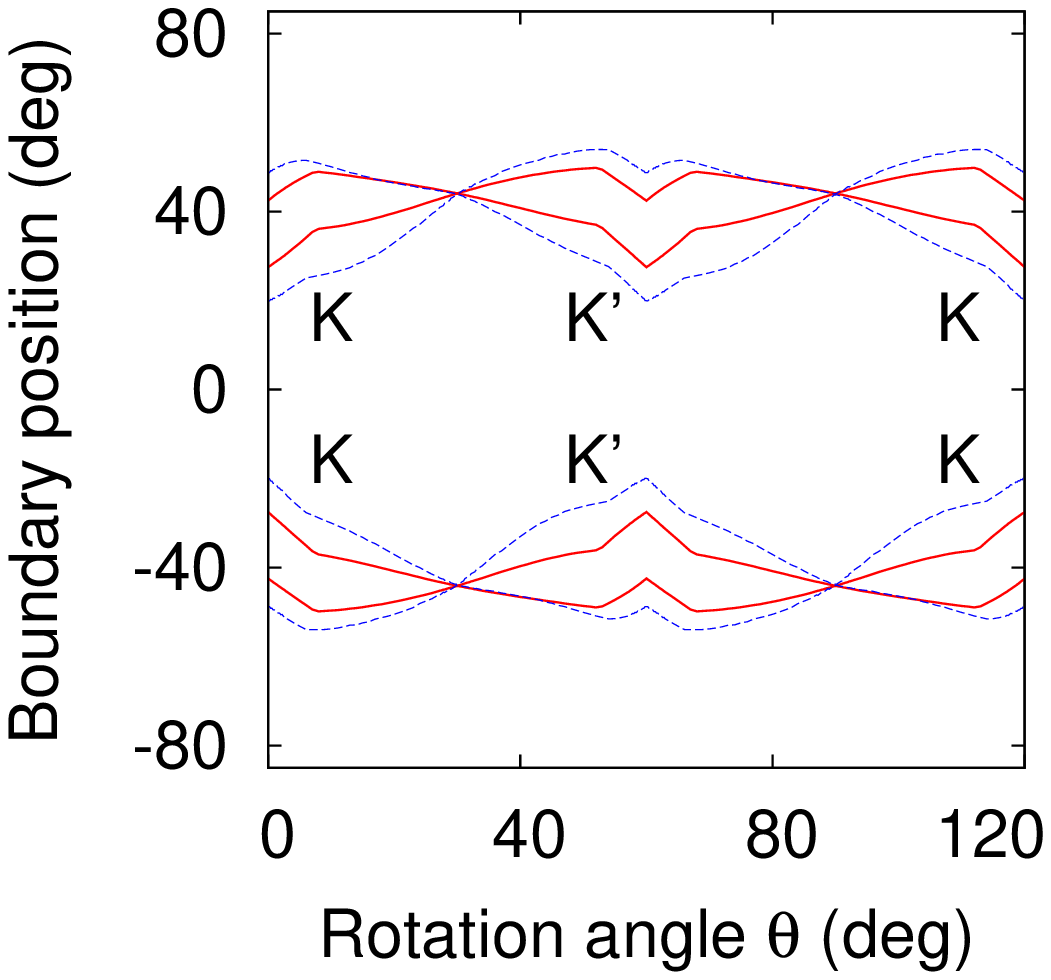}
  \includegraphics[width=4.4cm, angle=-90]{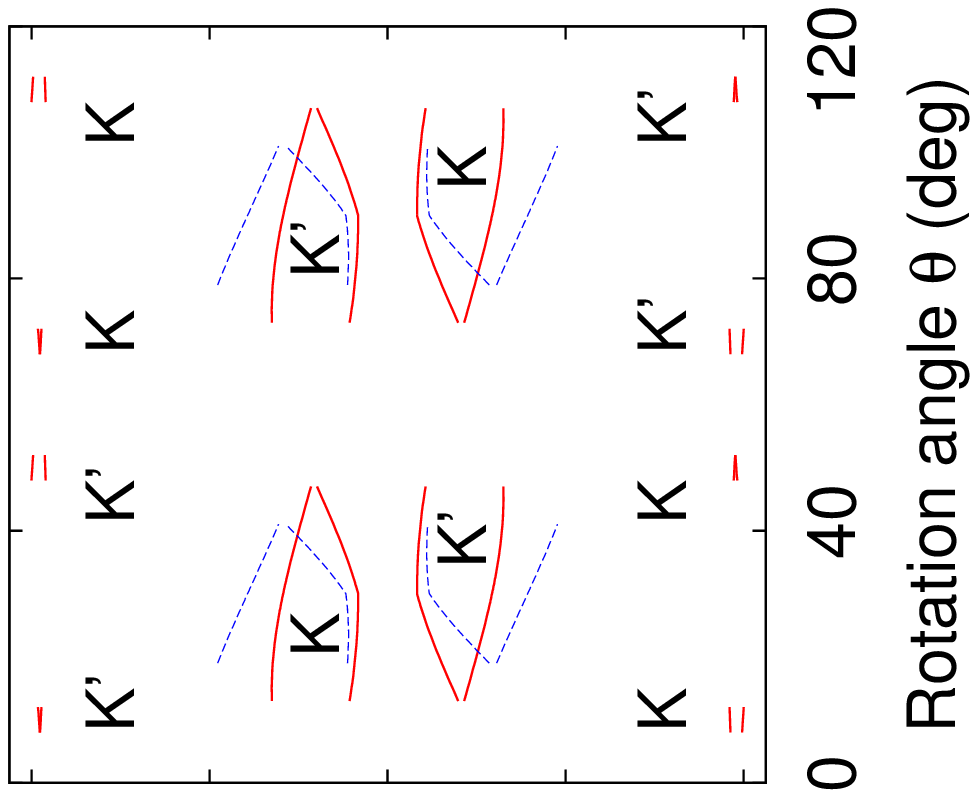}
  \caption{(Color online). Optimized single valley regions in BLG.
    Solid red lines: region
    boundaries computed with parameter set 3 and $E$ = 56 meV; 
    dashed blue lines: boundaries computed with $E$ = 19 meV.
    Left: same-valley case. Right: different-valley case.
    See Table~\ref{BLGParameterTable} for parameter values.}
\label{svtp3fig}
\end{center}
\end{figure}

Fig.~\ref{svtp2fig} shows region boundaries computed with parameter sets 1
and 2. The boundaries computed with parameter set 1 are identical to those
in Fig.~\ref{svtfig} and the tiny regions of different-valley behavior,
which are omitted from Fig.~\ref{svtfig} also shown. When parameter set 2
is used the range of single valley region widths becomes
$\sim 17.4 - 47.0^\circ$ which is similar to the range found with
parameter set 1. If parameter set 3 is used, the single valley range is
smaller, $\sim 7.5 -18.3^\circ$ (Fig.~\ref{svtp3fig}).
However the range width depends on energy and when the energy is decreased
to 19 meV, the range width becomes $\sim 9.6 - 29.0^\circ$.

\subsection{Sensitivity to Substrate}
\label{SubstrateSection}
There are reports that a small band gap occurs in MLG on an hBN substrate
\cite{mlggap}. If a similar gap occurred in BLG on hBN it could affect the
single valley regions, however there is experimental and theoretical
evidence that this effect is either small or experimentally controllable.
In particular, the authors of ref. \cite{Varlet14} were able to explain
their data on Fabry-Perot resonances in a BLG potential barrier without
considering interactions with the substrate. This is consistent with
the \textit{ab-initio} density functional calculations in ref.
\cite{Ramasubramaniam11}. The authors of this reference
computed the band structure of 3 different BLG-hBN heterostructures and found
that a substrate-induced gap occurs only in one case where the
heterostructure is asymmetric. The gap in this case is about 40 meV.

If a gap of this magnitude occurs, its effect can be compensated for by
adjusting the gate voltage. To show this, the optimization calculations
leading to Fig. \ref{svtfig} are repeated with substrate
interactions included. In ref. \cite{Ramasubramaniam11} the 40 meV gap results
from a $\sim +13$ meV shift of the conduction band edge and a $\sim -27$
meV shift of the valence band edge. The gap and shifts are modeled by
adding the mass terms $\mathrm{diag}(13, -13, 27, -27)$ to the
Hamiltonian. These terms are similar to the mass terms given by other
authors \cite{McCann13,Predin16} but are made asymmetric to model the
asymmetric shifts of the band edges in ref. \cite{Ramasubramaniam11}.

\begin{figure}
\begin{center}  
  \includegraphics[width=4.4cm, angle=-90]{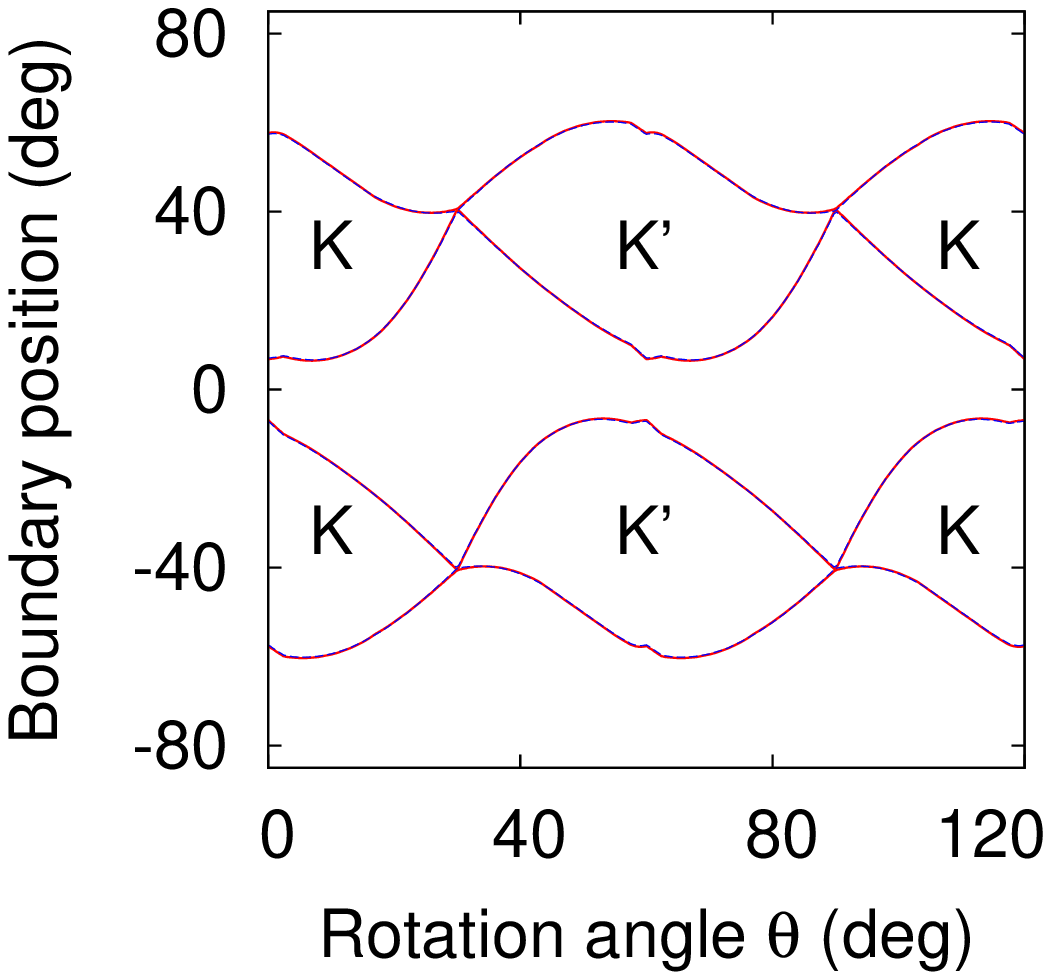}
  \includegraphics[width=4.4cm, angle=-90]{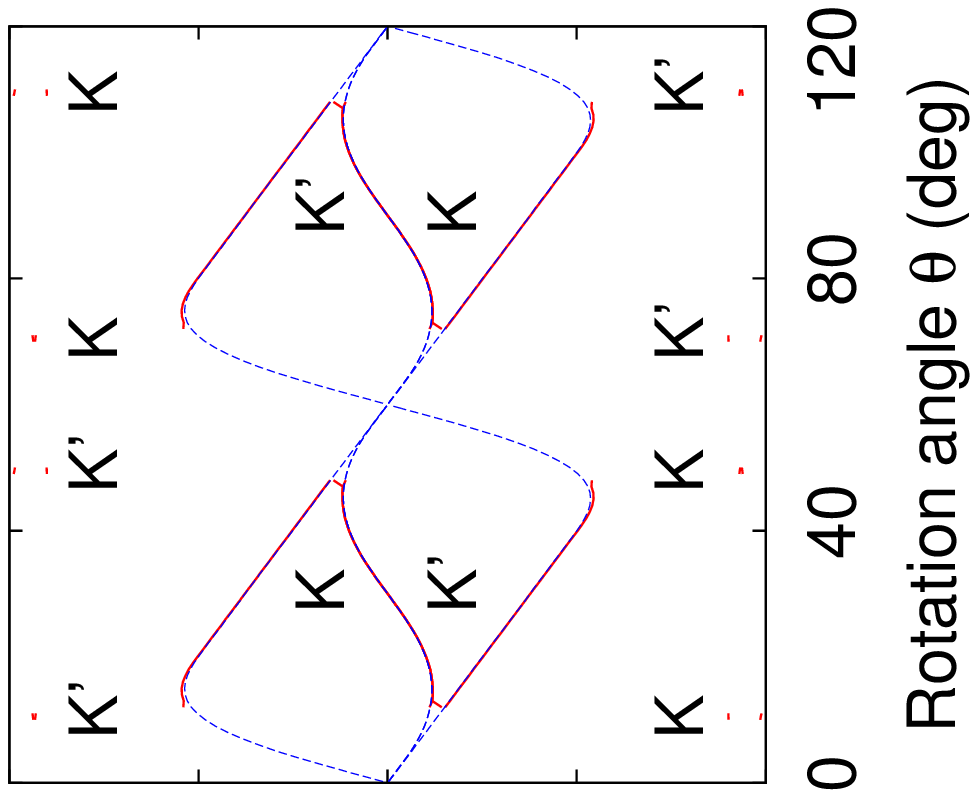}
  \caption{(Color online). Effect of gap on optimized single valley
    regions in BLG. Solid red lines: $K$ and $K'$ region
    boundaries computed with parameter set 1 as in Fig.~\ref{svtfig}; dashed
    blue lines: $K$ and $K'$ boundaries computed with gap included.
    Left: same-valley case. Right: different-valley case.}
\label{massgapfig}
\end{center}
\end{figure}

The optimized region boundaries with the gap included
(Fig.~\ref{massgapfig}) are almost identical to those computed without
including the gap. The only differences are that the tiny regions of single
valley transmission in the different-valley case are absent and the main
regions shrink down to the cut-offs which
occur within about $0.2^\circ$ of $0, 60$ and $120^\circ$. These
differences are too small to be of practical significance and are
probably caused by minor changes to the energy contours.
The optimal potentials with the gap are shifted by about 5-13 meV
from those shown in Fig.~\ref{optvfig}.

\subsection{Experimental Feasibility}
\label{BLGFeasibilitySection}

There are three important questions about the feasibility of observing the
predicted valley asymmetric transmission: Is the necessary bias in the
experimentally feasible range? Is the system in the ballistic transport
regime? and Is the trigonal warping strong enough?

The bias opens up a gap and to check whether the necessary bias is feasible
the predicted gaps are compared with experimentally observed gaps.
Fig.~\ref{optvfig} shows that the gap is at most $\sim 45$ meV. This is
significantly smaller than the largest reported transport gaps in bilayer
graphene which are up to $80-130$ meV \cite{Xia10,Miyazaki10,Yan10}. So it
is likely that the required bias can be achieved.

Ballistic transport in BLG occurs at sufficiently high carrier density
\cite{Cobaleda14,Nam17} and there are reports of operation of potential
barrier \cite{Varlet14} and antidot lattice \cite{Oka19} devices in the
ballistic regime. The electron density and temperature are $1.9$ -
$2.7\times 10^{12}$ cm$^{-2}$ and 1.6 K for the barrier device and $1$ -
$3\times 10^{12}$ cm$^{-2}$ and 4.2 K for the antidot device. These
densities can be compared with the density for the barrier device described
in section \ref{BLGVSection}. It is difficult to determine the density
accurately because of the uncertainty in the BLG Hamiltonian parameters
however with the parameters in Table~\ref{BLGParameterTable} the densities
are in the range $\sim 1.4$ - $1.7\times 10^{12}$ cm$^{-2}$ while without
trigonal warping the density for the same energy is $\sim 2.1\times
10^{12}$ cm$^{-2}$. These densities are similar to the experimental ones
and this suggests that the barrier device described in section
\ref{BLGVSection} would operate in the ballistic regime at low
temperature. In addition, the device described in ref. \cite{Varlet14} was
used to observe Fabry-Perot interference and this clearly shows that
experiments on barrier transmission in the ballistic regime are feasible.

There is less clarity about the trigonal warping.
Table~\ref{BLGParameterTable} shows that the value of the trigonal
warping parameter, $\gamma_3$, is not known accurately. If the actual value
lies between 380 and 300 meV as in parameter sets 1 and 2, the trigonal
warping is strong enough. If $\gamma_3$ is significantly smaller, the
predicted effects would be more difficult to observe but it is possible to
work at a lower energy to compensate for reduced trigonal warping (Section
\ref{BLGParameterSection}). In addition it may be possible to make a
collimator of narrower beamwidth.

\section{Single valley transmission in TMDs}
\label{TMDSection}

\subsection{Overview}
\label{OvtmdSection}

In this section we show that valley asymmetric transmission occurs in all
the semiconducting monolayer TMDs and in the most favorable case,
MoTe$_2$, the single valley region widths are similar to those in BLG.
However there are two important differences between BLG and TMDs.  First,
the most favorable carriers are holes as trigonal warping in TMDs is
strongest in the valence band. Secondly, spin-valley locking \cite{Xiao12}
ensures that the valence bands at $K$ and $K'$ are of definite and opposite
spin. Consequently, valley polarized currents are also spin polarized
provided that the Fermi level is above the top of the lower spin-split
valence band.

We start by explaining the TMD Hamiltonian (section \ref{TMDHSection}). To
ensure the trigonal warping is described to sufficient accuracy we
calculate the transmission coefficients with a 4-band
$\mathbf{k}\cdot\mathbf{p}$ Hamiltonian \cite{Kormanyos13}. However the
parameters of this Hamiltonian are not in the literature and we have
obtained them by fitting to \textit{ab-initio} band structures
(\ref{TMDparamSection}). The valley asymmetric transmission and single
valley regions for MoTe$_2$ are detailed in section \ref{Vamote2Section}.
Single valley regions for all the semiconducting TMDs are compared
in section \ref{TMDsvSection} and the requirements for observing the predicted
valley asymmetry are discussed in section \ref{TMDFeasibilitySection}.

\subsection{TMD Hamiltonians}
\label{TMDHSection}

\begin{table}
\begin{tabular}{lrrrrrr}
\hline
& MoS$_2$ & MoSe$_2$ & MoTe$_2$ & WS$_2$ & WSe$_2$ & WTe$_2$ \\\hline
$k$-range & $\pm$0.06 & $\pm$0.06 & $\pm$0.05 & $\pm$0.03 & $\pm$0.03 & $\pm$0.03 \\\hline
$\epsilon_v$ & 0.0 & 0.0 & 0.0 & 0.0 & 0.0 & 0.0 \\
$\epsilon_c$ & 1657.9 & 1429.3 & 1071.7 & 1806.2 & 1541.2 & 1066.8 \\
$\epsilon_{v-3}$ & -3500.0 & -2897.0 & -3670.0 & -3370.0 & -3220.0 & -3180.0 \\
$\epsilon_{c+2}$ & 3512.6 & 3003.4 & 2483.8 & 3990.6 & 3419.1 & 2805.9 \\
$\gamma_2$ & 185.3 & 179.1 & 88.2 & 154.3 & 157.3 & 3.1 \\
$\gamma_3$ & 309.2 & 274.9 & 243.5 & 322.2 & 342.7 & 269.7 \\
$\gamma_4$ & -275.1 & -250.9 & -189.4 & -436.9 & -294.3 & -352.5 \\
$\gamma_5$ & -401.9 & -333.8 & -470.0 & -608.3 & -469.9 & -592.4 \\
$\gamma_6$ & 44.6 & 52.6 & -97.2 & 52.3 & 61.2 & -74.1 \\
$\lambda$ & 74.0 & 92.0 & 107.5 & 215.0 & 233.0 & 243.0 \\
\hline
\end{tabular}
\caption{TMD Hamiltonian parameters. $k$-ranges are in nm$^{-1}$, band and
  SO energies are in meV and the $\gamma$ parameters are in meV nm.}
\label{TMDParameterTable}
\end{table}

The total Hamiltonian is the sum of the band Hamiltonian, the SO
Hamiltonian and the external potential. A
$\mathbf{k}\cdot\mathbf{p}$ Hamiltonian is appropriate for computing
transmission coefficients because the potential barrier has a soft wall
that varies slowly on an atomic scale. A 2-band $\mathbf{k}\cdot\mathbf{p}$
Hamiltonian is available \cite{Kormanyos13,Kormanyos15} but we have found
it does not reproduce trigonal warping well in the required energy range.
Instead we use the 4-band $\mathbf{k}\cdot\mathbf{p}$ Hamiltonian given in
the same references. However the parameters of this Hamiltonian are not in
the literature. We obtain them by fitting to \textit{ab-initio} band
structures (Table~\ref{TMDParameterTable} and section
\ref{TMDparamSection}).

The SO Hamiltonian is taken from ref. \cite{Kormanyos13} but only the
lowest order contributions are included as in ref. \cite{Liu13}.
This leads to the sum of band and SO Hamiltonians given in Eq.~(\ref{htmd}).
The $\lambda$ parameter (Table~\ref{TMDParameterTable}) is taken to be
1/2 of the SO splitting reported in ref. \cite{Liu13}. 

The external potential is taken to be a scalar function, $V(x')$.
As a function of lateral position, $V(x')$ is constant in the
barrier and at the barrier edges it decreases to zero with the same wall
function, $F(x')$, as used to model the BLG potential.
The parameters of $F(x')$ are also the same as for BLG.

The sum of the 4-band $\mathbf{k}\cdot\mathbf{p}$ Hamiltonian and $V(x')$
is used to compute transmission coefficients and single valley regions. In
addition the single valley regions are computed with the tight binding
Hamiltonian in ref. \cite{Liu13} which includes interactions up to third
neighbors. This gives excellent agreement with \textit{ab-initio} bands
and is used to check the accuracy of the single valley regions computed
with the 4-band $\mathbf{k}\cdot\mathbf{p}$ Hamiltonian.

\subsection{4-band $\mathbf{k}\cdot\mathbf{p}$ Hamiltonian parameters}
\label{TMDparamSection}

The 4-band $\mathbf{k}\cdot\mathbf{p}$ Hamiltonian is derived by
symmetry arguments in ref. \cite{Kormanyos13}. In the $K$ valley, and
without SO coupling, the 4-band Hamiltonian exactly as stated in
ref. \cite{Kormanyos13} is 
\begin{equation}
  H_{0K} = 
    \left( \begin{array}{cccc}
    \epsilon_v  & \gamma_3q_- & \gamma_2q_+  & \gamma_4q_+\\
    \gamma_3q_+ & \epsilon_c & \gamma_5q_- & \gamma_6q_-\\
    \gamma_2q_-  & \gamma_5q_+ & \epsilon_{v-3} & 0\\
    \gamma_4q_- & \gamma_6q_+ & 0 & \epsilon_{c+2}\\
    \end{array} \right), \label{htmdfit}
 \end{equation}
where $q_\pm = q_x \pm iq_y$ and $\mathbf{q}$ is the $\mathbf{k}$-vector
relative to the $K$ point. When crystallographic co-ordinates are chosen
as in ref. \cite{Kormanyos13}, $E(\mathbf{q})$ is a symmetric function of
$q_y$. Consequently, the characteristic polynomial cannot contain any terms
of odd order in $q_y$ and this implies that all the $\gamma$ parameters can
be taken to be real. To prove this, notice that a unitary transformation can
be used to make $\gamma_2$, $\gamma_3$ and $\gamma_4$ real. Then after
multiplying out the secular determinant it can be seen that the
characteristic polynomial contains no terms of odd order in $q_y$ provided
that $\gamma_5$ and $\gamma_6$ are also real.

We require the values of the parameters $\gamma_i = \hbar c_i$ and obtain
them by fitting to \textit{ab-initio} band structures. The third neighbor
tight-binding Hamiltonian in ref. \cite{Liu13} is used to generate
\textit{ab-initio} data for fitting and the tight binding parameters used
are the GGA parameters in Table III of this reference.  However this
Hamiltonian does not include the $v-3$ band. The value of $\epsilon_{v-3}$
is taken from the Materials Project database \cite{Jain13}.

The $\gamma_i$ are fitted with non-linear least squares. It is only
necessary to fit on the $\Gamma K M$ line because the 4-band Hamiltonian is
based on symmetry and gives the correct interpolation of $E(\mathbf{k})$
away from this line. This has been confirmed with numerical tests. The fit
is restricted to the valence and conduction bands because the 4-band
Hamiltonian does not reproduce the remote bands, $v-3$ and $c+2$, well.
100 $k$-points on a uniform grid are sampled from each band.
The $k$-range used for the fitting has to be chosen carefully. If it is
too small, $E(\mathbf{k})$ is not reproduced well at the desired energies.
However if it is too large, artifacts appear in the form of extra peaks in
$E(\mathbf{k})$; presumably because the $\mathbf{k}\cdot\mathbf{p}$
approximation breaks down far from the $K$ point. To minimize these
difficulties the fitting range is made as a large as possible without
introducing artifacts.

The $k$-ranges, band edge energies and fitted $\gamma$ parameters
are given in Table~\ref{TMDParameterTable}. The signs of the
$\gamma$ parameters are determined only to within a unitary
transformation. For example, the unitary transformation
$\mathrm{diag}(1, 1, 1, -1)$ can be used to change the signs of $\gamma_4$
and $\gamma_6$.

\subsection{Valley asymmetry in MoTe$_2$}
\label{Vamote2Section}

\begin{figure}
\begin{center}  
  \includegraphics[width=4.4cm, angle=-90]{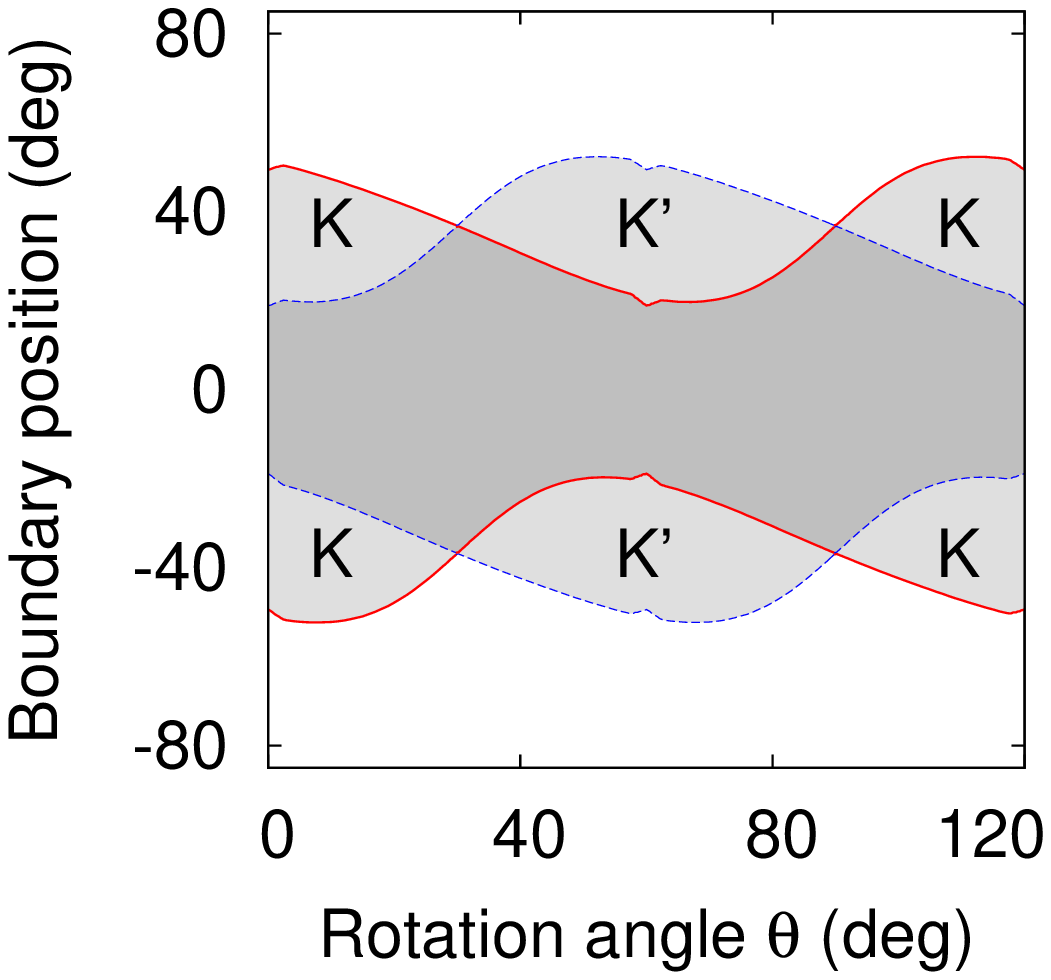}
  \includegraphics[width=4.4cm, angle=-90]{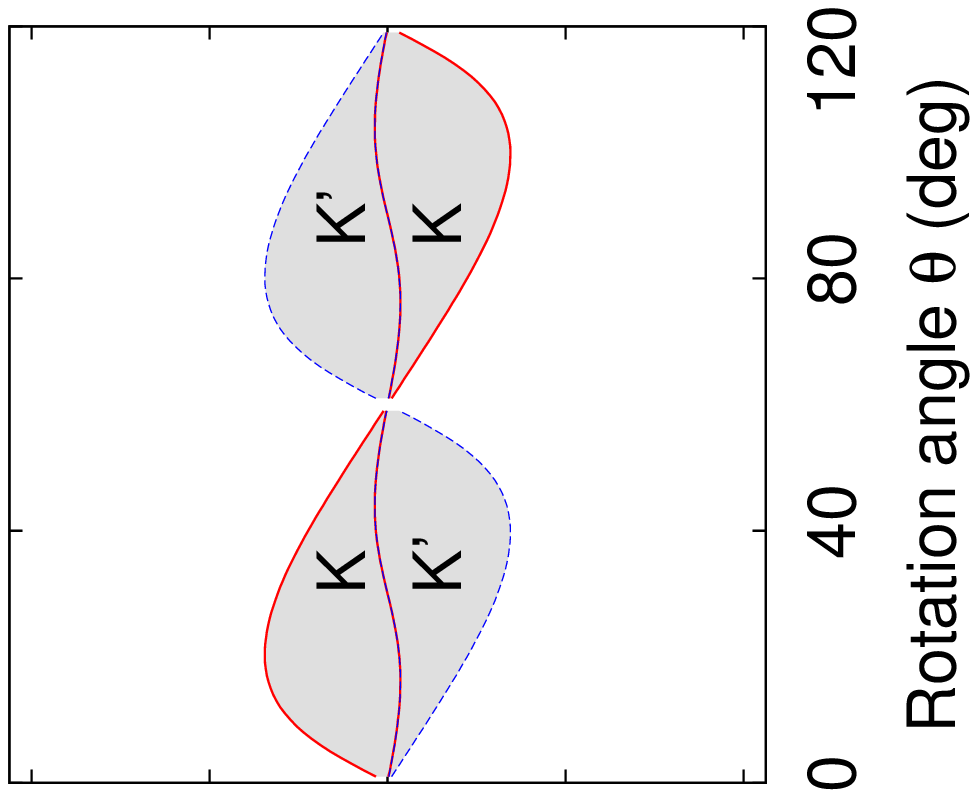}
  \caption{(Color online). Optimized single valley regions for 116.9 meV
    holes in MoTe$_2$.
    Solid red lines: $K$ transmission boundaries, dashed blue lines: $K'$
    boundaries. Light fill: single valley transmission, dark
    fill: two valley transmission.
    Left: same valley at $\pm$ incidence. Right: different valleys
    at $\pm$ incidence; in this case cut-offs similar to those in
    Fig.~\ref{svtfig} occur very close to  $\theta = 0$, $60$ and
    $120^\circ$.}
\label{tmdsvtfig}
\end{center}
\end{figure}

Fig.~\ref{tmdTfig} shows single valley transmission of holes in
MoTe$_2$. $K$ and $K'$ are defined as in ref.~\cite{Kormanyos13} and spin
up holes are transmitted in the $K$ valley. The barrier width is 300 nm, as
for BLG, and the transmission coefficients are qualitatively similar to
those for BLG. However the cut-offs at the critical angles are much
sharper than for BLG. The reason is that the evanescent wave decay lengths
in TMDs are typically an order of magnitude smaller than in BLG.
Consequently the transmission coefficients in the total external
reflection regime are much smaller, typically $<10^{-27}$ at 0.1$^\circ$
into this regime in MoTe$_2$. 

MoTe$_2$ is the most favorable TMD as it has the largest single valley
region widths of all the TMDs. Fig.~\ref{tmdsvtfig} shows that the
optimized region widths, $\sim 16.0 - 30.6^\circ$, are similar to those in
BLG. The region widths of all the semiconducting TMDs are compared in the
next sub-section.

\subsection{Comparison of Single Valley Regions in semiconducting TMDs}
\label{TMDsvSection}

\begin{table}
\begin{tabular}{lrrrrrr}
\hline
& MoS$_2$ & MoSe$_2$ & MoTe$_2$ & WS$_2$ & WSe$_2$ & WTe$_2$ \\\hline
$\mathbf{k}\cdot\mathbf{p}$ & 121.7 & 112.2 & 116.9 & 97.3 & 95.2 & 114.2 \\
\textit{ab-initio} & 125.3 & 116.9 & 117.7 & 108.3 & 108.1& 129.2 \\
\hline
\end{tabular}
\caption{Hole Fermi energies in meV corresponding to hole densities of
  $3\times 10^{13}$ cm$^{-2}$ in MoX$_2$ and
    $1.5\times 10^{13}$ cm$^{-2}$ in WX$_2$.}
\label{EnergyTable}
\end{table}

\begin{figure*}
  \includegraphics[width=3.7cm, angle=-90]{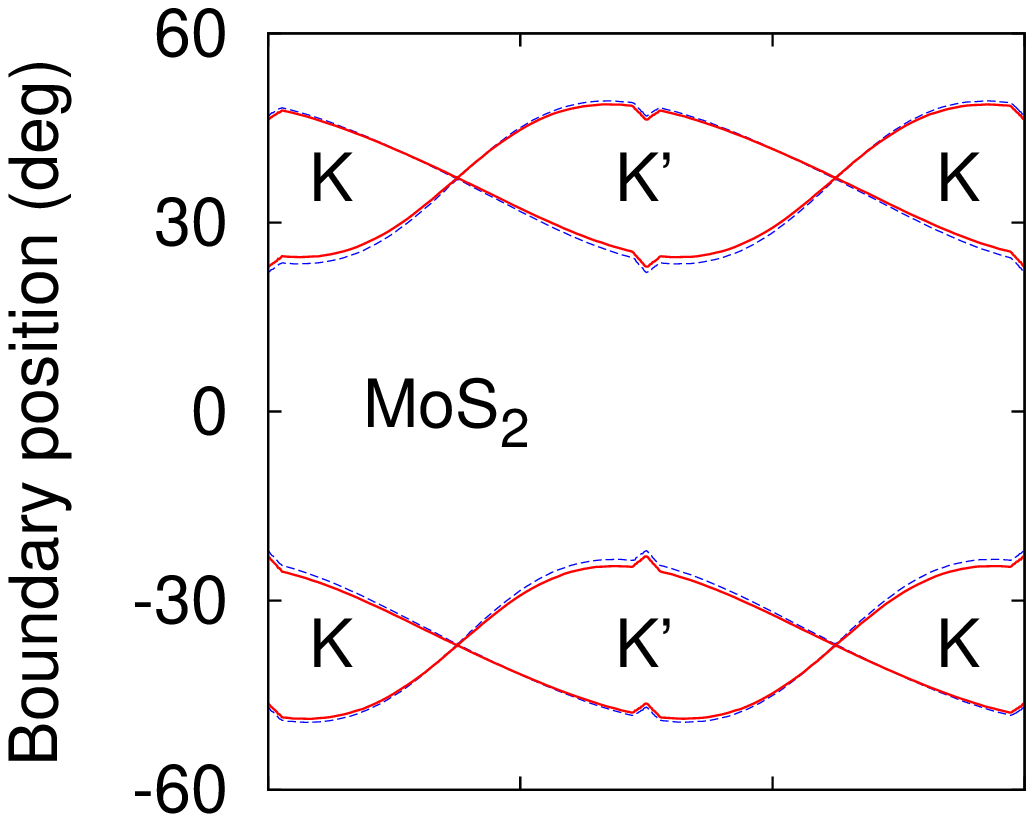}\hspace{0.5mm}
  \includegraphics[width=3.7cm, angle=-90]{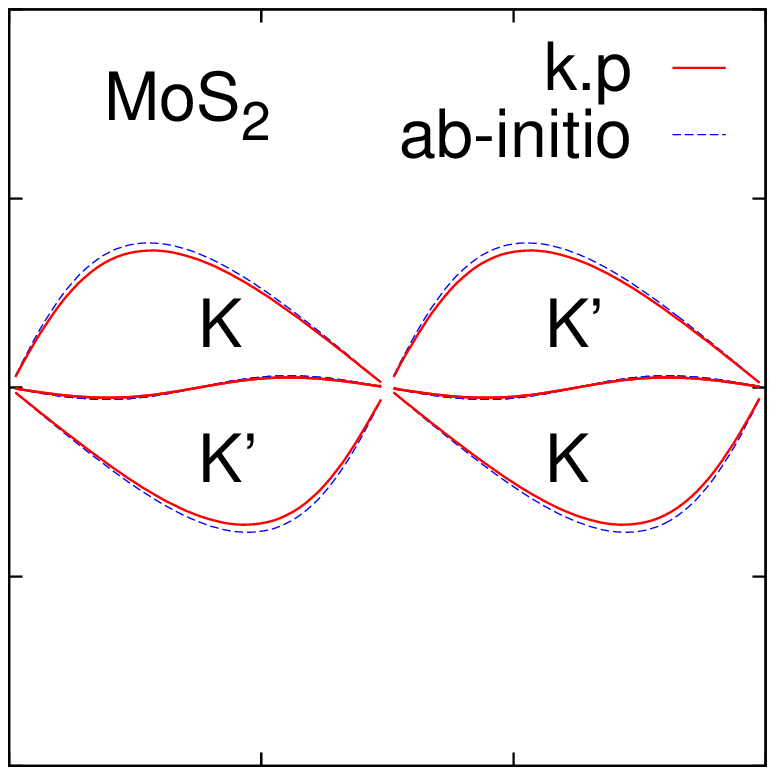}\hspace{6mm}
  \includegraphics[width=3.7cm, angle=-90]{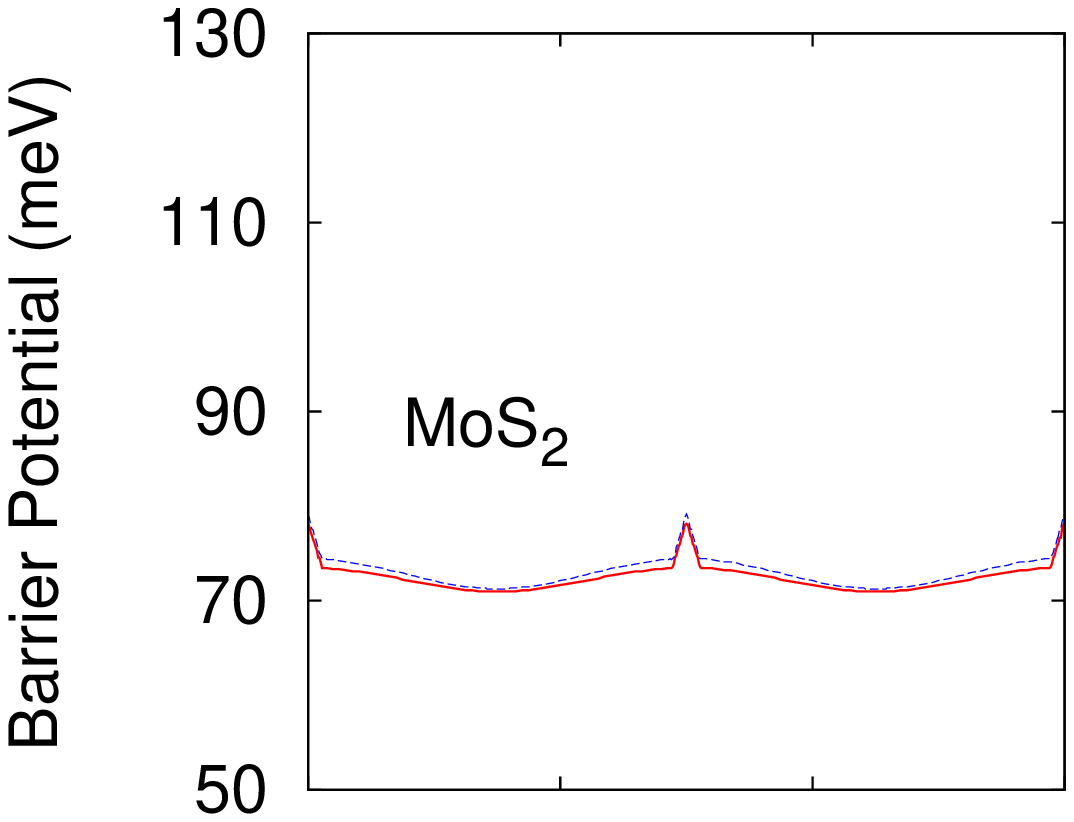}\hspace{0.5mm}
  \includegraphics[width=3.7cm, angle=-90]{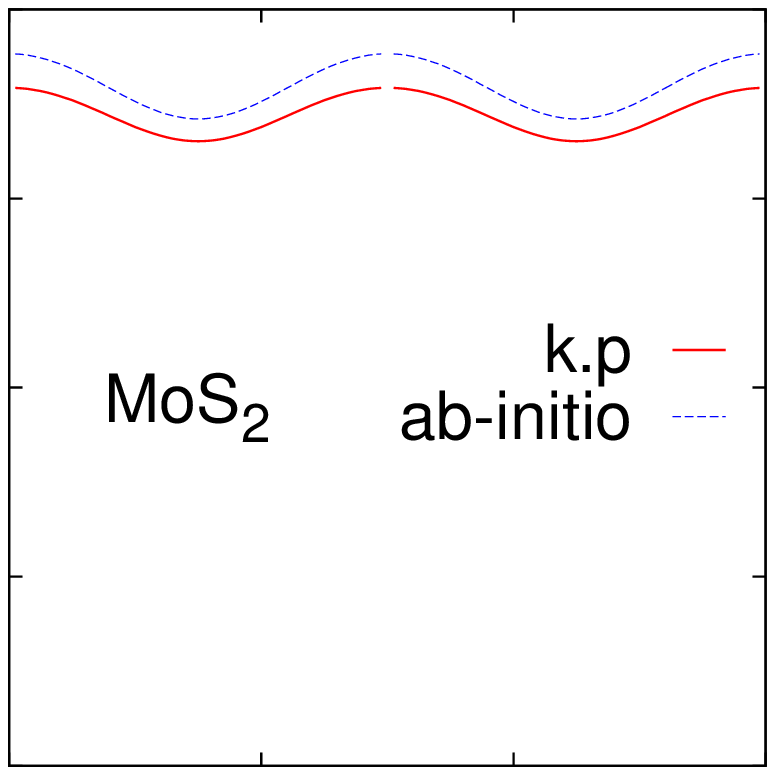}\\[1ex]
  
  \includegraphics[width=3.7cm, angle=-90]{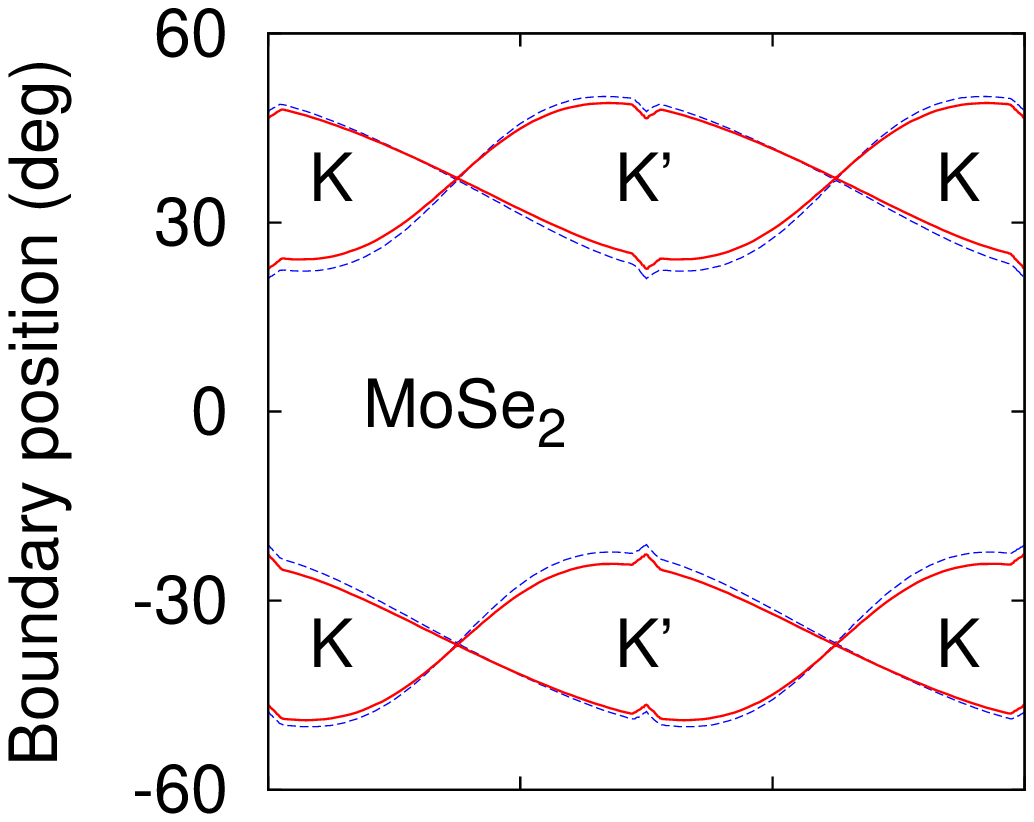}\hspace{0.5mm}
  \includegraphics[width=3.7cm, angle=-90]{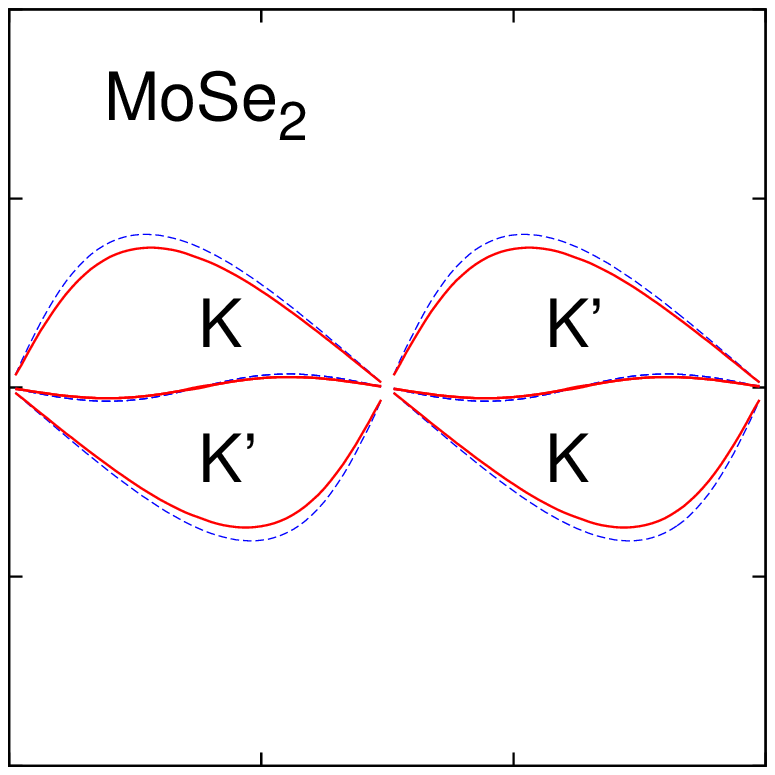}\hspace{6mm}
  \includegraphics[width=3.7cm, angle=-90]{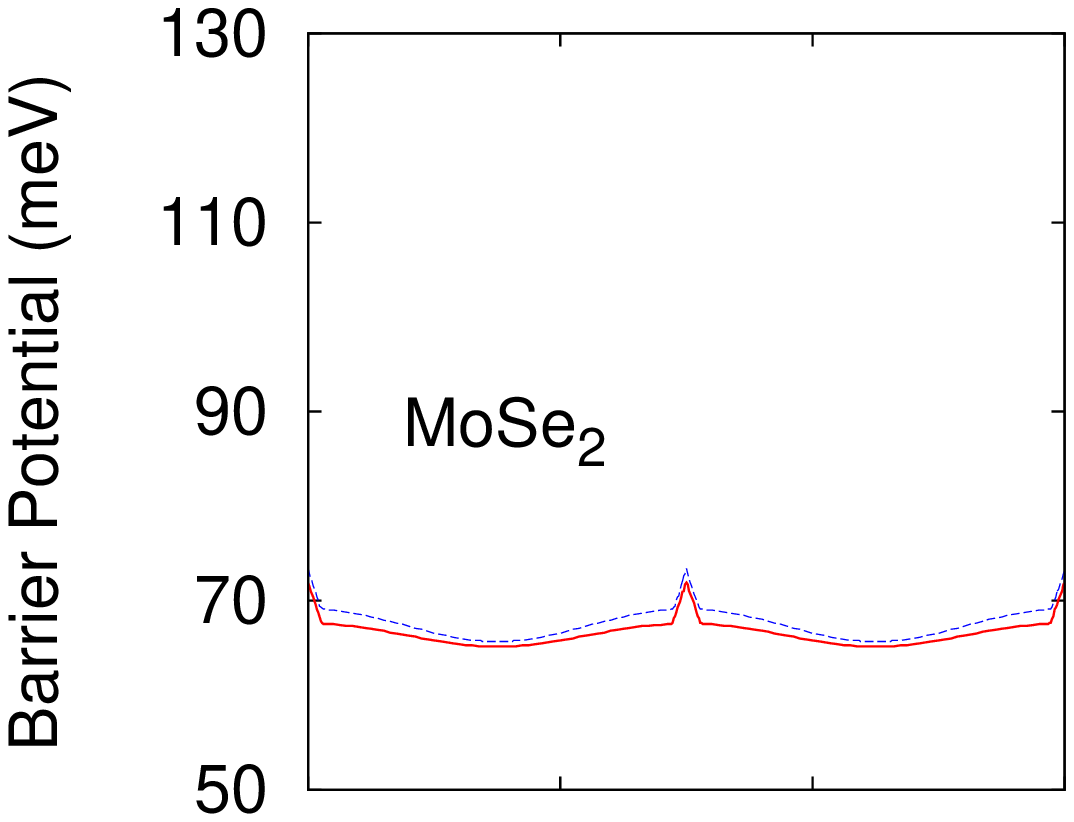}\hspace{0.5mm}
  \includegraphics[width=3.7cm, angle=-90]{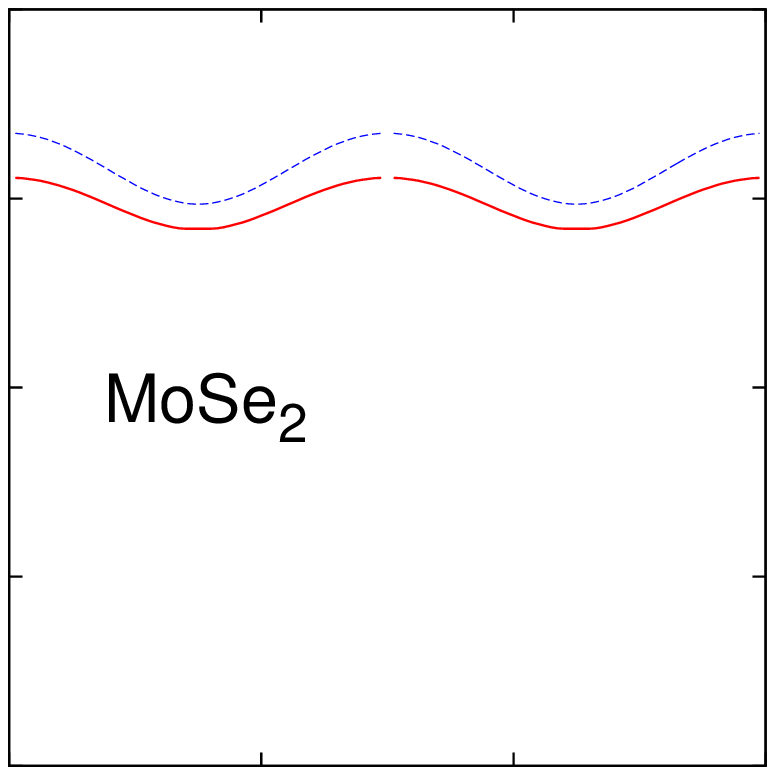}\\[1ex]
  
  \includegraphics[width=3.7cm, angle=-90]{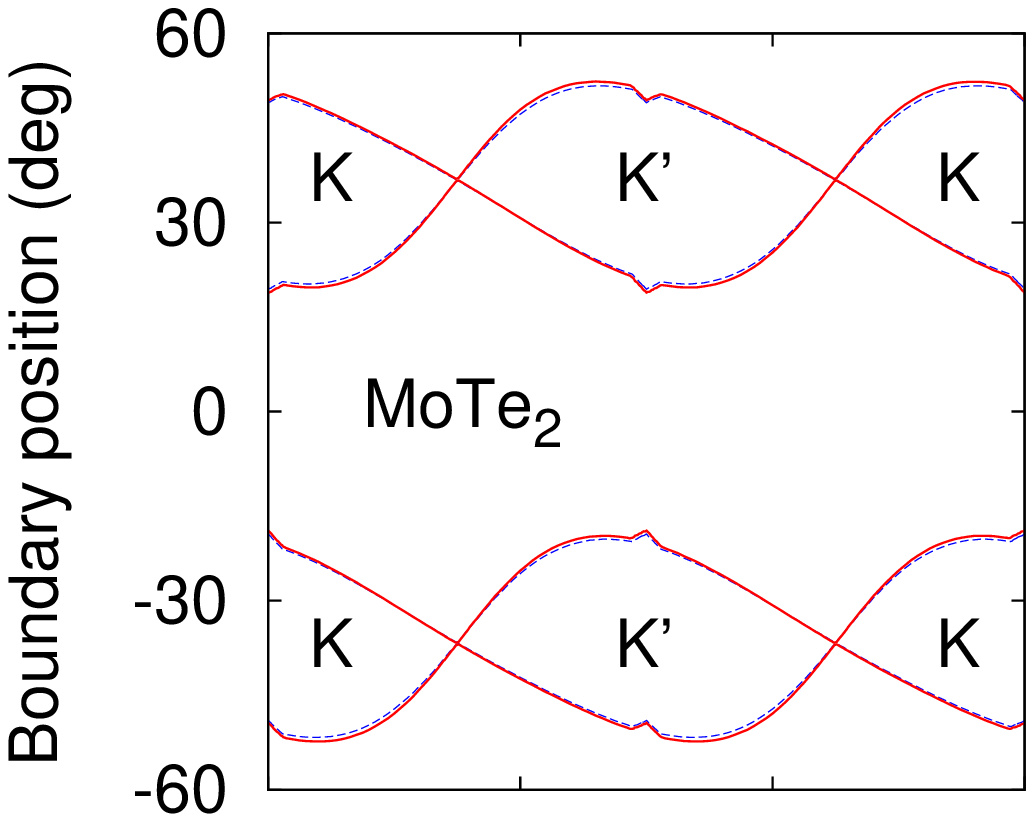}\hspace{0.5mm}
  \includegraphics[width=3.7cm, angle=-90]{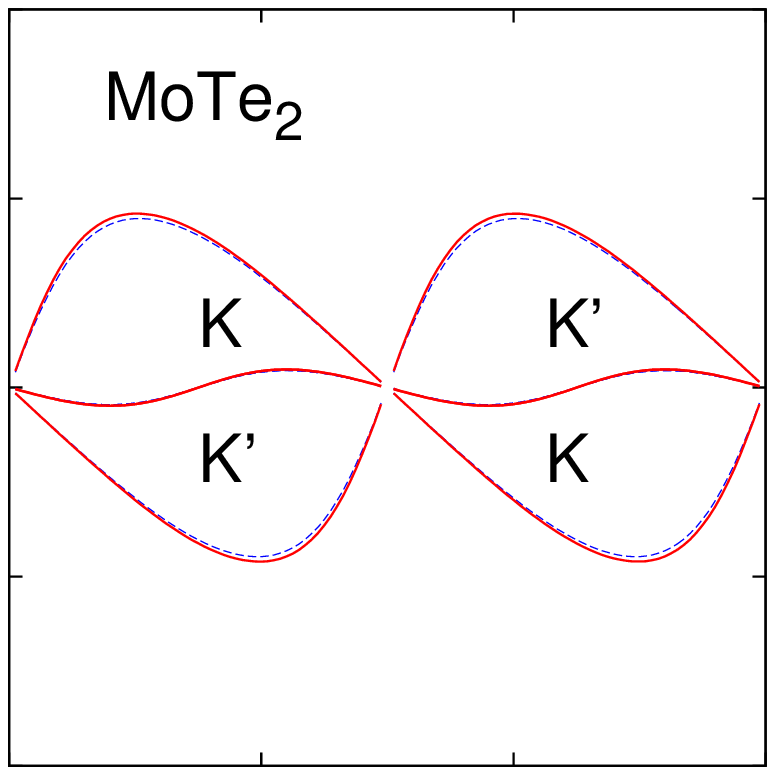}\hspace{6mm}
  \includegraphics[width=3.7cm, angle=-90]{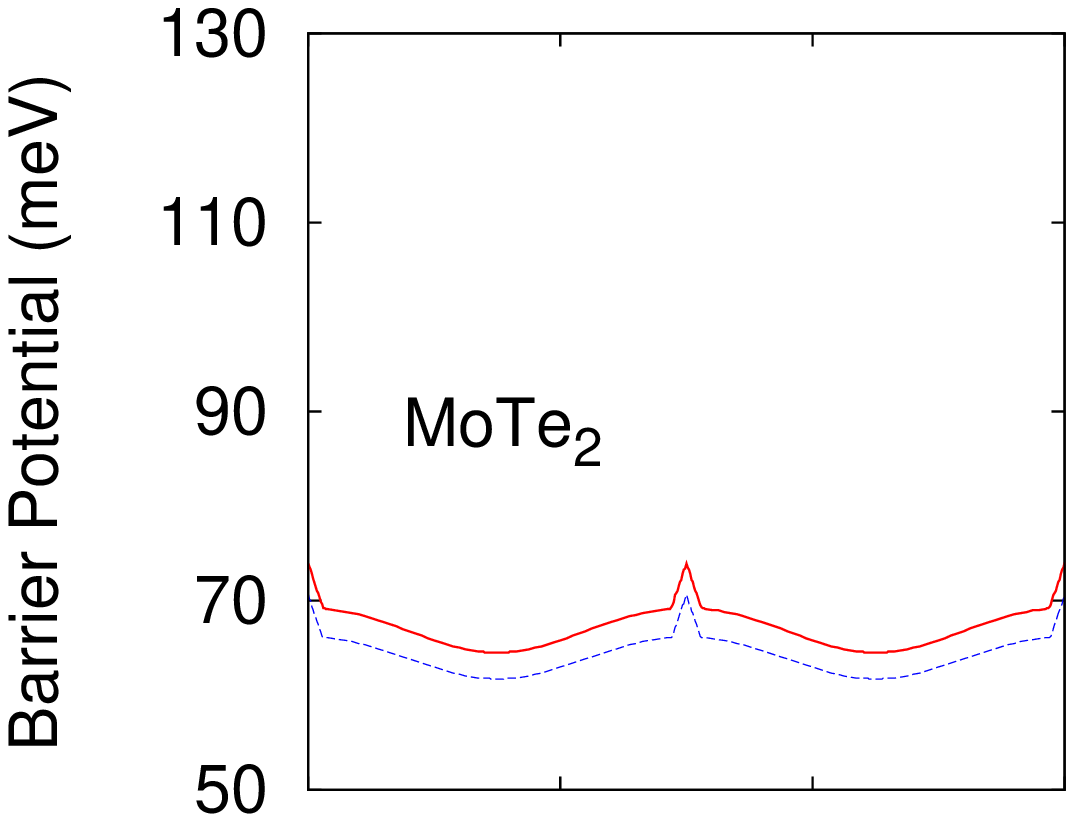}\hspace{0.5mm}
  \includegraphics[width=3.7cm, angle=-90]{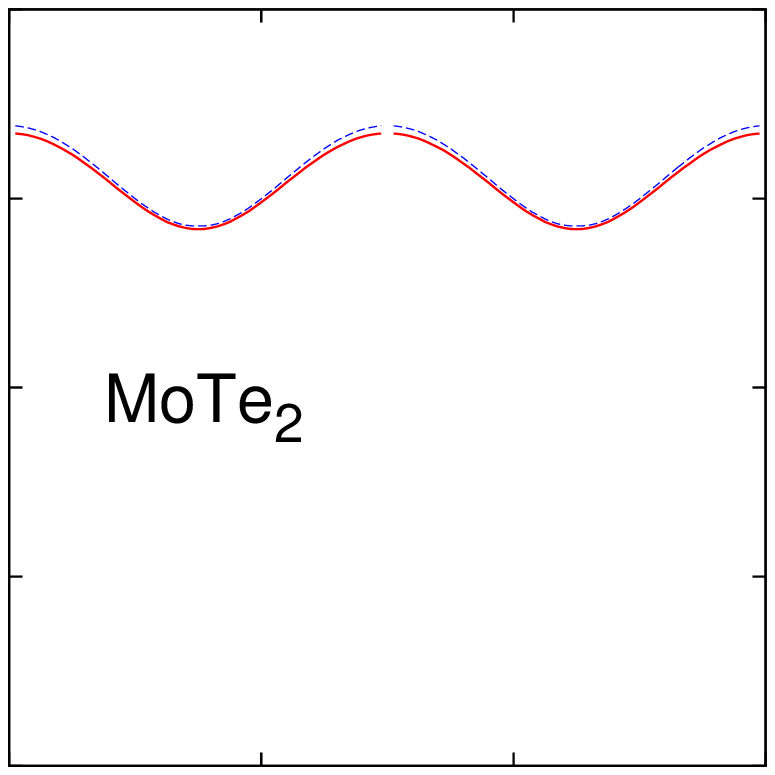}\\[1ex]

  \includegraphics[width=3.7cm, angle=-90]{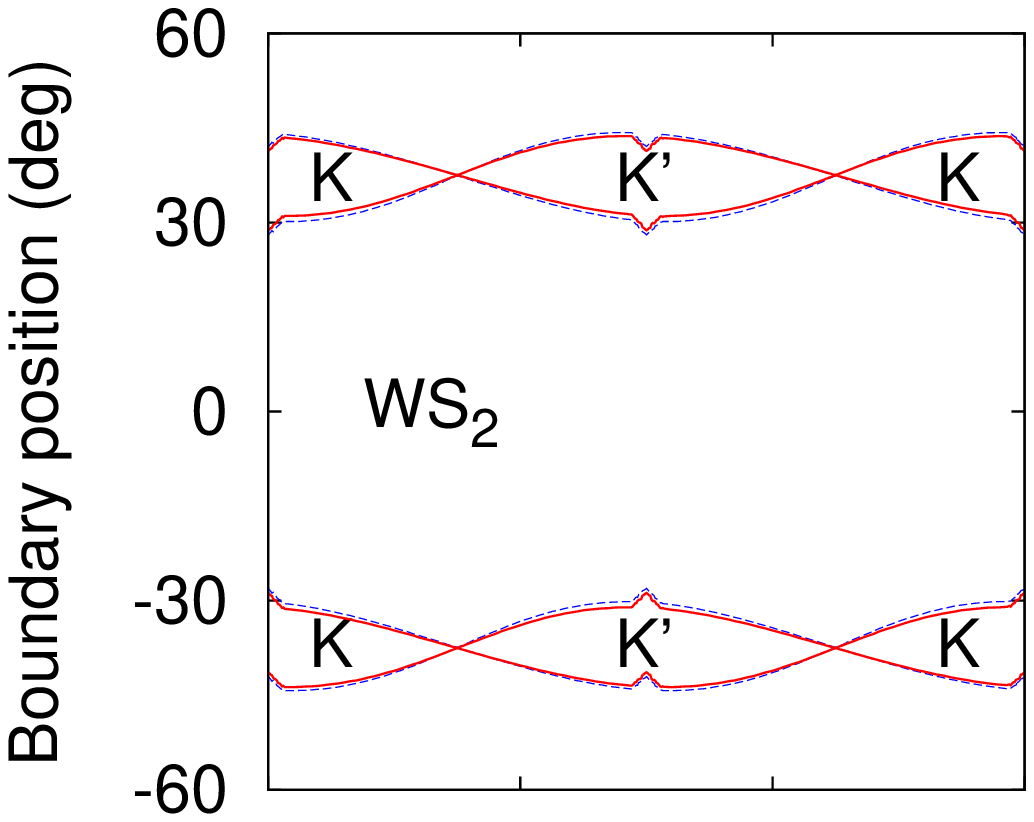}\hspace{0.5mm}
  \includegraphics[width=3.7cm, angle=-90]{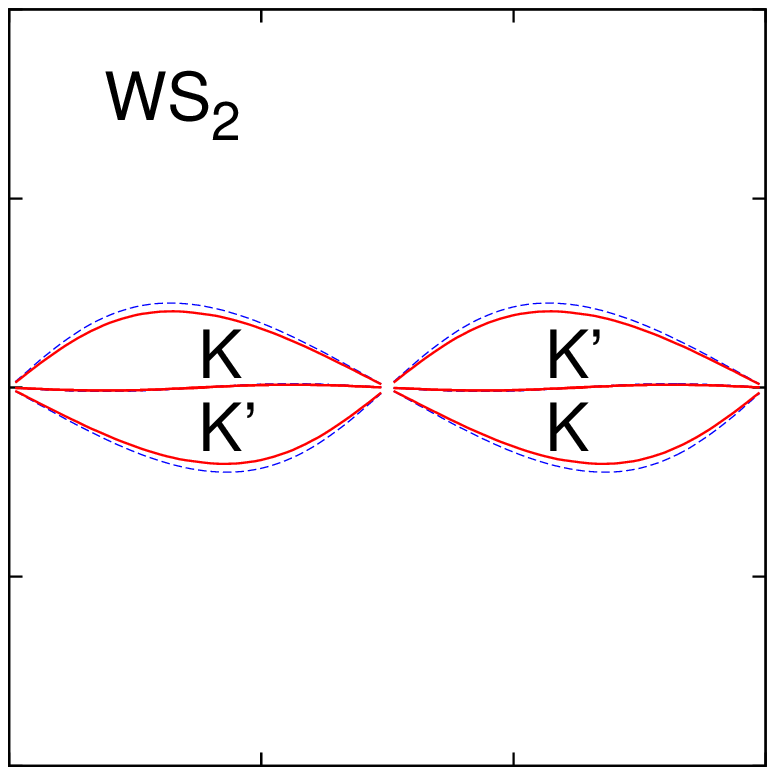}\hspace{6mm}
  \includegraphics[width=3.7cm, angle=-90]{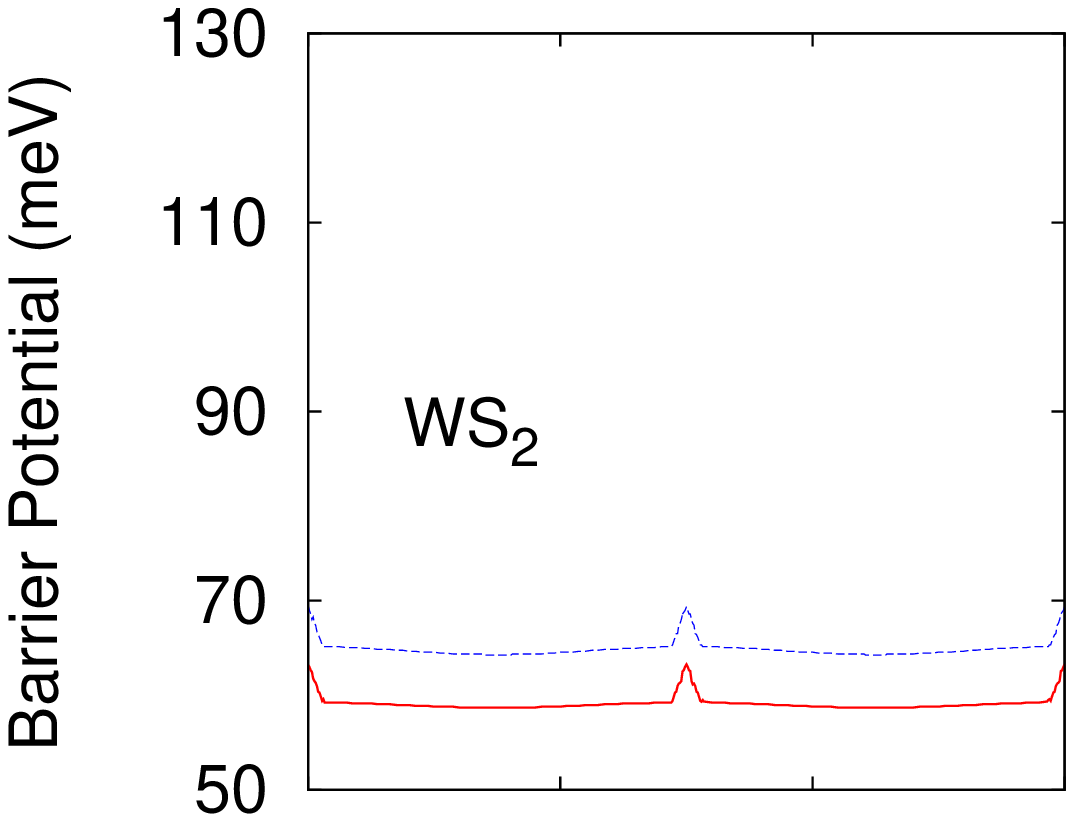}\hspace{0.5mm}
  \includegraphics[width=3.7cm, angle=-90]{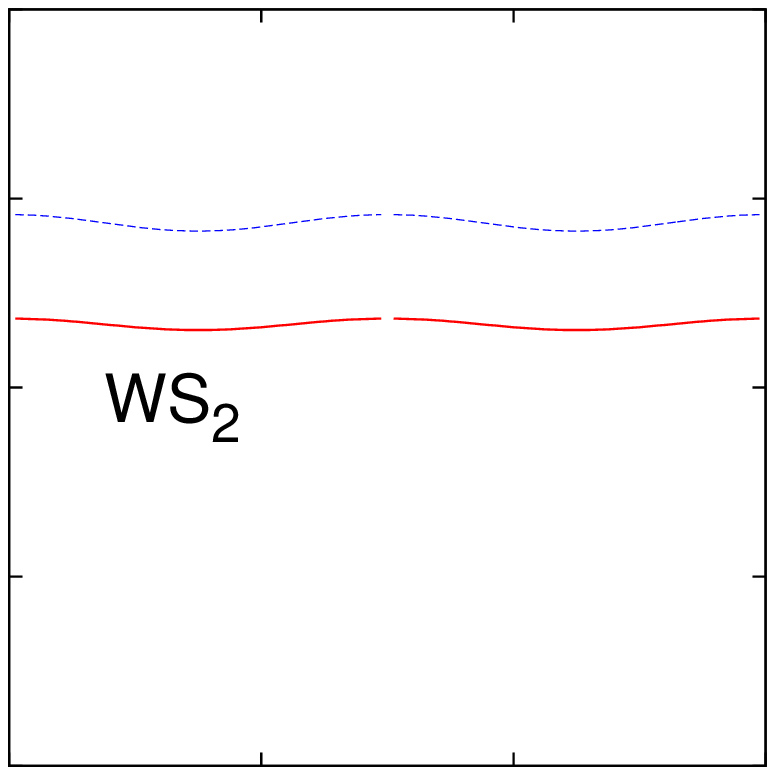}\\[1ex]

  \includegraphics[width=3.7cm, angle=-90]{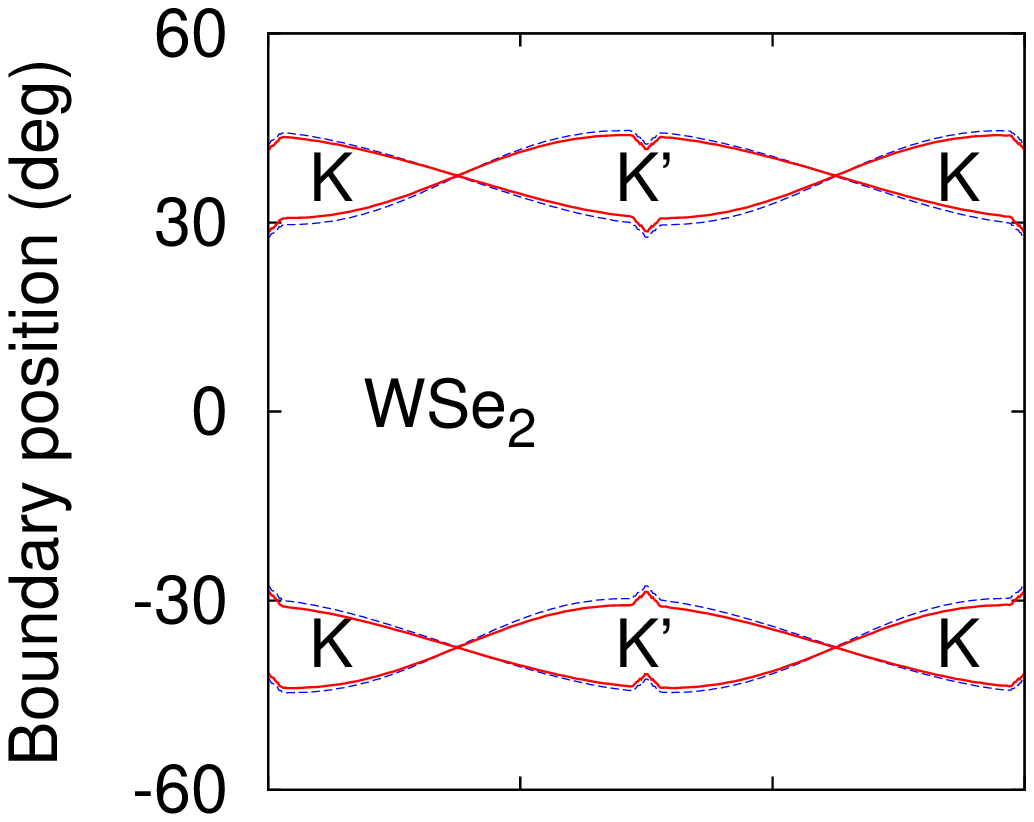}\hspace{0.5mm}
  \includegraphics[width=3.7cm, angle=-90]{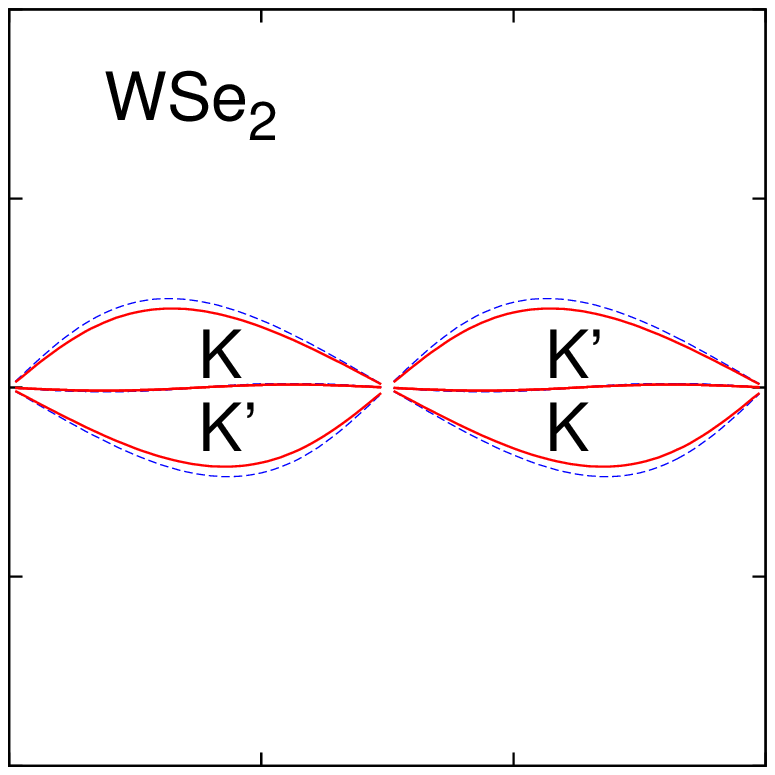}\hspace{6mm}
  \includegraphics[width=3.7cm, angle=-90]{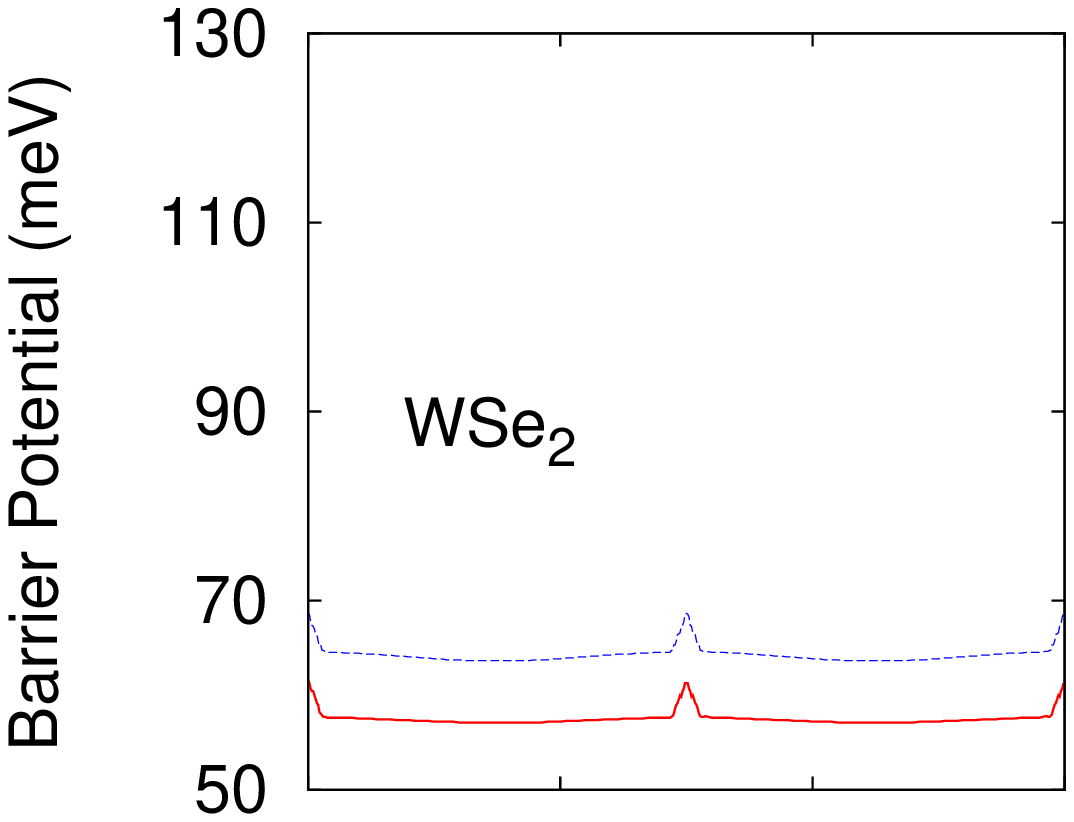}\hspace{0.5mm}
  \includegraphics[width=3.7cm, angle=-90]{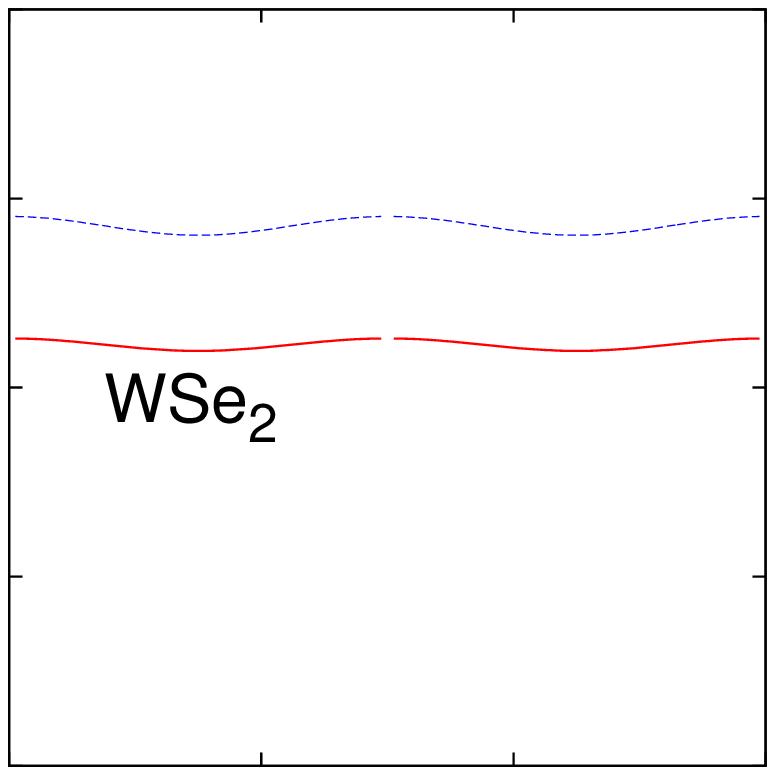}\\[1ex]

  \includegraphics[width=3.7cm, angle=-90]{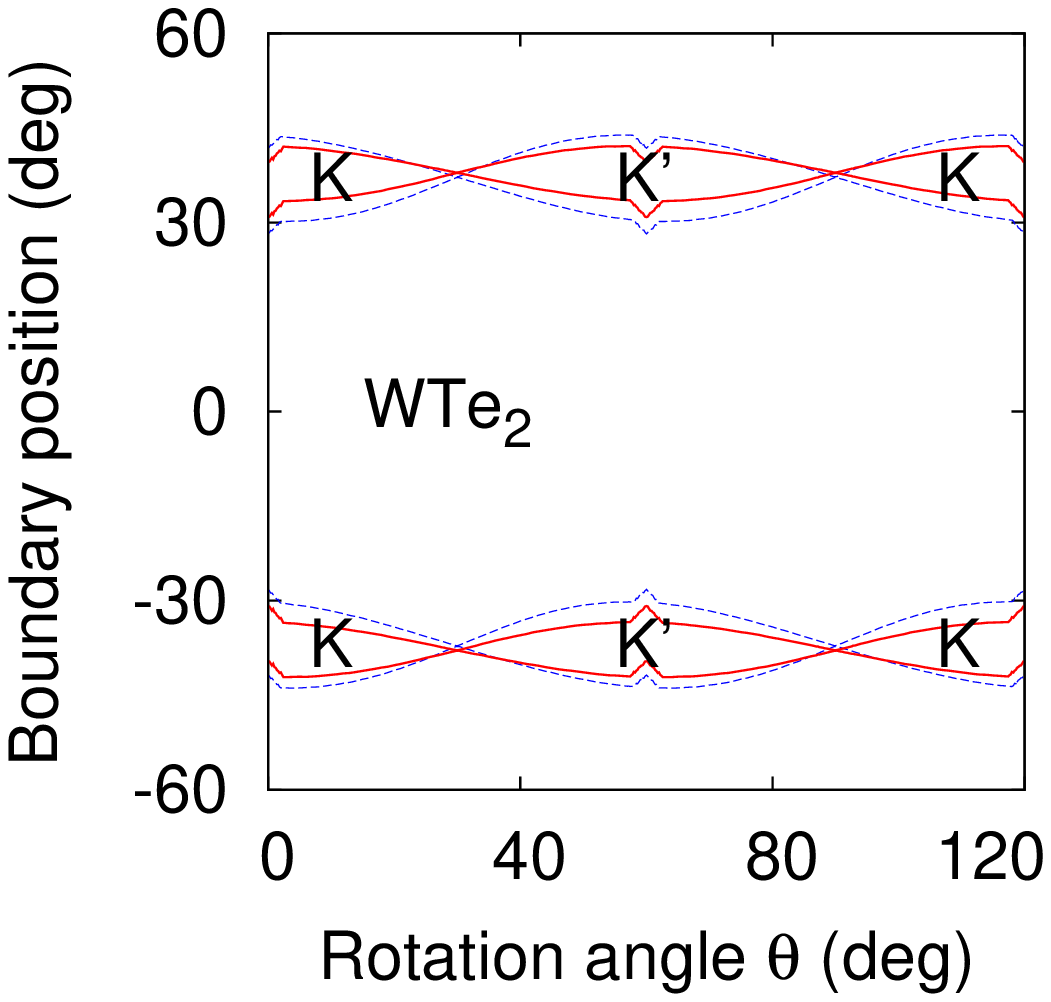}\hspace{0.5mm}
  \includegraphics[width=3.7cm, angle=-90]{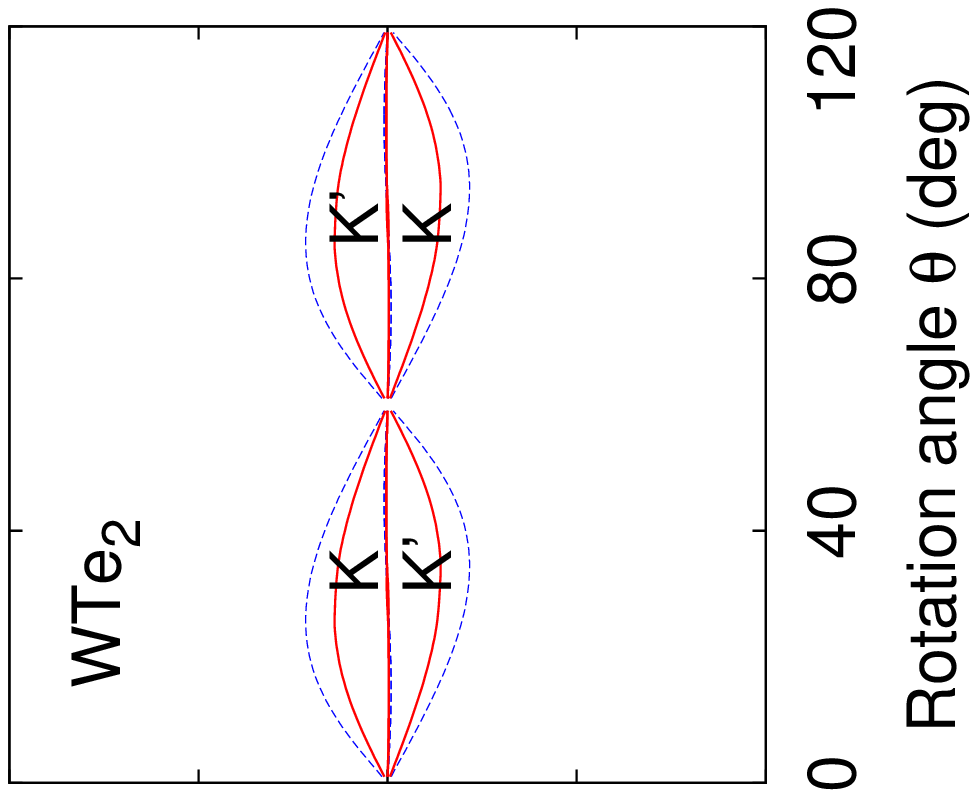}\hspace{6mm}
  \includegraphics[width=3.7cm, angle=-90]{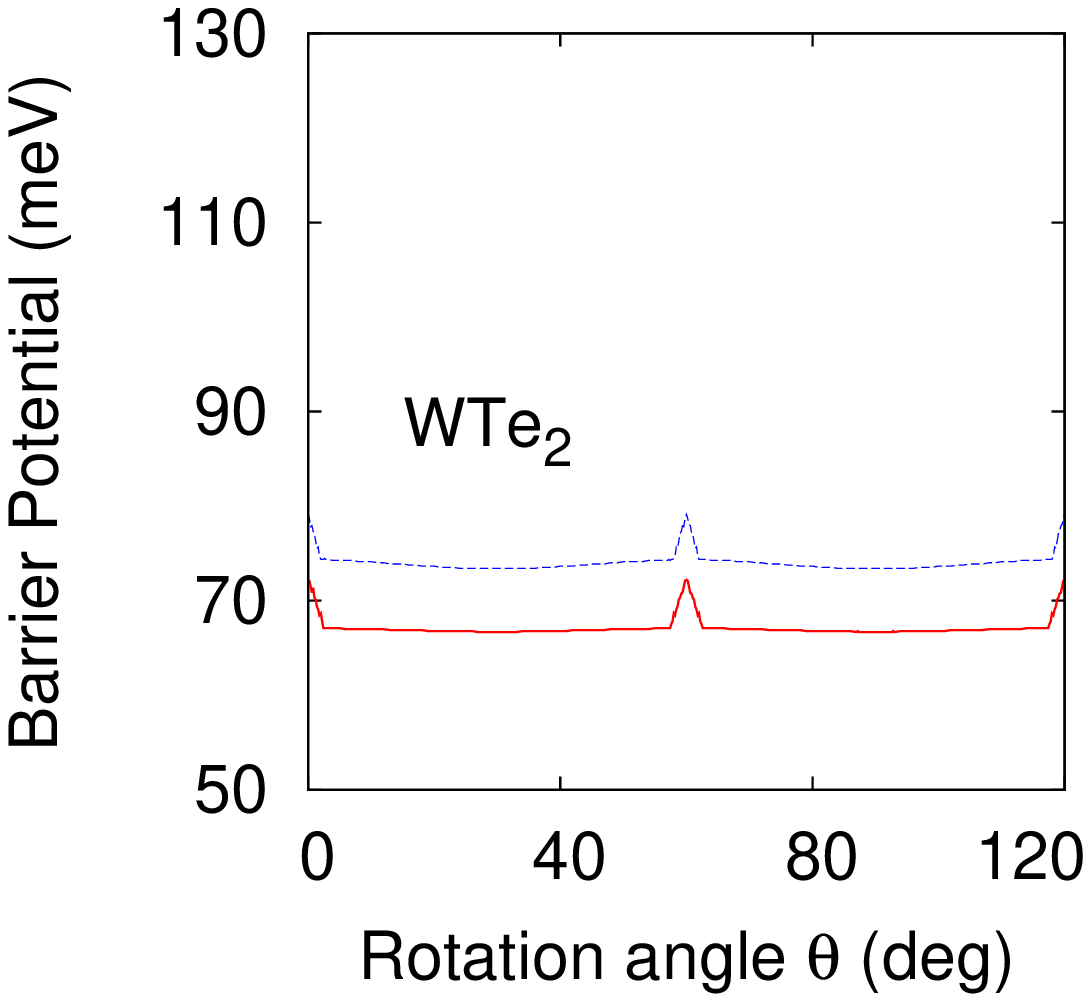}\hspace{0.5mm}
  \includegraphics[width=3.7cm, angle=-90]{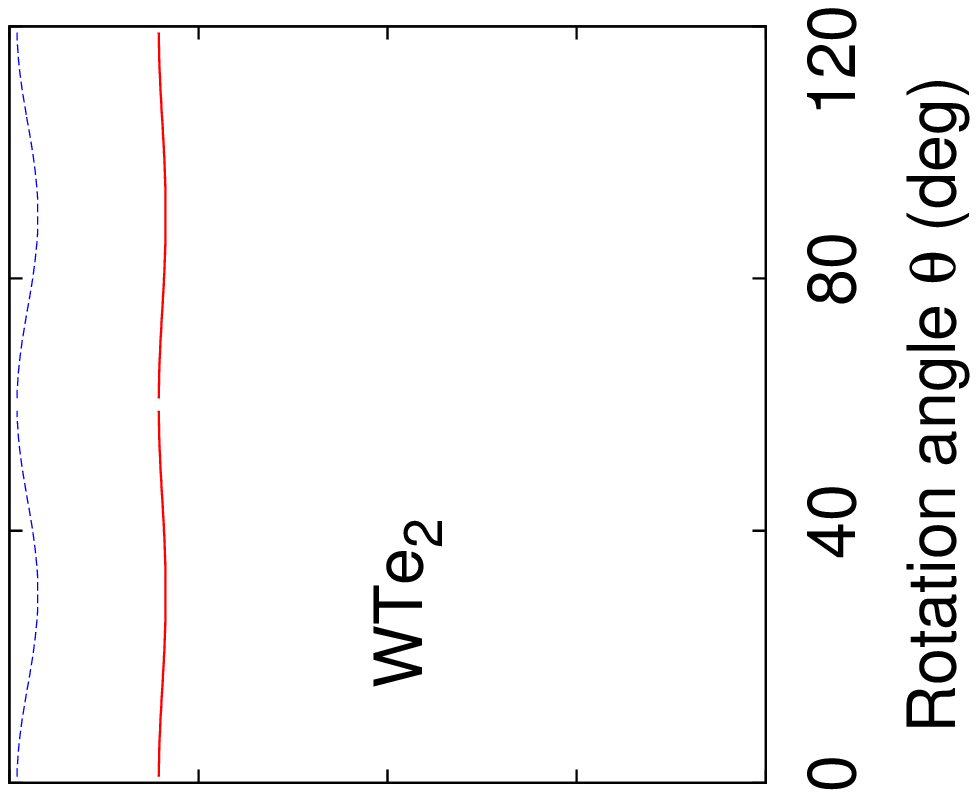}\\\vspace{7mm}
  \caption{(Color online). Left two columns: single valley regions in the
    semiconducting TMDs
    (left side, same valley; right side, different valleys). Transmission
    of holes occurs between the $K$ and $K'$ boundary lines in the indicated
    valleys. Right two columns: corresponding optimal barrier potentials
    (left side, same valley; right side, different valleys).}
  \label{allsvtfig}
\end{figure*}

Region widths obtained from the $\mathbf{k}\cdot\mathbf{p}$ and
\textit{ab-initio} tight binding Hamiltonians are compared at constant hole
density. The reason for working at constant density is that the density is
proportional to the area enclosed by a constant energy contour. So when the
comparison is done at constant density, differences in the region widths
may be attributed to differences in the shape of the contour. This allows
one to assess whether the $\mathbf{k}\cdot\mathbf{p}$ Hamiltonian
reproduces trigonal warping accurately.

The hole density is taken to be $3\times 10^{13}$ cm$^{-2}$ in the Mo
materials. However $E(\mathbf{k})$ varies more rapidly in the W materials 
so a significantly larger gate bias would be needed to achieve the same
hole density as in the Mo materials. For this reason the density is taken
to be $1.5\times 10^{13}$ cm$^{-2}$ in the W materials. The hole Fermi energies
at which these densities occur are given in Table~\ref{EnergyTable},
relative to the edge of the upper spin split valence band.

Fig.~\ref{allsvtfig} (left two columns) shows single valley region boundaries
for all six semiconducting TMD monolayers. The boundaries computed with the
$\mathbf{k}\cdot\mathbf{p}$ and \textit{ab-initio} tight binding
Hamiltonians typically agree to within 1.0-2.7$^\circ$ for all materials
except WTe$_2$. This suggests that the $\mathbf{k}\cdot\mathbf{p}$
Hamiltonian is reliable except in the case of WTe$_2$ so the transmission
coefficients for MoTe$_2$ shown in Fig.~\ref{tmdTfig} should also be
reliable. In addition, Fig.~\ref{allsvtfig} shows that the single valley
region widths are largest in MoTe$_2$ so, as stated section
\ref{Vamote2Section}, MoTe$_2$ is the most favorable TMD. This is
consistent with ref.~\cite{Habe15} in which the authors suggest the use
of MoTe$_2$ to observe spin-dependent refraction, an effect that also
depends on trigonal warping.

In the different-valley case, cut-offs occur as for BLG but only within
about $1^\circ$ of $\theta = 0$, $60$ and $120^\circ$. Hence single valley
transmission in different valleys occurs in a much wider $\theta$ range
than in BLG. The reason for this difference is that in TMDs the typical
radial size of the barrier contour relative to the size of the contact
contour is much smaller than in BLG. (For example, near the cut-off closest
to $\theta = 0^\circ$, the ratio of the barrier contour size to the contact
contour size on the positive $k_x$-axis is 0.017 in MoTe$_2$ and 0.27
in BLG.)  Hence in TMDs a larger rotation away from $\theta = 30^\circ$ or
$\theta = 90^\circ$ is needed to take the end point of a propagating part
around a corner of a contour. Thus the different-valley regions persist
over a wider $\theta$ range in TMDs, for the energies considered here.

Fig.~\ref{allsvtfig} (right two columns) shows the optimal potential barrier
heights used to compute the single valley region boundaries shown in
Fig.~\ref{allsvtfig}.  As in BLG, the potentials needed for single valley
regions in same and different valleys do not overlap and the potential has
to be adjusted to get single valley regions of large width for all
$\theta$. In addition, and as in BLG, the region widths calculated with a
fixed $\theta$-independent potential, equal to the mid-range optimal
potential are up to $\sim 20\%$ smaller than the optimized regions.

\subsection{Experimental Feasibility}
\label{TMDFeasibilitySection}

The necessary experimental conditions are the same as for BLG: the material
must be in the ballistic regime, the incident hole beam must be collimated
and gates are needed to set the hole density and provide a barrier
potential. The ballistic regime in monolayer TMDs has not yet been reached;
the current experimental situation is detailed in section
\ref{BallisticSection}. The other two conditions are probably close to
being satisfied. The MLG collimator \cite{Barnard17} simply consists of
suitably shaped gates deposited on hBN encapsulated graphene. There seems
to be no reason why similar gates should not be deposited on insulated
TMDs, although two top gates or two collimators may be needed as the
same-valley and different-valley cases occur at different angles of
incidence. The bottom and top gates, that are needed to control the hole
density and provide the barrier, resemble the gates used to make FETs and
TMD FETs have been fabricated. For example n-FETs have been made from
monolayer MoS$_2$ \cite{Baugher13,Radisavljevic13}, p-FETs from monolayer
WSe$_2$ \cite{Fang12} and ambipolar FETs from monolayer MoTe$_2$
\cite{Larentis17}.

However the question of whether the hole density of $3\times10^{13}$ cm$^{-2}$
used here can be achieved in MoTe$_2$ is open as the hole density in the
ambipolar MoTe$_2$ FET has not been reported. Typical carrier densities in
TMD FETs exceed about $10^{12}$ - $10^{13}$ cm$^{-2}$ and the hole density
used here is slightly less than the maximum electron
density reported in monolayer MoS$_2$ ($3.6\times 10^{13}$ cm$^{-2}$,
\cite{Radisavljevic13}). If this density cannot be achieved it would be possible to
use a lower hole density which would require a lower hole Fermi level.
However this would lead to reduced trigonal warping and narrower single
valley region widths and hence require an incident hole beam of narrower
width.

\section{Possible realization of a valley polarizer}
\label{PolariserSection}

In sections \ref{BLGSection} and \ref{TMDSection} we have shown that
transmission of carriers through potential barriers in BLG and TMDs is
valley asymmetric and single valley transmission occurs over a wide range
of incidence angles. In this section we suggest these effects can be used
to realize a valley polarizer.

We detail the minimum requirements for this device in section
\ref{RequirementSection}. Then in section \ref{AccuracySection} we examine
factors which may affect the operating temperature and the accuracy of
valley polarization. The maximum operating temperature is likely to be the
maximum temperature at which ballistic transport occurs
(\ref{BallisticSection}). Thermally excited minority carriers could affect
the polarization accuracy but only in BLG and their effect can be
suppressed by raising the back gate voltage(\ref{MinorityCarrierSection}).
Because of the thermal spread of energies in the incident beam, the
same-valley regime is most favorable for higher temperature operation
(\ref{SvtendepSection}). The effect of in-plane electric fields is
likely to be small (\ref{InplaneESection}).

\subsection{Minimum requirements for a valley polarizer}
\label{RequirementSection}

The main requirement is a collimated beam of carriers in the ballistic
regime. If a macroscopic contact was used instead of a collimator, it would
probably supply carriers with valley symmetric and equal probability at
each point on each energy contour. Then time reversal symmetry would ensure
that the conductance is valley symmetric. However a collimator operating in
the ballistic regime can be arranged to supply carriers only in the range of
velocities where single valley transmission occurs and thus make a valley
polarizer. The necessary collimator has been demonstrated in graphene
\cite{Barnard17} and its beamwidth is $18^\circ$, similar to the minimum
range widths in Figs.~\ref{svtfig} and \ref{tmdsvtfig}. Another requirement
is to dispose of the reflected carriers which are in the undesired valley
and could be backscattered from the edges of the 2D material and pass
through the barrier. This can be done by putting grounded electrodes at the
edges to absorb the undesired carriers. A similar absorber has been
demonstrated as a key part of the collimator in ref.~\cite{Barnard17}. The
ballistic regime has been reached in BLG
\cite{Cobaleda14,Nam17,Varlet14,Oka19} hence a BLG valley polarizer can be
realized from components that have been demonstrated. In TMDs, the hole
regime is experimentally accessible in monolayer MoTe$_2$ \cite{Larentis17}
but ballistic transport in this material has not yet been investigated.

\subsection{Factors affecting temperature of operation and polarization
  accuracy}
\label{AccuracySection}

\subsubsection{Ballistic Transport}
\label{BallisticSection}

Ballistic transport in BLG at low temperature is well established experimentally
\cite{Cobaleda14,Nam17,Varlet14,Oka19} but the maximum temperature for
ballistic transport is not known. The authors of ref. \cite{Cobaleda14}
investigated the temperature dependence of transport in hBN encapsulated
BLG and found that ballistic transport occurs above a
temperature-independent critical carrier density of $2.5\times 10^{11}$
cm$^{-2}$ up to 50 K, the maximum temperature used in the experiment. The
authors of ref. \cite{Nam17} investigated transport in suspended BLG and
found that ballistic transport occurs above a temperature-dependent critical
density. The maximum experimental temperature was 70 K and the
corresponding critical density is $\sim 2\times 10^{11}$ cm$^{-2}$. Hence
the available experimental evidence suggests that ballistic transport in BLG
occurs at least up to $\sim 50$ - $70$ K but further work is needed to
determine the upper limit.

In the case of TMDs ballistic transport has been investigated only for
electrons in MoS$_2$ \cite{English16}. The authors of this work observed
the onset of ballistic transport at a device temperature of 175 K and
suggested that the ballistic limit can be achieved. As the electron and
hole masses are similar ($\sim 0.5$) in all the
semiconducting TMDs \cite{Kormanyos15}, it is possible that ballistic transport of holes can be
achieved. However there is no relevant data and further
experimental investigations are needed.

In summary, the temperature dependence of ballistic transport may limit the
maximum operating temperature of a valley polarizer but there is
insufficient experimental evidence to estimate this temperature.

\subsubsection{Minority Carriers}
\label{MinorityCarrierSection}

Thermally excited minority carriers in the contacts could affect the valley
polarization but the physics is different in BLG and TMDs.

In the case of BLG and the device model in \ref{BLGVSection}, the
electron Fermi level is 56 meV and the layer potentials in the contacts are
$\pm 14$ meV.  The physics depends on the alignment of the bands in the
contacts and underneath the top gate. From Fig.~\ref{optvfig} it can be
seen that for almost all $\theta$, the layer 1 potential under the top gate
is $> 14$ meV and the layer 2 potential is $> -14$ meV. Hence the top gate
generates a barrier for electrons and a well for holes. This means that
thermally excited holes in both valleys could flow underneath the top gate,
leading to a reduction in valley polarization.

The magnitude of this effect depends on the thermal distribution of the holes.
The Fermi function is equal to 0.01 when $E - E_F \sim 4.6
k_B T$, where $E_F$ is the Fermi level, $T$ is the absolute temperature
and $k_B$ is Boltzmann's constant. This condition should give a rough
approximation to the temperature at which the valley polarization is
affected by a few \%. For the device model in section \ref{BLGVSection},
the energy needed to create a hole is 70 meV and the corresponding
temperature is $\sim 177$ K.

It should be possible to reduce the effect of the holes by increasing the
back gate voltage. In the device model detailed in section
\ref{BLGVSection}, the electron Fermi level becomes 105 meV if the
back gate voltage is raised to 4000 mV, and the layer potentials in the
contact become $\pm 31$ meV. Then the hole creation energy increases to 136
meV and the corresponding temperature is $\sim 343$ K. This shows it should
be possible to overcome the effects of holes in BLG with a suitable device
design.

In the case of TMDs, the band gap exceeds 1 eV so thermal excitation of
carriers across the gap is unlikely to be significant at room temperature
and beyond. However the effect of excitation across the spin split valence
bands needs to be considered.

The holes in both of the spin split bands are subjected to the same potential
barrier. In addition, $E(\mathbf{k})$ is similar for both spins. Hence the
single valley regions for both spins are similar. Consequently the valley
polarization should not be affected by minority spin holes. However the
spin polarization could be affected.

Minority spin holes can be transmitted through the barrier only if their
energy relative to the bottom of the minority spin band exceeds the barrier
height. Creation of holes of this energy requires a thermal excitation of
energy $2|\lambda| - E_F + V$ where $E_F$ is the hole Fermi energy in the
majority spin band and $V$ is the barrier height. With $E_F = 116.9$ meV
and $V = 66.55$ meV as for Fig.~\ref{tmdTfig}, this gives an energy
of 164.65 meV and the corresponding temperature, obtained with the same
criterion as for BLG, is 415 K.

In summary, minority carriers are unlikely to affect the spin and valley
polarizations in TMDs and their effect on the valley polarization in BLG
can be suppressed by increasing the back gate voltage.

\subsubsection{Thermal Spread of Energies in Incident Beam}
\label{SvtendepSection}

\begin{figure}
\begin{center}  
  \includegraphics[width=4.4cm, angle=-90]{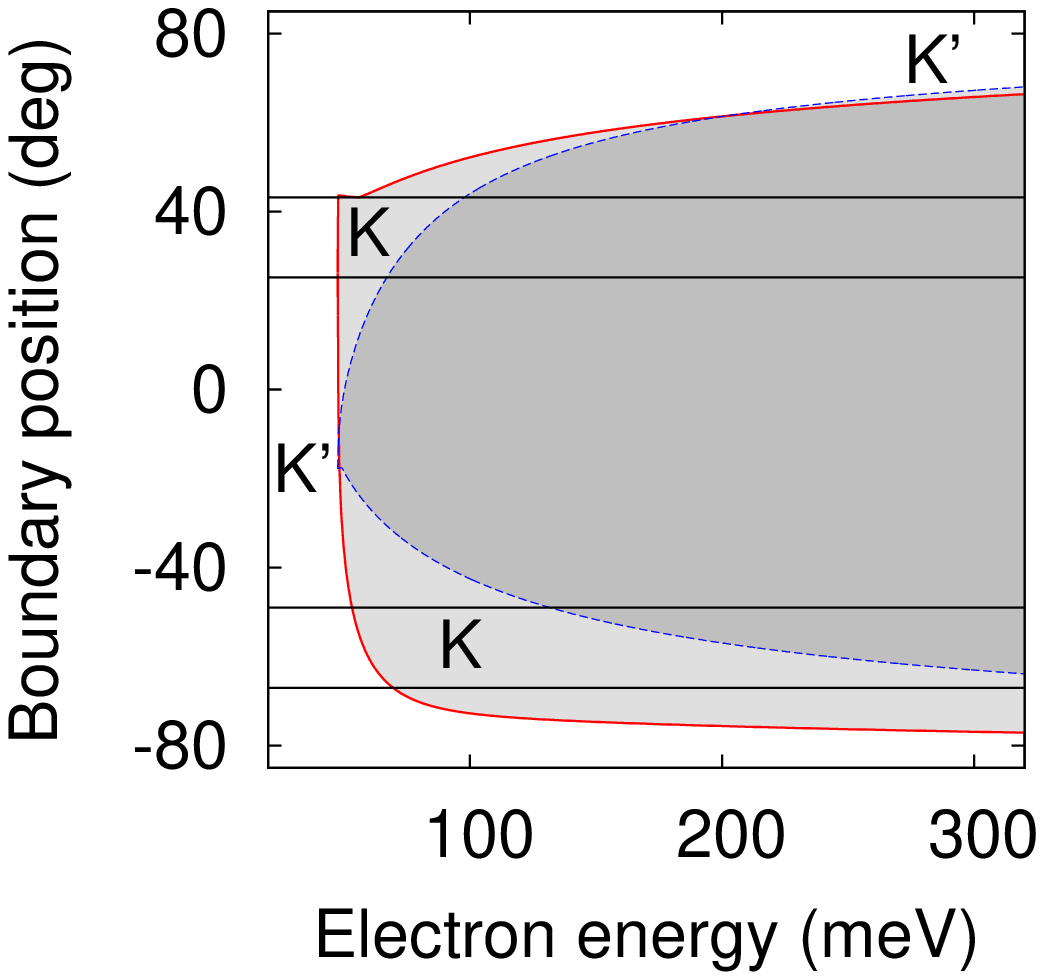}\hspace{1mm}
  \includegraphics[width=4.4cm, angle=-90]{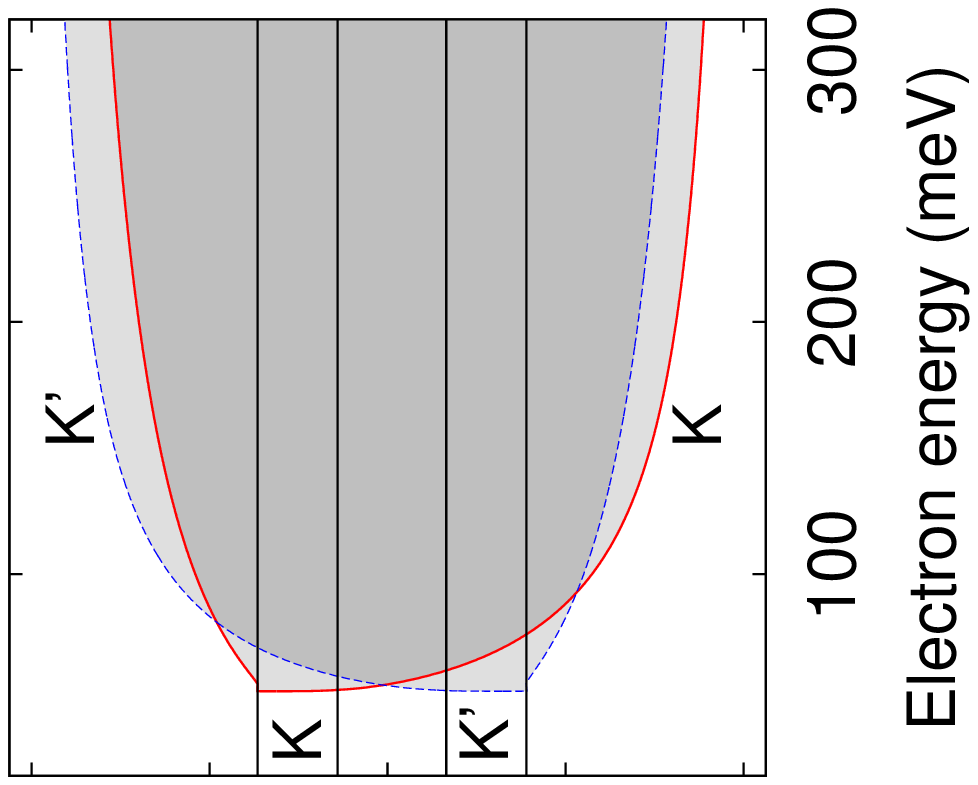}
  \caption{(Color online). Energy dependence of single valley regions in BLG.
    The potentials and top gate width are as for Fig.~\ref{bgTfig}.
    Solid red lines: $K$ transmission boundaries, dashed blue lines: $K'$
    boundaries. Horizontal lines indicate the beam extent. Light fill:
    single valley transmission, dark fill: two valley transmission.
    Left: $\theta = 17^\circ$ (same-valley case at $E=56$ meV).
    Right: $\theta = 31^\circ$ (different-valley case at $E=56$ meV).}
\label{bgendepfig}
\end{center}
\end{figure}

\begin{figure}
\begin{center}  
  \includegraphics[width=4.4cm, angle=-90]{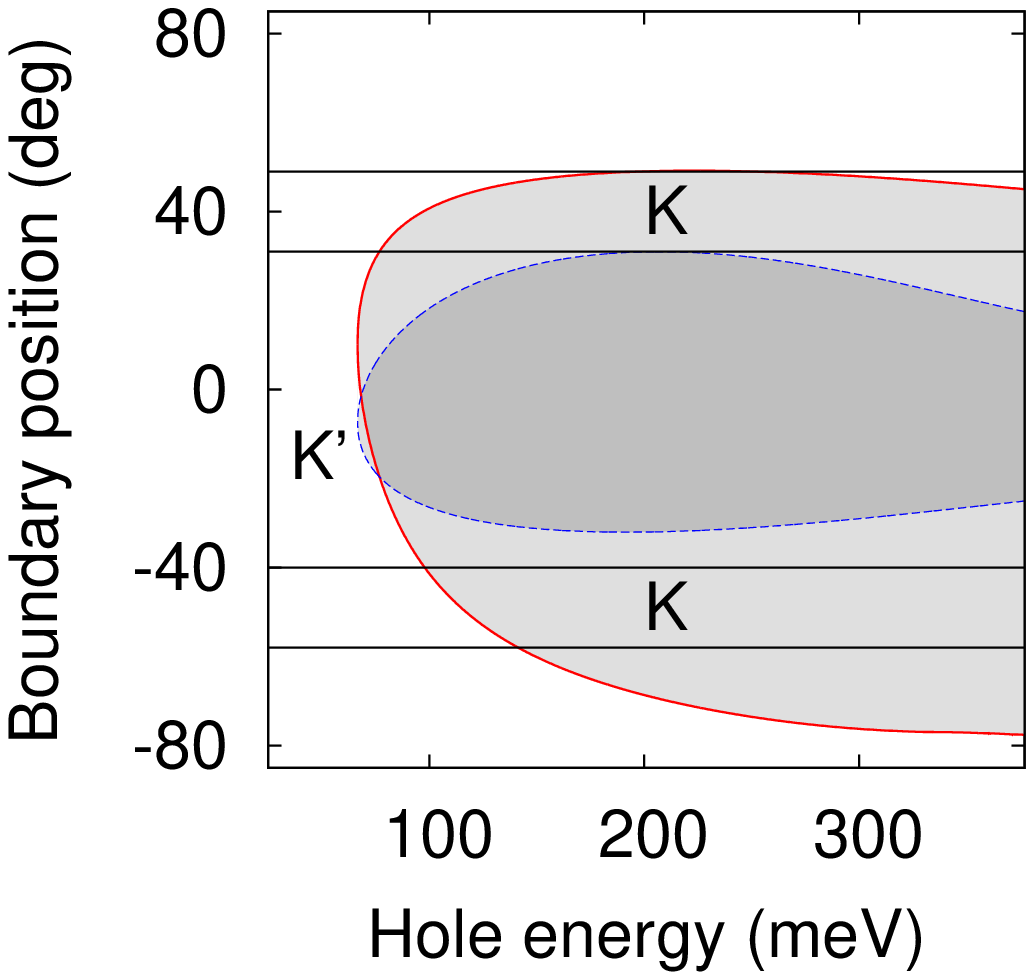}\hspace{1mm}
  \includegraphics[width=4.4cm, angle=-90]{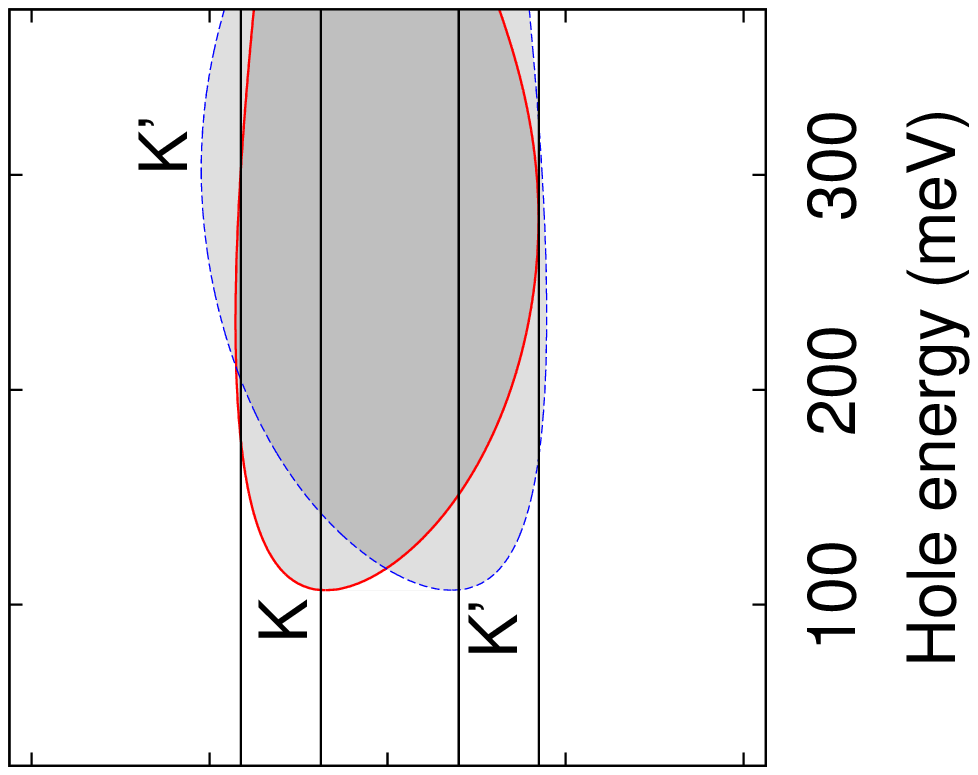}
  \caption{(Color online). Energy dependence of single valley regions in
    MoTe$_2$. The potentials and top gate width are as for Fig.~\ref{tmdTfig}.
    Solid red lines: $K$ transmission boundaries, dashed blue lines: $K'$
    boundaries. Horizontal lines indicate the beam width. Light fill:
    single valley transmission, dark fill: two valley transmission.
    Left: $\theta = 17^\circ$ (same-valley case at $E=116.9$ meV).
    Right: $\theta = 31^\circ$ (different-valley case at $E=116.9$ meV).}
\label{tmdendepfig}
\end{center}
\end{figure}

The single valley regions depend significantly on energy. Therefore at
finite temperature the spread of energies in the incident beam could affect
the valley polarization. To investigate this, the single valley region
boundaries are computed as a function of energy at $\theta = 17^\circ$ and
$31^\circ$. Figs.~\ref{bgTfig} and \ref{tmdTfig} show that same-valley
transmission occurs at $\theta = 17^\circ$ when $E=56$ meV in BLG and
$E=116.9$ meV in MoTe$_2$ while different-valley transmission occurs at the
same energies when $\theta = 31^\circ$. However Figs.~\ref{bgendepfig}
(BLG) and~\ref{tmdendepfig} (MoTe$_2$) confirm that the form of the
transmission is energy-dependent.

In both materials there is a threshold energy equal to the
barrier height. If $\phi_c$ is fixed and only the energy is varied, then
in BLG at both angles and MoTe$_2$ at $31^\circ$, there is a critical
energy where single valley transmission changes to transmission
in both valleys (light fill changes to dark fill when the energy increases).
This limits the maximum operating temperature.

To quantify this, a beam of width $18^\circ$ is indicated by the parallel,
horizontal lines in the figures. Each pair of lines is centered on the angle
that makes the threshold energy approximately equal to the Fermi energy, 56
meV for electrons in BLG and 116.9 meV for holes in MoTe$_2$. With this
choice, carriers whose energy is significantly less than the Fermi energy
are below the first threshold and are not transmitted. Then the maximum
operating temperature is determined by the carrier population above the
critical energy.

For example, in BLG at $31^\circ$ the critical energy at $\phi_c
= -13.2^\circ$ is 61.7 meV and with the criterion used in
section \ref{MinorityCarrierSection} this corresponds to a temperature of
14.4 K. For the $\phi_c = 11.2^\circ$ line the temperature is 8.8 K. In
MoTe$_2$ the equivalent temperatures are 86.8 K and 64.6 K. This suggests
that the regime where single valley transmission occurs in different
valleys at positive and negative incidence is not very suitable for high
temperature operation.

The regime where the single valley transmission occurs in the same valley
is much more suitable. In BLG at $\theta = 17^\circ$, the critical energy
at $\phi_c = -49^\circ$ corresponds to a temperature of 190 K. However the
critical energy at
$\phi_c = 25.2^\circ$ corresponds to 27.5 K and generally in BLG the second
threshold only occurs at high energy in one of the single valley
regions. In MoTe$_2$ at $\theta = 17^\circ$ there is no crossover to
transmission in both valleys up to an energy of at least 377.5 meV. This
is very favorable for high temperature operation. The physical reason for
the different behavior of
BLG and MoTe$_2$ is that in the energy range considered here, trigonal
warping weakens with energy in BLG but strengthens with energy in TMDs.

In summary, the regime where single valley transmission occurs in the same
valley at both positive and negative angles of incidence is very favorable
for high temperature operation. However in BLG this is the case for only
one of the single valley regions. Which one it is depends on $\theta$ as
consequence of Eq.~(\ref{symrel2}).

\subsubsection{In-plane Electric Fields}
\label{InplaneESection}

In-plane electric fields should deflect a collimated carrier beam and change
the angle of incidence. This could cause loss of polarization if the
incident beam is shifted away from a single valley region into a two
valley region. However we estimate that this effect is likely to be small.

The magnitude of the effect depends on the experimental voltages and device
dimensions. The Fabry-Perot interference experiments described in ref.
\cite{Varlet14} were done with a source-drain bias of around 1 mV over a
distance of around 1-3 $\mu$m while the collimation experiments described
in ref. \cite{Barnard17} probably involved smaller fields. Hence
1000 Vm$^{-1}$ is taken to be an upper limit to the in-plane field.

To estimate the deflection, the field is taken to be normal to the barrier
and classical trajectories for a charged particle with energy-momentum
relation $E(\mathbf{p}/\hbar)$ are computed for each valley, where
$E(\mathbf{k})$ is the band energy. The results show that the incident beam
can undergo a small deflection towards the two valley region. The
deflection angle depends on $\theta$ but is only $\sim 0.1$ - $0.2^\circ$
for BLG and only $\sim 0.05$ - $0.14^\circ$ for MoTe$_2$. This is small
compared to single valley region widths and suggests the effect of in-plane
electric fields will be small under typical experimental conditions.
 
\section{Detection of valley polarization}
\label{DetectionSection}

Valley polarization can be detected via the valley Hall effect
\cite{Xiao07,Sui15,Shimazaki15} and it has been suggested that two valley
polarizers of opposite polarity can block current \cite{Rycerz07}. When the
polarizers are made from barriers, the blockage is exact
because of the symmetry relation, Eq.~(\ref{symrel1}), between the transmission 
coefficients of two barriers with a relative rotation angle of $\pi / 3$.

\begin{figure}
\begin{center}  
  \includegraphics[width=8cm, angle=0]{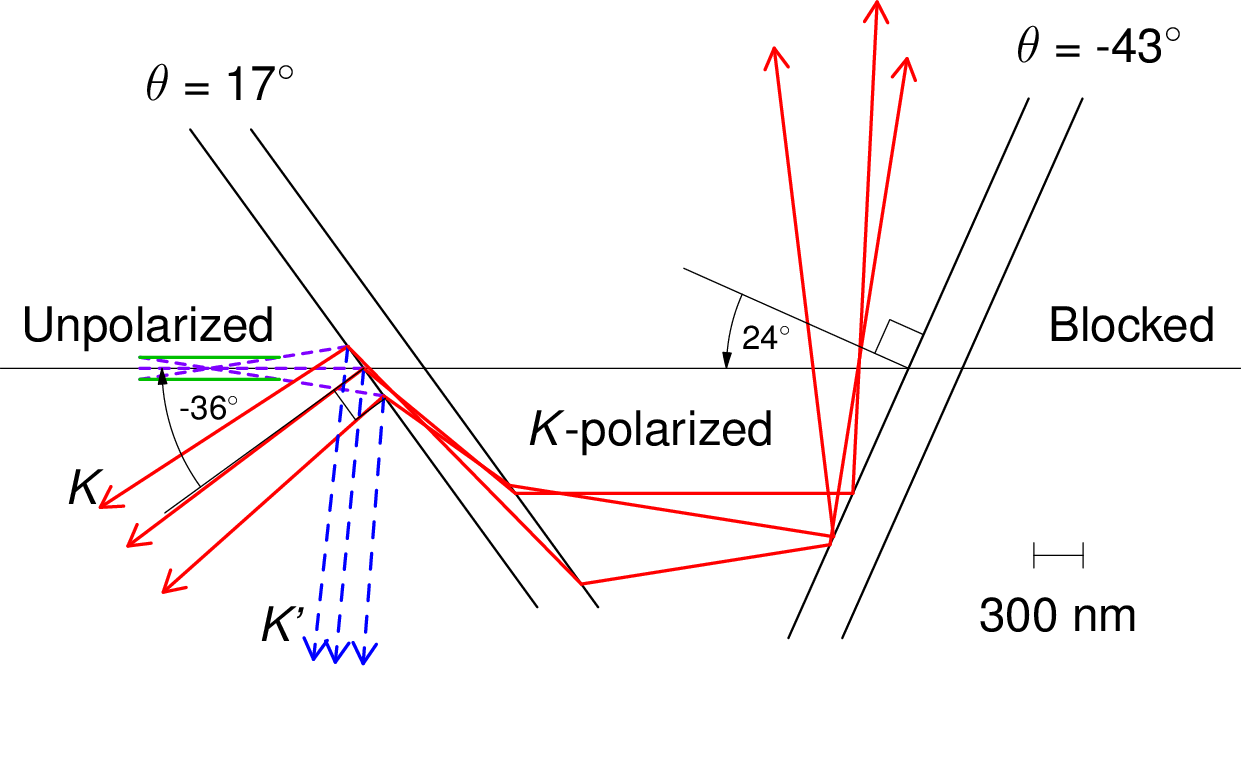}
  \caption{(Color online). Example of current blocking by two barriers.
    Sloping lines indicate barrier edges, the horizontal feint line is the
    optic axis and the short parallel lines (green) represent the collimator.
    Other lines show the current directions at the center and edges of each
    electron beam. Short dashed lines (purple): unpolarized incident
    current; solid lines (red): $K$ polarized; long dashed lines (blue):
    $K'$ polarized. For clarity, current paths that enter the collimator
    from the left are not shown. Beam center reflection coefficients at
    the first barrier: 100\% in $K'$ and 5.5\% in $K$.}
\label{twobarfig}
\end{center}
\end{figure}

This relation allows a polarization detector to be made from two identical
and inversion symmetric barriers in series, with a relative rotation of
$\pm \pi/3$ (Fig.~\ref{twobarfig}). When the two single valley regions are
in the same valley at positive and negative $\phi_c$, as in
Fig.~\ref{bgTfig} (left), Eq.~(\ref{symrel1}) guarantees that the second
barrier transmits in the opposite valley to the first barrier. Hence the
barrier pair blocks current and can be used like a pair of Polaroid filters
to demonstrate valley polarization.

In practice, this requires that two more conditions are satisfied. The first
is that the current that is transmitted through the first barrier should be
incident on the second barrier at an angle within the single valley range for
that barrier. As the two barriers must have a relative rotation of $\pm
60^\circ$ to swap the valleys, the angles of incidence on the two barriers
differ by $\mp 60^\circ$. To satisfy this condition, the 
angles of incidence in the two single valley regions should differ by about
$60^\circ$. This is the case only in part of the $\theta$ range.

The second condition is that the reflected current from the front edge of
the second barrier is not incident on the back edge of the first
barrier. If this condition is not satisfied, multiple reflection between
the two barriers could occur and this could change the transmission
characteristics of the barrier combination. This can be prevented by
adjusting the barrier lengths and separation so that the current reflected
from the second barrier does not reenter the first barrier.

To demonstrate that the two conditions can be satisfied, ray tracing is
used to compute the current paths through the two barriers for the case of
BLG and $\theta = 17^\circ$. The angles of reflection and refraction are
obtained from the BLG band structure and the incident beam width is taken
to be $18^\circ$. The current paths are shown in Fig.~\ref{twobarfig} and
it is clear that the two conditions are satisfied. The angles of incidence
on each barrier fall within the single valley ranges as can be checked by
looking at Figs.~\ref{bgTfig} and \ref{svtfig}. In addition, the current
reflected from the second barrier clearly passes out of the region between
the barriers. This suggests that the current blocking is experimentally
observable, at least at one value of $\theta$.

The full $\theta$ range in which current blocking should be observable is
probably somewhat smaller than the $\theta$ range of the same valley
regions (Fig.~\ref{svtfig}, left). These regions become narrower and
vanish as $\theta$ approaches $30^\circ \pmod{120^\circ} $ and
$90^\circ \pmod{120^\circ}$. A significant fraction of the $\theta$
range should still be available but how much depends on the experimental
conditions and extensive ray tracing calculations for the full range of
$\theta$ angles and beam widths would be needed to determine this.

The current paths in Fig.~\ref{twobarfig} differ qualitatively from the
paths of optical rays passing through a refractive medium. In particular,
the order of the paths reverses at the first barrier, for example the top
path on the entrance side becomes the bottom path on the exit side. The
reason is that the $\mathbf{k}$-vectors of the states involved are by
chance close to points of inflection on the barrier energy contour. Between
the points of inflection, $\phi_v$ increases when $\phi_k$ decreases (see
Fig.~\ref{janglefig} (lower) for an example) and this leads to the reversed
order. Another difference is that the angle of reflection is not equal to
the angle of incidence when the energy contours are warped. This has a
significant effect on the current paths reflected from the second barrier.

\section{Discussion}
\label{DiscussionSection}

Valley asymmetric transmission through a potential barrier in BLG and TMDs
inevitably occurs because of the low symmetry of the total
Hamiltonian. However it may be necessary to use additional physics to make
the valley asymmetry large. We have suggested the use of total external
reflection but this is not the only approach. For example, the valley
asymmetry is enhanced in barriers with broken inversion symmetry
\cite{Maksym18}.

The large valley asymmetry found in this work occurs because trigonal
warping leads to a large difference in the critical angles for total
external reflection in the two valleys. This results in single valley
transmission over a wide range of incidence angles and enables a valley
polarized incident current to be split into reflected and transmitted
currents with opposite valley polarization.  A valley polarizer can be
realized in BLG by arranging for a collimated beam of carriers to be
incident in one of the single valley regions. The same arrangement in TMDs
forms a spin and valley polarizer. The barrier potential can be adjusted to
ensure that single valley region widths are similar to or exceed the beam
width of a MLG electron collimator that has already been fabricated
\cite{Barnard17}.  In addition, we have shown that the transmitted valley
swaps when a barrier is rotated by $\pm\pi / 3^\circ$ with respect to the
crystallographic axes. This allows two barriers with a relative rotation of
$\pm\pi / 3^\circ$ to be used like Polaroid filters to demonstrate valley
polarization.

Our investigations show that the proposed valley polarizer appears to be
experimentally feasible and should have some advantages. First, the
polarizer is relatively immune to the crystallographic orientation of the
barrier because the top gate voltage can be adjusted to optimize the single
valley region widths. Secondly, the current on the exit side flows only in
the desired valley so there is no need for additional components to collect
the desired current stream. However some uncertainty about the feasibility
remains because the trigonal warping parameters in BLG are not known
reliably and accurate experimental values are desirable. Experimental
studies to determine the conditions for ballistic transport in BLG and
TMDs, particularly the temperature range, are also needed. Further
theoretical work should await these experimental developments.

The inevitability of valley asymmetry is expected to be relevant to other
applications and materials. It may be possible to use switchable pairs of
spin filters to inject spin-polarized holes into a TMD pn-junction and
hence make a polarized light emitting diode with electrically controllable
photon polarization. In addition, the strong $\theta$ dependence of the
transmission may be useful for determining the crystallographic orientation
of the 2D material. Beyond BLG and TMDs, the total Hamiltonian of any 2D
material in the presence of a gate should have low symmetry and
transmission through a gate-induced barrier should be strongly
$\theta$-dependent when the constant energy contours are not
circular. Further afield, the present work may be relevant to valley
photonic metamaterials \cite{Garcia08,Dong17}.

\begin{acknowledgments}
We thank M. Tanaka, Y. Shimazaki, I. V. Borzenets, M. Yamamoto, S. Tarucha
and A. Slobodeniuk for very useful discussions. PAM thanks
Prof. S. Tsuneyuki for hospitality at the Department of Physics, University
of Tokyo. The computations were done on the ALICE high performance
computing facility at the University of Leicester. This work was supported
by the ImPACT Program of the Council for Science, Technology and
Innovation, Cabinet Office, Government of Japan, Grant No. 2015-PM12-05-01,
and JSPS KAKENHI Grant Nos. JP25107005 and JP17H06138.
\end{acknowledgments}


\begin{thebibliography}{99}
\bibitem{StateNote} Outside this range four propagating states or four
  evanescent states may occur.
\end{thebibliography}
\end{document}